\documentclass[12pt]{article}
\usepackage{graphicx}
\usepackage{fullpage}
\begin{document}

\title {Ferromagnet - Superconductor Hybrids}
\author{I. F. Lyuksyutov and V. L. Pokrovsky\\
Department of Physics, Texas A\&M University}
\date{\today}

\maketitle

\begin{abstract}
 A new class of phenomena discussed in this review is based on interaction
between spatially separated, but closely located ferromagnets and
superconductors. They are called Ferromagnet-Superconductor
Hybrids (FSH). These systems include coupled smooth and textured
Ferromagnetic and Superconducting films, magnetic dots, wires etc.
The interaction may be provided by the magnetic flux from magnetic
textures and supercurrents. The magnetic flux from magnetic
textures or topological defects can pin vortices or create them,
changing drastically the properties of the superconductor. On the
other hand, the magnetic field from supercurrents (vortices)
strongly interacts with the magnetic subsystem leading to
formation of coupled magnetic-superconducting topological defects.
We discuss possible experimental realization of the FSH. The presence
of ferromagnetic layer can change dramatically the properties of
the superconducting film due to proximity effect. We discuss
experimental and theoretical studies of the proximity effect in
the FSH including transition temperature, order parameter
oscillations and triplet superconductivity.

\end{abstract}
\maketitle

\newpage

\tableofcontents

\newpage

\section{Introduction}

\indent In this review we discuss a new avenue  in solid state physics:
studies of physical phenomena which appear when two mutually
exclusive  states of matter, superconductivity and ferromagnetism,
are combined in an unified Ferromagnet-Superconductor Hybrid (FSH)
system.
In the hybrid systems fabricated from materials with different and
even mutually exclusive properties, a strong mutual interaction
between subsystems can dramatically change properties of the
constituent materials. This approach offers vast opportunities  for
science and technology.
The interplay  of superconductivity and ferromagnetism has been
thoroughly studied experimentally and  theoretically
\cite{121,bul1} for {\sf homogeneous} systems. In such  systems,
both order parameters are homogeneous in space and suppress each
other. As a result, one or both the orderings are weak.
 A natural way to avoid the mutual suppression of
the order parameter of the superconducting (S) and ferromagnetic
(F) subsystems is to separate 
them by a thin but impenetrable
insulator film. In such systems the S and F subsystems interact
via magnetic field induced by the nonuniform magnetization of the
F textures penetrating into the superconductor. If this field is
strong enough, it can generate vortices in the superconductor. The
textures can be either artificial (dots, wires) or topological
like Domain Walls (DW). The inverse effect is also important: the
S currents generate magnetic field interacting with the
magnetization in F subsystem.

First experimental works on FSH were focused on  pinning properties of
magnetic dot arrays covered by a thin superconducting film
\cite{parallel,parallel1,parallel2,shull,ketterson}. The effect of
commensurability on the transport properties was reported in
\cite{parallel,parallel1,parallel2,shull}. This effect is not specific for
magnets interacting with superconductors and was first  observed
in textured superconducting films.
First experiments with such films were performed in
seventies.  In these
experiments the periodicity of the vortex lattice fixed by
external magnetic field competed with the periodicity of an
artificial array created by experimenters.
Martinoli {\it et al.} \cite{martinoli1,martinoli2,martinoli} 
used grooves and  
Hebard  {\it et al.} \cite{hebard1,hebard2} used arrays of holes.
This approach was further developed by experimentalists 
in nineties \cite{metlu}-\cite{a24}.  
Theoretical analysis was also performed in the last century
\cite{pt,pt1,beck}.
First observation of the dependence of vortex pinning
by magnetic dots array on the magnetic field direction
was presented by  Morgan and Ketterson \cite{ketterson}.
This was first direct indication of new physics in FSH.
New insight into the FSH physics has been provided by 
Magnetic Force Microscope (MFM) and Scanning Hall Probe Microscope (SHPM).
By using such imaging technique the group at the University 
of Leuven has elucidated several pinning mechanisms 
in FSH \cite{mosch}-\cite{mosch2}. 

Different mesoscopic
magneto-superconducting systems were proposed and studied
theoretically: arrays of magnetic dots on the top of a SC film
\cite{dotlp,spie1,spie,rdot}, 
Ferromagnet-Superconductor Bilayer (FSB) 
\cite{spie,vmlp,ELPV,EKLP,PW,KP}, embedded
magnetic nanowires combined with bulk superconductor 
\cite{ln,ln2} or superconductor film \cite{ln3,ln4}, a layer of
magnetic dipoles between two bulk superconductors \cite{frey}, an
array of magnetic dipoles mimicking the FM dots on SC film
\cite{hwa},
``giant'' magnetic dot which generates several vortices in bulk
superconductor \cite{marmorkos}, 
single magnetic dots on a thin superconducting
film \cite{peeters1,peeters2,peeters3,kayali,erdin}, 
thick magnetic film combined with thick 
\cite{bulchud,sonin1,sonin2,sonin3}  or thin superconducting
film \cite{helseth,aladyshkin}.

The characteristic scale scale of the magnetic field and current
variation in all mentioned systems significantly exceeds the
coherence length $\xi$. It means that they can be considered in
London approximation with good precision. In the next section we
derive basic equations describing FSH. Starting from
London-Maxwell equations, we derive a variational principle
(energy) containing only the values inside either S or F
components. These equations allowed us to study single
magnetic dots coupled with superconducting film (Sec. \ref{dot}) as
well as arrays of such dots (Section \ref{dots}). The simplest possible
FSH system - sandwich formed by Ferromagnetic and Superconducting
layers, divided by ultrathin insulating film ( FSB),- 
can demonstrate unusual behavior: spontaneous formation 
of coupled system of vortices and
magnetic domains. These phenomena are discussed in Section \ref{fsb}. We
also discuss the influence  of the thick magnetic film on the bulk
superconductor.

 An alternative approach to heterogeneous SC/FM systems is just to
employ the proximity effects instead of avoiding them. The
exchange field existing in the ferromagnet splits the Fermi
spheres for up and down spins. Thus, the Cooper pair acquires a
non-zero total momentum and its wave function oscillates in space.
This effect first predicted by Larkin and Ovchinnikov \cite{LO}
and by Ferrel and Fulde \cite{FF} will be cited further as LOFF
effect. One of its manifestation is the change of sign of the
Cooper pair tunneling amplitude in space. At some conditions the
Josephson current through a
superconductor-ferromagnet-superconductor (S/F/S) junction has 
sign opposite to $\sin\varphi$, where $\varphi$ is the phase
difference between right and left superconducting layers. This
type of junctions was first proposed theoretically long time ago 
by Bulaevsky {\it et al.} \cite{BulKuz},
\cite{BulBuz}  and was called $\pi$-junction in contrast to
standard or 0-junction. It was first
reliably realized in the experiment by Ryazanov and coworkers in 
2001 \cite{ryazanov01-1, ryazanov01-2}
and a little later by Kontos {\it et al.} \cite{kontos}. The 
experimental findings of these groups have generated an extended 
literature. A large exhausting review on this  topic was published
in the beginning of 2002 \cite{izyumov}. A more special survey was
published at the same time by Garifullin \cite{garifullin-review}.
We are not going to repeat what was already done in this reviews 
and will focus presumably on works which appeared after its
publication. Only basic notions and ideas necessary for
understanding will be extracted from previous works.

 Most of the proximity phenomena predicted theoretically
and found experimentally are based on the oscillatory behavior of
the Cooper pair wave function. These are the oscillations of the
transition temperature (first predicted in \cite{BuzKup,
radovic1}), and the critical current  vs. the
thickness of ferromagnetic layer which are seen as oscillatory
transitions from 0- to $\pi$-junctions \cite{BulBuz}. Other
proximity effects besides the usual suppression of the order
parameters include the preferential antiparallel orientation of
the F-layers in a F/S/F trilayer, the so-called spin-valve effect
\cite{deMelo, BuzVed, tagirov1}.

 More recently a new idea was proposed by Kadigrobov 
{\it et al.} \cite{KSJ} and by Bergeret, Efetov and Volkov
\cite{BEV1}: they have predicted that the magnetization varying
its direction in space transforms singlet Cooper pairs into
triplet ones. The triplet pairing is not suppressed by the
exchange field and can propagate in the ferromagnet on large
distances thus providing the long-range proximity between
superconductors in S/F/S junctions.

 The proximity effects may have technological applications as elements
of high-speed magnetic electronics based on the spin valve action
\cite{tagirov1} and also as elements of quantum computers
\cite{feigelman}. Purely magnetic interaction between
ferromagnetic and superconducting subsystems can also be used to
design magnetic field controlled superconducting devices.
A magnetic field controlled
Josephson interferometer in a thin magnetic F/S bilayer has been
demonstrated by Eom and Johnson \cite{18}.

In the next Section we derive basic equations. Third Section is 
focused on phenomena in FSH which are based on only magnetic interaction
between ferromagnetic and superconducting subsystem. Recent
results on proximity based phenomena in bi- and tri-layer
FSH are presented in the last Section.

\section{Basic Equations \label{basic}}

In the proposed and experimentally realized FSH a magnetic texture
interacts with the supercurrent. First we assume that
ferromagnetic and superconducting subsystems are separated by thin
insulating layer which prevents proximity effect, focusing on
magnetic interaction only. Inhomogeneous magnetization generates
magnetic field outside the ferromagnets. This magnetic field
generates screening currents in superconductors which, in turn,
change the magnetic field. The problem  must be solved
self-consistently. The calculation of the vortex and magnetization
arrangement for interacting, spatially separated superconductors
and ferromagnets is based on the static London-Maxwell equations
and corresponding energy. This description includes possible
superconducting vortices. Londons approximation works satisfactory
since the sizes of all structures in the problem exceed
significantly the coherence length $\xi$. We remind that in the
Londons approximation the modulus of the order parameter is
constant and the phase varies in space. Starting from the
London-Maxwell equation in all the space, we eliminate the
magnetic field outside their sources and obtain equations for the
currents, magnetization and fields inside them. This is done in
the subsection 2.1. In the subsection 2.2 we apply this method to
the case of very thin coupled ferromagnetic and superconducting
films. When proximity effects dominate, the Londons approximation
is invalid. The basic equations for this case will be described in
subsection 2.3.

\subsection{Three-Dimensional Systems.}
The total energy of a stationary F-S system reads: 

\begin{equation}
H  = \int \bigl [\frac{{\bf B}^2}{8 \pi}  + \frac{m_s n_s {\bf
v}_{s}^{2}}{2} - {\bf B}\cdot {\bf M} \bigr ]dV \label{en}
\end{equation}
\noindent where ${\bf B}$ is the magnetic induction,  ${\bf M}$ is
the magnetization, $n_s$ is the density of S-electrons, $m_s$ is
their effective mass and ${\bf v}_s$ is their velocity. We assume
that the SC density $n_s$ and the magnetization ${\bf M}$ are
separated in space. We assume also that the magnetic field ${\bf
B}$ and its vector-potential  ${\bf A}$  asymptotically turn to
zero at infinity. Employing static Maxwell equation ${\bf
\nabla}\times {\bf B} = \frac{4 \pi}{c} {\bf j}$ , and ${\bf B} =
{\bf \nabla}\times {\bf A}$, the magnetic field energy can be
transformed as follows:

\begin{equation}
\int \frac{{\bf B}^2}{8 \pi}  dV=
\int  \frac{{\bf j}\cdot  {\bf A}} {2 c}dV
\label{en2}
\end{equation}
\noindent Though the vector-potential enters explicitly in the
last equation, it is gauge invariant due to the current
conservation ${\rm div}{\bf j}=0$. When integrating by part, we
neglected the surface term. This is correct if the field,
vector-potential and the current decrease sufficiently fast at
infinity. This condition is satisfied for simple examples
considered in this article. The current ${\bf j}$ can be
represented as a sum: ${\bf j}={\bf j}_s + {\bf j}_m$ of the SC
and magnetic currents, respectively:
\begin{equation}
{\bf j}_s  = \frac{n_s\hbar e}{2m_s} \bigl ( \nabla \varphi -
\frac{2\pi}{\Phi_0}{\bf A}\bigr ) \label{scurr}
\end{equation}
\begin{equation}
{\bf j}_{m} = c {\bf \nabla} \times {\bf M}.
\label{mcurr}
\end{equation}

\noindent We consider contributions from magnetic and S-currents
into the integral (\ref{en2}) separately. We start with the
integral:
\begin{equation}
\frac{1}{2c}\int {\bf j}_m{\bf A}dV=\frac{1}{2}\int\bigl (
\nabla\times{\bf M}\bigr ) \cdot{\bf A}dV \label{magn}
\end{equation}
Integrating by part and neglecting the surface term again, we
arrive at a following result:
\begin{equation}
\frac{1}{2c}\int {\bf j}_m{\bf A}dV=\frac{1}{2}\int {\bf
M}\cdot{\bf B}dV \label{magn1}
\end{equation}
We have omitted the integral over a remote surface $\oint \bigl
({\bf n}\times{\bf M}\bigr ) \cdot{\bf A}dS$. Such an omission is
valid if the magnetization is confined to a limited volume. But
for infinite magnetic systems it may be wrong even in simplest
problems. We will discuss such a situation in the next section.

Next we consider the contribution of the superconducting current
${\bf j}_s$ to the integral (\ref{en2}). In the gauge-invariant
equation \ref{scurr} $\varphi$  is the phase of the S-carriers (Cooper
pairs) wave-function and $\Phi_0=hc/2e$  is the flux quantum. Note
that the phase gradient ${\nabla\varphi}$ can be included into
${\bf A}$ as a gauge transformation with exception of vortex
lines, where ${\varphi}$ is singular. We employ equation
(\ref{scurr}) to express the vector-potential ${\bf A}$ in terms
of the supercurrent and the phase gradient:
\begin{equation}
{\bf A}=\frac{\Phi_0}{2\pi}\nabla\varphi - \frac{m_sc}{n_se^2}{\bf
j}_s \label{A}
\end{equation}
Plugging  equation (\ref{A}) into  equation (\ref{en2}), we find:
\begin{equation}
\frac{1}{2c}\int {\bf j}_s{\bf A}
dV=\frac{\hbar}{4e}\int\nabla\varphi\cdot{\bf j}_s
dV-\frac{m_s}{2n_se^2}\int j_s^2dV \label{ensup}
\end{equation}
Since ${\bf j}_s=en_s{\bf v}_s$, the last term in this equation is
equal to the kinetic energy taken with the sign minus. It exactly
compensates the kinetic energy in the initial expression for the
energy (\ref{en}). Collecting all remaining terms, we obtain a
following expression for the total energy:
\begin{equation}
H  = \int \bigl[\frac{n_s\hbar^2}{8m_s}({\nabla\varphi})^2 -
\frac{ n_s\hbar e}{4m_sc}{\nabla\varphi} \cdot{\bf A} - \frac{{\bf
B}\cdot {\bf M}}{2}\bigr] dV \label{en3}
\end{equation}
We remind again about a possible surface term for infinite
magnetic systems. Note that integration in the expression for
energy (\ref{en3}) proceeds over the volumes occupied either by
superconductors or by magnets. Equation  (\ref{en3}) allows to separate
the energy of vortices from the energy of magnetization induced
currents and fields and their interaction energy. Indeed, as we
noted earlier, the phase gradient can be ascribed to the
contribution of vortex lines only. It is representable as a sum of
independent integrals over different vortex lines. The
vector-potential and the magnetic field can be represented as a
sum of magnetization induced and vortex induced parts: ${\bf
A}={\bf A}_{m}+{\bf A}_{v}$, ${\bf B}={\bf B}_{m}+{\bf B}_{v}$,
where ${\bf A}_{k}$, ${\bf B}_{k}$ (the index $k$ is either $m$ or
$v$) are determined as solutions of the Londons-Maxwell equations:
\begin{equation}
\nabla\times\left( \nabla\times{\bf
A}_{k}\right)=\frac{4\pi}{c}{\bf j}_{k}, \label{LM3d}
\end{equation}
The effect of the screening of magnetic field generated by
magnetization by superconductor is included into the vector fields
${\bf A}_{m}$ and ${\bf B}_{m}$. Applying such a separation, we
present the total energy (\ref{en3}) as a sum of terms containing
only vortex contributions, only magnetic contributions and the
interaction terms. The purely magnetic part can be represented as
a nonlocal quadratic form of the magnetization. The purely
superconducting part is representable as a non-local double
integral over the vortex lines. Finally, the interaction term is
representable as a double integral which proceeds over the vortex
lines and the volume occupied by the magnetization and is
bi-linear in magnetization and vorticity. To avoid cumbersome
formulas, we will not write these expressions explicitly.

\subsection{Two-Dimensional Systems.}\label{2d}

Below we perform a more explicit analysis for the case of two
parallel films, one F, another S, both very thin and very close to 
each other. Neglecting their thickness, we assume that both films
are located approximately at $z = 0$. In some cases we need a more
accurate treatment. Then we introduce a small distance $d$ between
films which in the end will be put zero. Though the thickness of
each film is assumed to be small, the 2-dimensional densities of
S-carriers $n_s^{(2)}=n_sd_s$ and magnetization 
${\bf m}={\bf M}d_m$ remain finite. Here we introduced 
the thickness of the S film $d_s$ and the F film $d_m$. 
The 3d super-carrier density
$n_s({\bf R})$ can be represented as $n_s({\bf R}) = \delta (z)
{n}^{(2)}_s({\bf r})$ and the 3d magnetization  ${\bf M}({\bf R})$
can be represented as ${\bf M}({\bf R}) = {\delta(z-d)}{\bf
m}({\bf r})$, where  ${\bf r}$ is the two-dimensional
radius-vector and $z$-direction is perpendicular to the films. In
what follows ${n}^{(2)}_s$  is assumed to be a constant and the
index (2) is omitted.  The energy (\ref{en3}) can be rewritten for
this special case:

\begin{equation}
H  = \int \bigl[\frac{n_s\hbar^2}{8m_s}({\nabla\varphi})^2 -
\frac{ n_s\hbar e}{4m_sc}{\nabla\varphi} \cdot{\bf a} - \frac{{\bf
b}\cdot {\bf m}}{2}\bigr] d^2 {\bf r}
\label{en4}
\end{equation}
where ${\bf a} = {\bf A}({\bf r},z = 0)$ and ${\bf b} = {\bf
B}({\bf r},z = 0)$. The vector-potential satisfies Maxwell-Londons
equation:

\begin{eqnarray}
\nabla\times(\nabla\times {\bf A}) &=& - \frac{1}{\lambda} {\bf A}
\delta (z) + \frac{ 2\pi\hbar  n_se}{m_s
c}{\nabla\varphi}\delta(z)\\
\nonumber & +& 4\pi\nabla\times ({\bf m}\delta(z))
\label{vec}
\end{eqnarray}

\noindent Here $\lambda = \lambda_L^2/d_S$ is the effective
screening length for the S film, $\lambda_L$ is the London
penetration depth and $d_s$ is the S-film
thickness\cite{abrikosov}.

According to our general arguments, the term proportional to
${\nabla\varphi}$ in equation (\ref{vec}) describes vortices.  A plane
vortex characterized by its vorticity $q$ and by the position of
its center on the plane ${\bf r}_0$ contributes a singular term to
${\nabla\varphi}$:
\begin{equation}
\nabla\varphi_0({\bf r,r}_0)=q\frac{\hat z\times ({\bf r}-{\bf
r}_0) } {\vert {\bf r}-{\bf r}_0 \vert^2} \label{vortphase}
\end{equation}
and generates a standard vortex vector-potential:
\begin{eqnarray}
{\bf A}_{v0}({\bf r}-{\bf r}_0,z)
&=&\nonumber
\frac{q \Phi_0}{2 \pi}
\frac{\hat z\times ({\bf r}-{\bf r}_0) }{\vert {\bf r}-{\bf r}_0
\vert}
\\&\times&
\int_{0}^{\infty} \frac{J_1 ( k\vert{\bf r} - {\bf
r}_0\vert) e^{-k| z |}} {1 + 2 k\lambda}dk \label{vec2}
\end{eqnarray}
Different vortices contribute independently into the
vector-potential and magnetic field. A peculiarity of this problem
is that the usually applied gauge ${\rm div}{\bf A}=0$ becomes
singular in the limit $d_s,d_m\rightarrow 0$. Therefore, it is
reasonable to apply another gauge $A_z = 0$. The calculations are
much simpler in Fourier-representation. Following the general
procedure, we present the Fourier-transform of the
vector-potential ${\bf A_k}$ as a sum ${\bf A_k}={\bf A}_{m{\bf
k}}+{\bf A}_{v{\bf k}}$. Equation for the magnetic part of the
vector-potential reads:
\begin{equation}
  {\bf k}({\bf qA}_{m{\bf k}})-k^2{\bf A}_{m{\bf k}}=\frac{{\bf
  a}_{m{\bf q}}}{\lambda}-4\pi i{\bf k}\times{\bf m_q}e^{ik_zd}
  \label{A-mag}
\end{equation}
where ${\bf q}$ is projection of the wave vector ${\bf k}$ onto
the plane of the films: ${\bf k}=k_z{\hat z}+{\bf q}$. An
arbitrary vector field ${\bf V_k}$ in the wave-vector space can be
represented by its local coordinates:

\begin{equation}\label{local}
  {\bf V_k}= V_{\bf k}^z{\hat
z}+ V_{\bf k}^{\parallel}{\hat q} + V_{\bf k}^{\perp}({\hat
z}\times{\hat q})
\end{equation}

 In terms of these coordinates the solution of
equation (\ref{A-mag}) reads:

\begin{equation}\label{A-par}
  A_{m{\bf k}}^{\parallel}=-\frac{4\pi im_{\bf
  q}^{\perp}}{k_z}e^{ik_zd}
\end{equation}

\begin{equation}
 A_{m{\bf k}}^{\perp}=-\frac{1}{\lambda k^2}a_{\bf
 q}^{\perp}+\frac{4\pi i\left( k_zm_{\bf q}^{\parallel}-qm_{{\bf
 q}z}\right)}{k^2}e^{ik_zd}
 \label{A-perp}
\end{equation}
Integration of the latter equation over $k_z$ allows to find the
perpendicular component of ${\bf a_{\bf q}}^{(m)}$:
\begin{equation}
  a_{m{\bf q}}^{\perp}=-\frac{4\pi\lambda q(m_{\bf q}^{\parallel}+im_{{\bf
  q}z})}{1+2\lambda q}e^{-qd},
  \label{a-perp}
\end{equation}
whereas it follows from equation  (\ref{A-mag}) that $a_{m{\bf
q}}^{\parallel}=0$. Note that the parallel component of the
vector-potential $A_{m{\bf k}}^{\parallel}$ does not know anything
about the S film. It corresponds to the magnetic field equal to
zero outside the plane of F film. Therefore, it is inessential for
our problem.

The vortex part of the vector-potential also does not contain
$z$-component since the supercurrents flow in the plane. The 
vortex solution in a thin film was first found by Pearl 
\cite{pearl}. An explicit expression for the vortex-induced
potential is:
\begin{equation}
  {\bf A}_{v{\bf k}}=\frac{2i\Phi_0(\hat{z}\times\hat{q})F({\bf q})}
{{\bf k}^2(1+2\lambda
  q)},
  \label{A-vortex}
\end{equation}
where $F({\bf q})=\sum_{j}e^{i{\bf qr}_j}$ is the vortex
form-factor; the index $j$ labels the vortices and ${\bf r}_j$ are
coordinates of the vortex centers. The Fourier-transformation for
the vortex-induced vector-potential at the surface of the SC film
${\bf a}_{v{\bf q}}$ reads:
\begin{equation}
 {\bf a}_{v{\bf q}}=
\frac{i\Phi_0(\hat{z}\times\hat{q})F({\bf q})}{q(1+2\lambda q)}
 \label{a-vortex}
\end{equation}
The z-component of magnetic field induced by the Pearl vortex in
real space is:
\begin{equation}\label{Bzv-coord}
B_{vz}=\frac{\Phi_0}{2\pi}
\int_0^{\infty}\frac{J_0(qr)e^{-q|z|}}{1+2\lambda q}qdq 
\end{equation}
Its asymptotic at $z=0$ and $r\gg\lambda$ is $B_{vz}\approx
\Phi_0\lambda/(\pi r^3)$; at $r\ll\lambda$ it is $B_{vz}\approx
\Phi_0/(\pi\lambda r)$. Each Pearl vortex carries the flux quantum
$\Phi_0=\pi\hbar c/e$.\newline
The energy (\ref{en4}), can be expressed in terms of
Fourier-transforms:
\begin{equation}
  H=H_v+H_m+H_{vm},
  \label{en5}
\end{equation}
where purely vortex energy $H_v$ is the same as it would be in the
absence of the FM film:
\begin{equation}
H_v=\frac{n_s\hbar^2}{8m_s}\int \nabla\varphi_{-{\bf
q}}(\nabla\varphi_{\bf q}-\frac{2\pi}{\Phi_0}{\bf a}_{v{\bf
q}})\frac{d^2q}{(2\pi)^2}; \label{en-v}
\end{equation}
The purely magnetic energy $H_m$ is:
\begin{equation}
  H_m=-\frac{1}{2}\int{\bf m_qb}_{m{\bf q}}
  \label{en-m}
\end{equation}
It contains the screened magnetic field and therefore differs from
its value in the absence of the SC film . Finally the interaction
energy reads:
%\begin{equation}
\begin{eqnarray}
  H_{mv}&=&\nonumber
-\frac{n_s\hbar e}{4m_sc}\int (\nabla\varphi)_{-{\bf
  q}}{\bf a}_{m{\bf q}}\frac{d^2q}{(2\pi)^2}
\\&-&
\frac{1}{2}\int{\bf
  m}_{-{\bf q}}{\bf b}_{v{\bf q}}\frac{d^2q}{(2\pi)^2}
  \label{en-mv}
\end{eqnarray}
%\end{equation}
Note that the information on the vortex arrangement is contained
in the form-factor $F({\bf q})$ only.

To illustrate how important can be the surface term, let consider
a homogeneous perpendicularly magnetized magnetic film and one
vortex in superconducting film. The authors \cite{vmlp} have shown
that the energy of this system is
$\varepsilon_v=\varepsilon_v^0-m\Phi_0$, where $\varepsilon_v^0$
is the energy of the vortex in the absence of magnetic film, $m$
is the magnetization per unit area and $\Phi_0=hc/2e$ is the
magnetic flux quantum. Let consider how this result appears from
the microscopic calculations. The vortex energy (\ref{en-v}) is
just equal to $\varepsilon_v^0$. Purely magnetic term (\ref{en-m})
does not change in the presence of vortex and is inessential. The
first term in the interaction energy (\ref{en-mv}) is equal to
zero since the infinite magnetic film does not generate magnetic
field outside. The second term of this energy is equal to
$-m\Phi_0/2$. The second half of the interaction energy comes from
the surface term. Indeed, it is equal to
\begin{eqnarray}
(1/2)\lim_{r\rightarrow\infty}\int_0^{2\pi}m(\hat{r}\times\hat{z})\cdot{\bf
A}rd\varphi&=& -(1/2)\oint{\bf A}\cdot d{\bf
r}\nonumber\\&=&-m\Phi_0/2\nonumber
\end{eqnarray}

\subsection{Eilenberger and Usadel Equations}
The essence of proximity phenomena is the change of the order
parameter (Cooper pair wave function). Therefore, the London
approximation is not valid in this case and equations for the
order parameter must be solved. They are either
Bogolyubov-DeGennes equations \cite{Bogolyubov, DeGennes} for the
coefficients $u$ and $v$ or more conveniently the Gor'kov
equations \cite{Gor'kov} for Green functions. Unfortunately the
solution of these equations is not an easy problem in the
spatially inhomogeneous case combined with the scattering by
impurities and/or irregular boundaries. This is a typical
situation for the experiments with F/S proximity effects, since
the layers are thin, the diffusion delivers atoms of one layer
into another and the control of the structure and morphology is
not so strict as for 3d single crystals. Sometimes experimenters
deliberately use amorphous alloys as magnetic layers
\cite{ryazanovusp}. Fortunately, if the scale of variation for the
order parameter is much larger than atomic, the semiclassical
approximation can be applied. Equations for the superconducting
order parameter in semiclassical approximation were derived long
time ago by Eilenberger \cite{eilenberger} and by Larkin and 
Ovchinnikov \cite{larkinovch-eq}. They were further simplified in
the case of strong elastic scattering (diffusion approximation) by
Usadel \cite{usadel}. For the reader convenience and for the
unification of notations we demonstrate them here referring the
reader for derivation to original works or to the textbooks
\cite{tinkham, kopnin}.\newline

The Eilenberger equations are written for the electronic Green
functions integrated in the momentum space over the momentum
component perpendicular to the Fermi surface. Thus, they depend on
a point of the Fermi-surface characterized by two momentum
components, on the coordinates in real space and time. It is more
convenient in thermodynamics to use their Fourier-components over
the imaginary time, the so-called Matsubara representation
\cite{AGD}. The frequencies in this representation accept discrete
real values $\omega_n=(2n+1)\pi T$, where $T$ is the temperature.
The case of singlet pairing is described by two Eilenberger
anomalous Green functions $F(\omega,{\bf k},{\bf r})$ and
$F^{\dag}(\omega,{\bf k},{\bf r})$ (integrated along the normal to 
the Fermi-surface Gor'kov anomalous functions), where $\omega$ 
stays for $\omega_n$, $\bf{k}$ is the wave vector at the Fermi
sphere and $\bf{r}$ is the vector indicating at a point in real
space (the coordinate of the Cooper pair center-of-mass). The 
function $F$ generally is complex in contrast to the integrated
normal Green function $G(\omega,\bf{k},\bf{r})$, which is real.
Eilenberger has proved that the functions $G$ and $F$ are not
independent: they obey the normalization condition:
\begin{equation}\label{normalization}
[G(\omega,{\bf k},{\bf r})]^2+|F(\omega,{\bf k},{\bf r})|^2=1
\end{equation}
Besides, the Eilenberger  Green functions obey the following
symmetry relations:
\begin{eqnarray}
F(\omega,{\bf k},{\bf r})=F^*(-\omega,{\bf k},{\bf
r})=F^*(\omega,-{\bf k},{\bf r})
\label{Fsym}\\
G(-\omega,{\bf k},\bf{r})=-G^*(\omega,{\bf k},{\bf
r})=-G(\omega,-{\bf k},{\bf r}) \label{Gsym}
\end{eqnarray}
Eilenberger equations read:
\begin{eqnarray}
\left[ 2\omega+{\bf v}\left(\frac{\partial}{\partial{\bf
r}}-i\frac{2e}{c}{\bf A}({\bf r})\right)\right]
F(\omega,{\bf k},{\bf r})=2\Delta G(\omega,{\bf k},{\bf r})\nonumber\\ 
+\int d^2q \rho ({\bf q})W({\bf k,q})[G({\bf k})F({\bf q})-F({\bf
k})G({\bf q})]\label{EilenF}
\end{eqnarray}
where $\Delta ({\bf r})$ is the space (and time)- dependent order
parameter (local energy gap); ${\bf v}$ is the velocity on the
Fermi surface; $W({\bf k},{\bf q})$ is the probability of
transition per unit time from the state with the momentum ${\bf
q}$ to the state with the momentum ${\bf k}$ and $\rho({\bf q})$
is the angular dependence of the density of states normalized by
$\int d^2q \rho({\bf q})=N(0)$. Here $N(0)$ is the total density 
of states (DOS) in the normal state at the Fermi level. The 
Eilenberger equations have the structure of Boltzmann kinetic
equation, but they also incorporate quantum coherence effects.
They must be complemented by the self-consistency equation
expressing local value of $\Delta ({\bf r})$ in terms of the
anomalous Green function $F$:
\begin{equation}
\Delta ({\bf r})\ln(\frac{T}{T_c})+2\pi T
\sum_{n=0}^{\infty}\left[ \frac{\Delta ({\bf r})}{\omega_n}-\int 
d^2k \rho({\bf k})F(\omega_n,{\bf k},{\bf r})\right] = 0
\label{selfconsistency}
\end{equation}
In a frequently considered by theorists case of the isotropic 
scattering the collision integral in equation (\ref{EilenF}) is
remarkably simplified:
\begin{equation}\label{isotropic}
\int d^2q \rho ({\bf q})W({\bf k,q})[G({\bf k})F({\bf q})-F({\bf
k})G({\bf q})]=\frac{1}{\tau}\left[ G({\bf k})\langle F \rangle -
F({\bf k})\langle G \rangle\right],
\end{equation}
where the relaxation time $\tau$ is equal to inverse value of
angular independent transition probability $W$ and $\langle ...
\rangle$ means the angular average over the Fermi sphere.\newline
The Eilenberger equation is simpler than complete Gor'kov
equations since it contains only one function depending on by one
less number of arguments. It could be expected that in the limit
of very short relaxation time $T_{c0}\tau\ll 1$ ($T_{c0}$ is
transition temperature in the clean superconductor) the
Eilenberger kinetic-like equation will become similar to diffusion
equation. Such a diffusion-like equation was indeed derived by
Usadel \cite{usadel}. In the case of strong elastic scattering and
the isotropic Fermi surface (sphere) the Green function does not
depend on the direction on the Fermi sphere and depends only on
frequency and the spatial coordinate ${\bf r}$. The Usadel
equation reads (we omit both arguments):
\begin{equation}\label{usadel}
2\omega F - D\hat{\partial}\left(
G\hat{\partial}F-F\hat{\partial}G\right)=2\Delta G
\end{equation}
In this equation $D=v_F^2\tau/3$ is the diffusion coefficient for
electrons in the normal state and $\hat{\partial}$ stays for the
gauge-invariant gradient: $\hat{\partial}=\nabla - 2ie{\bf
A}/\hbar c$. The Usadel equations must be complemented by the same
self-consistency equation (\ref{selfconsistency}). It is also
useful to keep in mind expression for the current density in terms
of the function $F$:
\begin{equation}
{\bf j}=ie2\pi
TN(0)D\sum_{\omega_n>0}(F^*\hat{\partial}F-F\hat{\partial}F^*).
\label{current}
\end{equation}
One can consider the set of Green functions $G$, $F$, $F^{\dag}$
as elements of the 2x2 matrix Green function $\hat{g}$ where the
matrix indices can be identified with the particle and hole or
Nambu channels. This formal trick becomes rather essential when
the singlet and triplet pairing coexist and it is necessary to
take in account the Nambu indices and spin indices simultaneously.
Eilenberger in his original article \cite{eilenberger} has
indicated a way to implement the spin degrees of freedom in his
scheme. Below we demonstrate a convenient modification of this
representation proposed by Bergeret {\it et al.} \cite{BEV2003}.
Let us introduce a matrix $\check{g}({\bf r},t;{\bf
r}^{\prime},t^{\prime})$ with matrix elements
$g_{s,s^{\prime}}^{n,n^{\prime}}$, where $n,n^{\prime}$ are the
Nambu indices and $s,s^{\prime}$ are the spin indices, defined as
follows:
\begin{equation}\label{matrixgreen}
g_{s,s^{\prime}}^{n,n^{\prime}}({\bf r},t;{\bf
r}^{\prime},t^{\prime})=\frac{1}{\hbar}\sum_{n^{\prime\prime}}
(\hat{\tau}_3)_{n,n^{\prime\prime}}\int
d\xi \langle \psi_{n^{\prime\prime}s}({\bf
r},t)\psi^{\dag}_{n^{\prime}s^{\prime}}({\bf
r}^{\prime},t^{\prime})\rangle
\end{equation}
The matrix $\hat{\tau}_3$ in the definition (\ref{matrixgreen}) is
the Pauli matrix in the Nambu space. To clarify the Nambu indices
we write explicitly what do they mean in terms of the electronic
$\psi$-operators: $\psi_{1s}\equiv \psi_s; \psi_{2s}\equiv
\psi^{\dag}_{\bar{s}}$ and $\bar{s}$ means $-s$. The most general
matrix $\check{g}$ can be expanded in the Nambu space into a
linear combination of 4 independent matrices $\hat{\tau}_k;
k=0,1,2,3$, where $\hat{\tau}_0$ is the unit matrix and three
others are the standard Pauli matrices. Following \cite{BEV2003},
we accept following notations for the components of this
expansion, which are matrices in the spin space\footnote{Each time
when Nambu and spin matrices stay together we mean the direct
product.}:
\begin{equation}\label{tau-expansion}
    \check{g}=\hat{g}_0\hat{\tau}_0+\hat{g}_3\hat{\tau}_3+\check{f};
    \check{f}=\hat{f}_1i\hat{\tau}_1+\hat{f}_2i\hat{\tau}_2
\end{equation}
The matrix $\check{f}$ describes Cooper pairing since it contains
only anti-diagonal matrices in the Nambu space. In turn the spin
matrices $\hat{f}_1$ and $\hat{f}_2$ can be expanded in the basis
of spin Pauli matrices $\hat{\sigma_j; j=0,1,2,3}$. Without loss
of generality we can accept the following agreement about the
scalar components of the spin-space expansion:
\begin{equation}\label{spin-expansion}
    \hat{f}_1=f_1\hat{\sigma}_1+f_2\hat{\sigma}_2;
    \hat{f}_2=f_0\hat{\sigma}_0+f_3\hat{\sigma}_3.
\end{equation}
It is easy to check that the amplitudes $f_i; i=0...3$ are
associated with the following combinations of the wave-function
operators:
\begin{eqnarray}
  f_0 &\rightarrow& \langle\psi_{\uparrow}\psi_{\downarrow}
+\psi_{\downarrow}\psi_{\uparrow}\rangle\nonumber \\
  f_1 &\rightarrow& \langle\psi_{\uparrow}\psi_{\uparrow}
-\psi^{\dag}_{\downarrow}\psi^{\dag}_{\downarrow}\rangle\nonumber \\
  f_2 &\rightarrow& \langle\psi_{\uparrow}\psi_{\uparrow}
+\psi^{\dag}_{\downarrow}\psi^{\dag}_{\downarrow}\rangle\nonumber \\
  f_3 &\rightarrow& \langle\psi_{\uparrow}\psi_{\downarrow}
-\psi_{\downarrow}\psi_{\uparrow}\rangle\nonumber
\end{eqnarray}
Thus, the amplitude $f_3$ corresponds to the singlet pairing,
whereas three others are responsible for the triplet pairing. In
particular, in the absence of triplet pairing only the component
$f_3$ survives and the matrix $\check{f}$ is equal to
$$\left(%
\begin{array}{cc}
  0 & F \\
  F^{\dag} & 0 \\
\end{array}
\right)
$$
\newline Let us consider what modification must be
introduced into the Eilenberger and Usadel equations to take in
account the exchange interaction of Cooper pairs with the
magnetization in the ferromagnet. Neglecting the reciprocal effect
of the Cooper pairs onto the electrons of d- or f-shell
responsible for magnetization, we introduce the effective exchange
field $h({\bf r})$ acting inside the ferromagnet. It produces
pseudo-Zeeman splitting of the spin energy\footnote{In reality the 
exchange energy has quite different origin than the Zeeman
interaction, but at a fixed magnetization there is a formal
similarity in the Hamiltonians.}. In the case of the singlet
pairing the Matsubara frequency $\omega$ must be substituted by
$\omega +ih({\bf r})$. When the direction of magnetization changes
in space generating triplet pairing, the Usadel equation is
formulated in terms of the matrix $\check{g}$ \cite{BEV2003}:
\begin{equation}\label{usadel-exch}
    \frac{D}{2}\partial(\check{g}\partial\check{g})
-|\omega|[\hat{\tau}_3\hat{\sigma}_3,\check{g}]
    +\textrm{sign}\omega[\check{h},\check{g}]=-i[\check{\Delta},\check{g}],
\end{equation}
where the operators of the magnetic field $\check{h}$ and the
energy gap $\check{\Delta}$ are defined as follows:
\begin{equation}\label{field-oper}
    \check{h}=\tau_3 {\bf \sigma}\cdot {\bf h}
\end{equation}
\begin{equation}\label{Delta-oper}
    \check{\Delta}=\Delta i\tau_2\sigma_2
\end{equation}
\newline
To find a specific solution of the Eilenberger and Usadel
equations proper boundary conditions should be formulated. For the
Eilenberger equations the boundary conditions at an interface of
two metals were derived by Zaitsev \cite{zaitsev}. They are most 
naturally formulated in terms of the antisymmetric ($\check{g}^a$)
and symmetric ($\check{g}^s$) parts of the matrix $\check{g}$ with 
respect to reflection of momentum $p_z\rightarrow -p_z$ assuming
that $z$ is normal to the interface. One of them states that the
antisymmetric part is continuous at the interface ($z=0$):
\begin{equation}\label{Z1}
    \check{g}^a(z=-0)=\check{g}^a(z=+0)
\end{equation}
The second equation connects the discontinuity of the symmetric
part at the interface
$\check{g}^s_{-}=\check{g}^s(z=+0)-\check{g}^s(z=-0)$ with the 
reflection coefficient $R$ and transmission coefficient $D$ of the
interface and antisymmetric part $\check{g}^a$ at the boundary:
\begin{equation}\label{Z2}
D\check{g}^s_{-}(\check{g}^s_{+}-\check{g}^a\check{g}^s_{-})
=R\check{g}^a[1-(\check{g}^a)^2], 
\end{equation}
where $\check{g}^s_{+}=\check{g}^c(z=+0)+\check{g}^c(z=-0)$. If 
the boundary is transparent (R=0, D=1), the symmetric part of the
Green tensor $\check{g}$ is also continuous.
\newline
The boundary conditions for the Usadel equations, i.e. under the 
assumption that the mean free path of electron $l$ is much shorter
than the coherence length $\xi$, were derived by Kupriyanov and
Lukichev\cite{KuLu}. The first of them ensures the continuity of
the current flowing through the interface:
\begin{equation}\label{KuLu1}
\sigma_<\check{g}_<\frac{d\check{g}_<}{dz}=
\sigma_>\check{g}_>\frac{d\check{g}_>}{dz},
\end{equation}
where the subscripts $<$ and $>$ relate to the left and right
sides of the interface; $\sigma_{<,>}$ denote the conductivity of
the proper metal. The second boundary condition connects the
current with the discontinuity of the order parameter through the
boundary and its transmission and reflection coefficients
$D(\theta)$ and $R(\theta)$:
\begin{equation}\label{KuLu2}
    l_>\check{g}_>\frac{d\check{g}_>}{dz}=\frac{3}{4}\langle \frac{cos\theta
    D(\theta)}{R(\theta)}\rangle [\check{g}_<,\check{g}_>],
\end{equation}
where $\theta$ is the incidence angle of the electron at the
interface and $D(\theta)$, $R(\theta)$ are corresponding
transmission and reflection coefficients. This boundary condition
can be rewritten in terms of measurable characteristics:
\begin{equation}\label{KuLumodified}
    \sigma_>\check{g}_>\frac{d\check{g}_>}{dz}=\frac{1}{R_b}\rangle
    [\check{g}_<,\check{g}_>],
\end{equation}
where $R_b$ is the resistance of the interface. In the case of
high transparency ($R\ll 1$) the boundary conditions
(\ref{KuLu1},\ref{KuLu2}) can be simplified as follows
\cite{BalBuzVed}:
\begin{equation}\label{simple1}
    \check{f}_<=\check{f}_>; \frac{d\check{f}_<}{dz}=\gamma
    \frac{d\check{f}_>}{dz},
\end{equation}
where $\gamma$ is the ratio of normal state resistivities.

\section{ Hybrids Without Proximity Effect}

\subsection{Magnetic Dots \label{dot} }

\begin{figure}[t]
\begin{center}
\includegraphics[width=2.5in]{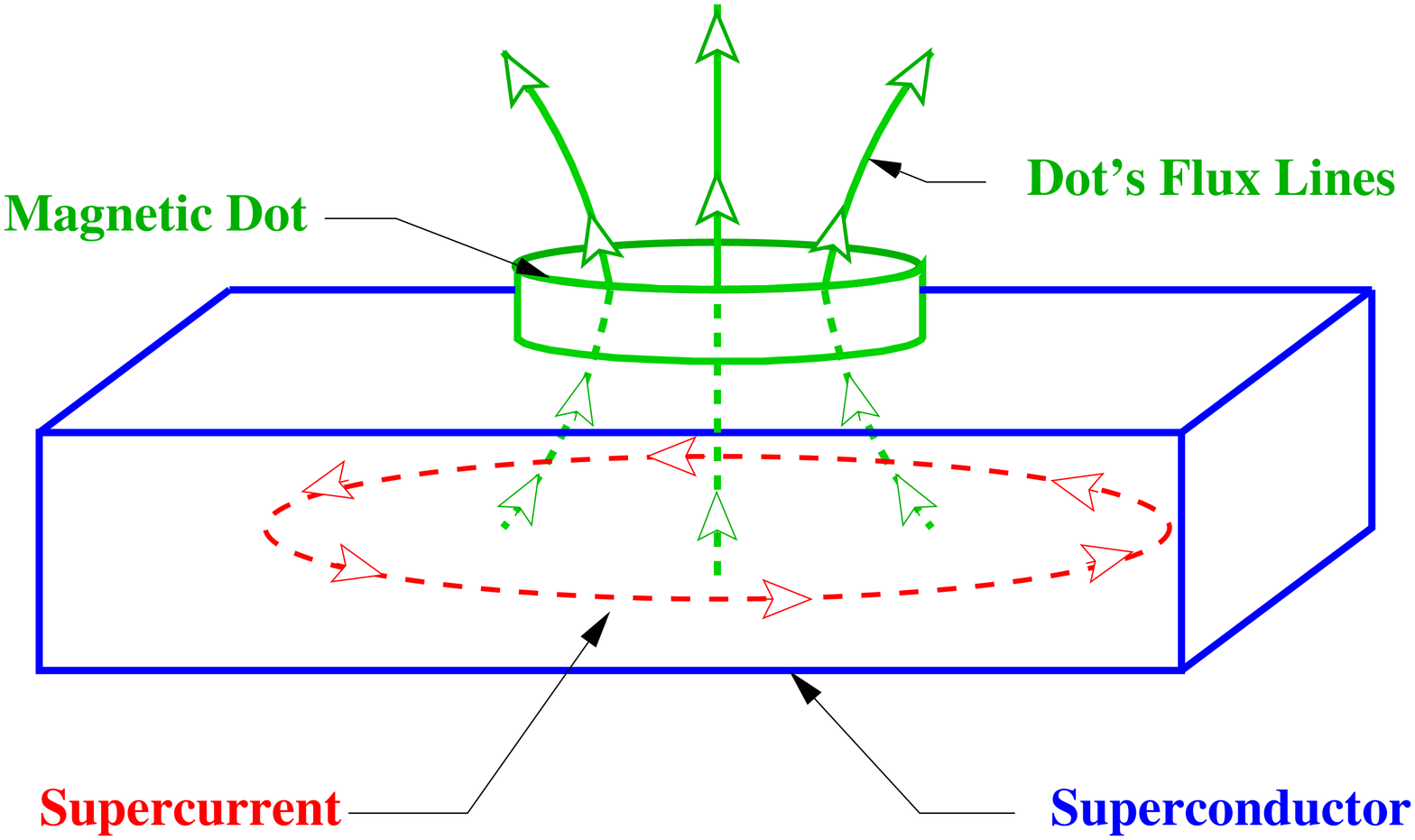}
\includegraphics[width=2.5in]{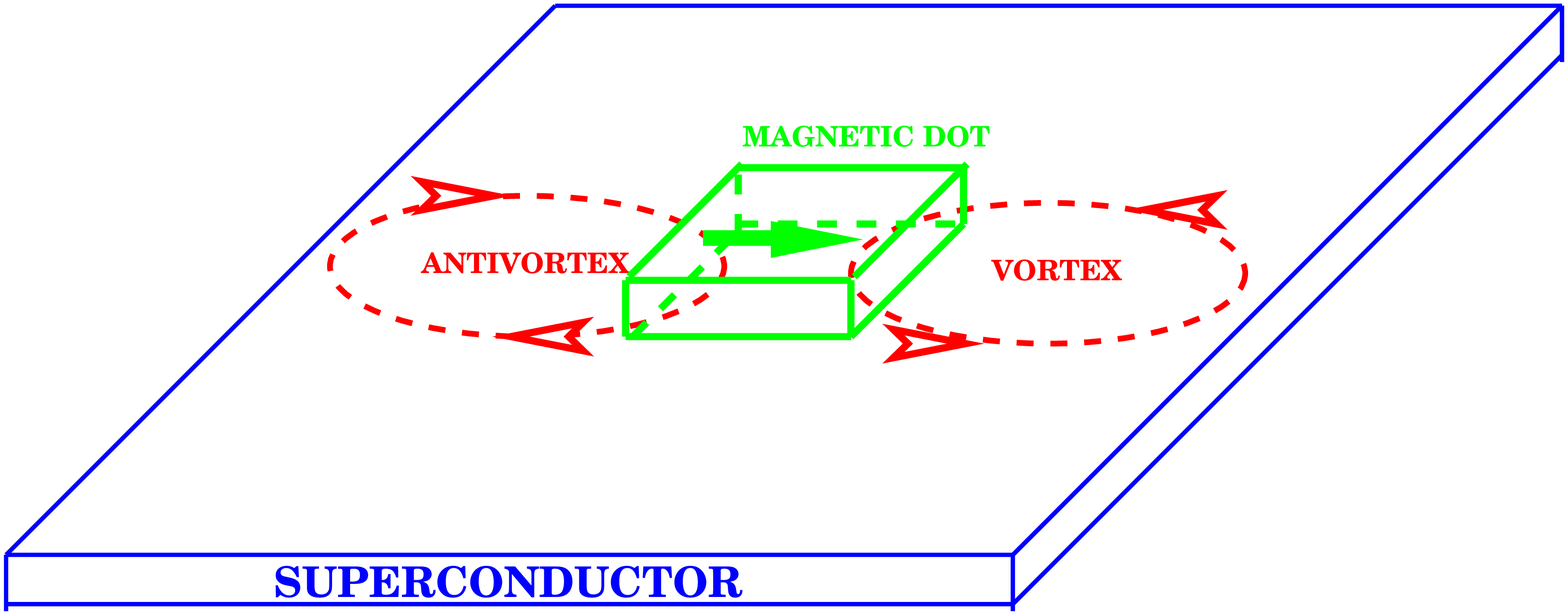}
\caption{\label{dots1}Magnetic dots with
out-of-plane and  in-plane magnetization and vortices.}
\end{center}
\end{figure}

In this subsection we consider the ground state of a SC film with
a circular very thin FM dot grown upon it. The magnetization will
be considered to be fixed, homogeneous inside the dot and directed
either perpendicular or parallel to the SC film (see figure
\ref{dots1}). This problem is basic one for a class of more
complicated problems incorporating arrays of magnetic dots.

We will analyze what are conditions for appearance of vortices in 
the ground state, where do they appear and what are magnetic
fields and currents in these states. The S film is assumed to be 
very thin, plane and infinite in the lateral directions. Since the
magnetization is confined inside the finite dot no difficulties
with the surface integrals over infinitely remote surfaces or
contours arise.

\subsubsection{Magnetic Dot: Perpendicular magnetization}

   For an infinitely thin circular  magnetic dot of the radius $R$
with 2d magnetization ${\bf m(r)} = m \hat{z}\Theta
(R-r)\delta(z-d)$ on the top of the SC film the magnetic field can
be calculated using equations  (\ref{A-perp},\ref{a-perp}). The
Fourier-component of magnetization necessary for this calculation
is:
\begin{equation}
  {\bf m_k}=\hat{z}\frac{2\pi mR}{q}J_1(qR)e^{ik_zd},
  \label{mperp-Fourier}
\end{equation}
where $J_1(x)$ is the Bessel function. The Fourier-transforms of
the vector-potential reads:
%\begin{equation}
\begin{eqnarray}
  A_{m{\bf k}}^{\perp}&=&\nonumber
-\frac{i8\pi^2 mRJ_1(qR)}{k^2}
\\&\times&
  \bigl( e^{-qd}\frac{2q\lambda}{1+2q\lambda}+(e^{ik_zd}-e^{-qd})
  \bigr)
  \label{Aperp}
\end{eqnarray}
Though the difference in the round brackets in equation
(\ref{Aperp}) looks to be always small (we remind that $d$ must be
put zero in the final answer), we can not neglect it since it
occurs to give a finite, not small contribution to the parallel
component of the magnetic field between the two films. From
equation (\ref{Aperp}) we immediately find the Fourier-transforms
of the magnetic field components:
\begin{equation}
B_{m{\bf q}}^z=iqA_{m{\bf q}}^{\perp};
\,\,\,B_{m{\bf q}}^{\perp}=-ik_zA_{m{\bf q}}^{\perp}
\label{Bperp}
\end{equation}
For the readers convenience we also present the Fourier-transform
of the vector-potential at the superconductor surface:
\begin{equation}
  a_{m{\bf q}}^{\perp}=
-\frac{i8\pi^2\lambda mR}{1+2q\lambda}J_1(qR)
  \label{aperp}
\end{equation}
In the last equation we have put $e^{-qd}$ equal to 1.

Performing inverse Fourier-transformation, we find the magnetic
field in real space:
\begin{equation}
B_m^z(r,z)=4\pi\lambda mR
\int_0^{\infty}\frac{J_1(qR)J_0(qr)e^{-q|z|}}{1+2q\lambda}q^2dq
\label{Bz-coord}
\end{equation}
\begin{eqnarray}
\label{Br-coord}
 &B_m^r&({\bf r},z) =2\pi mR
 \int_0^{\infty}J_1(qR)J_1(qr)e^{-q|z|}
\nonumber\\&\times& \bigl[\frac{2q\lambda
 }{1+2q\lambda}\Theta(z)+\Theta(z-d) - \Theta(z)\bigr]qdq,
\end{eqnarray}
where $\Theta(z)$ is the step function equal to $+1$ at positive
$z$ and $-1$ at negative $z$. Note that $B_m^r$ has
discontinuities at $z=0$ and $z=d$ due to surface currents in the
S- and F-films, respectively, whereas the normal component $B_m^z$ 
is continuous.

\begin{figure}[t]
\begin{center}
\includegraphics[angle=270,width=3.5in]{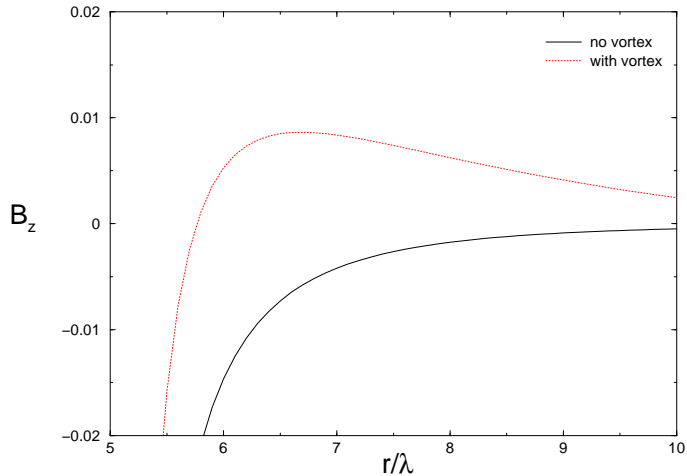}
\caption{\label{vor}Magnetic field of dot with and without
vortex for $R/\lambda = 5$ and $\Phi_0/8 \pi^2 m R = 4$ }
\end{center}
\end{figure}

A vortex, if appears, must be located at the center of the dot due
to symmetry. If $R \gg\lambda$, the direct calculation shows that
the central position of the vortex provides minimal energy. For
small radius of the dot the deviation of the vortex from the
central position seems even less probable. We have checked
numerically that central position is always energy favorable for
one vortex. Note that this fact is not trivial since the magnetic
field of the dot is stronger near its boundary. However, the gain
of energy due to interaction of the magnetic field generated by
the vortex with magnetization decreases when the vortex approaches
the boundary. The normal magnetic field generated by the Pearl vortex 
is given by equation  (\ref{Bzv-coord}). Numerical calculations based on
equations  (\ref{Bz-coord}, \ref{Bzv-coord}) for the case $R>\lambda$
shows that $B_z$ at the S-film $(z=0)$ changes sign at some $r=R_0$ 
(see figure \ref{vor}) in the presence of the vortex centered at
$r=0$, but it is negative
everywhere at $r > R $ in the absence of the vortex. 

The physical explanation of this fact is as follows. The dot
itself is an ensemble of parallel magnetic dipoles. Each dipole
generates magnetic field at the plane passing through the dot, 
which has the sign opposite to its dipolar moment. The fields from
different dipoles compete at $r<R$, but they have the same sign at 
$r>R$. The SC current resists this tendency. The field generated
by the vortex decays slower than the dipolar field ($1/r^3$ vs.
$1/r^4$ ). Thus, the sign of $B_z$ is opposite to the
magnetization at small values of $r$ (but larger than $R$) and 
positive at large $r$. The measurement of magnetic field near the
film may serve as a diagnostic tool to detect a
S-vortex confined by the dot. To our knowledge, so far there were no
experimental measurements of this effect.

In the presence of a vortex, energy of the system can be
calculated using equations (\ref{en5}-\ref{en-mv}). The appearance of
the vortex changes energy by the amount:
\begin{equation}\label{change}
  \Delta = \varepsilon_v + \varepsilon_{mv}
\end{equation}
where $\varepsilon_v = \varepsilon_0\ln (\lambda /\xi)$ is the
energy of the vortex without magnetic dot, $\varepsilon_0 =
{\Phi^2_0 /(16 \pi^2 \lambda)}$; $\varepsilon_{mv}$ is the energy
of interaction between the vortex and the magnetic dot given by
equation  (\ref{en-mv}). For this specific problem the direct
substitution of the vector-potential, magnetic field and the phase
gradient (see equations  (\ref{aperp},\ref{Bz-coord}))leads to a
following result:
\begin{equation}\label{epsilon-mv}
  \varepsilon_{mv} =
  -m\Phi_0 R\int_0^{\infty}\frac{J_1(qR)dq}{1+2\lambda q}
\end{equation}

\noindent The vortex appears when $\Delta$ turns into zero. This
criterion determines a curve in the plane of two dimensional
variables $R/\lambda$ and $m\Phi_0/\varepsilon_v$. This curve
separating regimes with and without vortices is depicted in
figure {\ref{phaseperp}. The asymptotic of $\varepsilon_{mv}$ for
large and small values of $R/\lambda$ can be found analytically:
\begin{eqnarray*}
\varepsilon_{mv}  \approx &-& m\Phi_0 \;\;\;\;\;\;\;
\bigl(\frac{R}{\lambda }\gg 1\bigr)\nonumber \\
\varepsilon_{mv}\approx &-& m\Phi_0\frac{R}{2\lambda}\;\;\;
\bigl(\frac{R}{\lambda}\ll 1\bigr)
\end{eqnarray*}
Thus, asymptotically the curve $\Delta=0$ turns into a horizontal
straight line $m\Phi_0/\varepsilon_v=1$ at large $R/\lambda$ and
logarithmically distorted hyperbola
$(m\Phi_0/\varepsilon_v)(R/\lambda)=2$ at
small ratio $R/\lambda$.

\begin{figure}[t]
\begin{center}
\includegraphics[angle=270,width=3.5in]{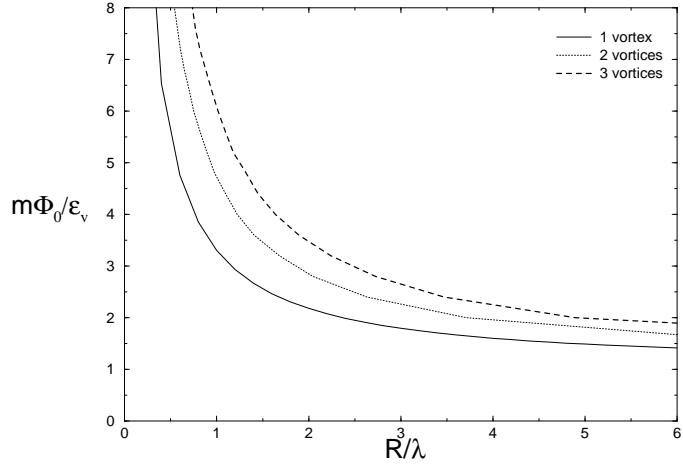}
\caption{\label{phaseperp}
Phase diagram of vortices induced by
a magnetic dot. The lines correspond to the appearance of 1,2 and
3 vortices, respectively.}
\end{center}
\end{figure}

At further increasing of either $m\Phi_0/\varepsilon_v$ or
$R/\lambda$ the second vortex becomes energy favorable. Due to
symmetry the centers of the two vortices are located on the
straight line including the center of the dot at equal distances
from it. The energy of the two-vortex configuration can be
calculated by the same method. The curve 2 on figure  
(\ref{phaseperp}) corresponds to this second phase transition. In
principle there exists an infinite series of such transitions.
However, here we limit ourselves with the first three since it is
not quite clear what is the most energy favorable configuration
for 4 vortices (for 3 it is the regular triangle). It is not yet
studied what is the role of configurations with several vortices
confined inside the dot region and antivortices outside.

\subsubsection{Magnetic Dot: Parallel Magnetization}

Next we consider an infinitely thin circular magnetic dot whose
magnetization ${\bf M}$ is directed in the plane and is
homogeneous inside the dot. An explicit analytical expression for
${\bf M}$ reads as follows:

\begin{equation}
{\bf{M}}=m_0 \theta (R-\rho) \delta(z) \hat{x} \label{Mpar}
\end{equation}
\noindent where $R$ is the radius of the dot, $m_0$ is the
magnetization per unit area and $\hat{x}$ is the unit vector along
the x-axis. The Fourier transform of the magnetization is:

\begin{equation}\label{Mparf}
  {\bf{M}}_{\bf{k}}= 2\pi m_0 R\frac{J_1 (qR)}{q} \hat{x}
\end{equation}
The Fourier-representation for the vector-potential generated by
the dot in the presence of magnetic film takes the form:
\begin{eqnarray}
{\bf{A}}_{m{\bf{k}}}^\bot&=&e^{ikd}\bigl[\frac{8\pi^2 m_0
R}{k_z^2
+q^2} J_1 (qR)cos(\phi_q)\bigr.\nonumber\\
\bigl.&\times& \left(\frac{\imath k_z e^{\imath k_z d}}{q}
-\frac{e^{-qd}}{1 +2\lambda q}\right)\bigr] \label{Apar}
\end{eqnarray}

Let introduce a vortex-antivortex pair with the centers of the
vortex and antivortex located at $x=+\rho_0$, $x=-\rho_0$,
respectively. Employing equations (\ref{en5}-\ref{en-mv}) to calculate
the energy, we find:

\begin{eqnarray}
E &=& \nonumber 2\epsilon_0 ln(\frac{\lambda}{\xi}) -4\epsilon_0 \lambda
\int_0^\infty \frac{J_0 (2q\rho_0)} {1 +2\lambda q}dq
\\&-& 2m_0
\Phi_0R \int_0^\infty \frac{J_1 (qR) J_1(q\rho_0)}{1+2\lambda q}dq
+ E_0 \label{Epar}
\end{eqnarray}
where $E_0$ is the dot self energy. Our numerical calculations 
indicate that the equilibrium value of $\rho_0$ is equal to $R$.
The vortex-anti-vortex creation changes the energy of the system
by:
\begin{eqnarray}
\Delta=2\epsilon_0 ln(\frac{\lambda}{\xi}) &-& 4\epsilon_0 \lambda
\int_0^\infty \frac{J_0 (2qR)}{1 +2\lambda q}dq\nonumber\\ &-&
2m_0\Phi_0R \int_0^\infty \frac{J_1 (qR) J_1(qR)}{1+2\lambda q}dq
\end{eqnarray}

\noindent The instability to the vortex-anti-vortex pair 
appearance develops when $\Delta$ changes sign. The curve that
corresponds to $\Delta=0$ is given by a following equation:

\begin{eqnarray}
\frac{m_0\Phi_0}{\epsilon_0}=\frac{2ln(\frac{\lambda}{\xi})
-4\lambda \int_0^\infty \frac{J_0 (2qR)}{1+2\lambda q} dq} {2R
\int_0^\infty \frac{J_1(qR)J_1(qR)}{1 +2\lambda q} dq}
\end{eqnarray}

\begin{figure}[t]
\begin{center}
\includegraphics[angle=270,width=3.5in]{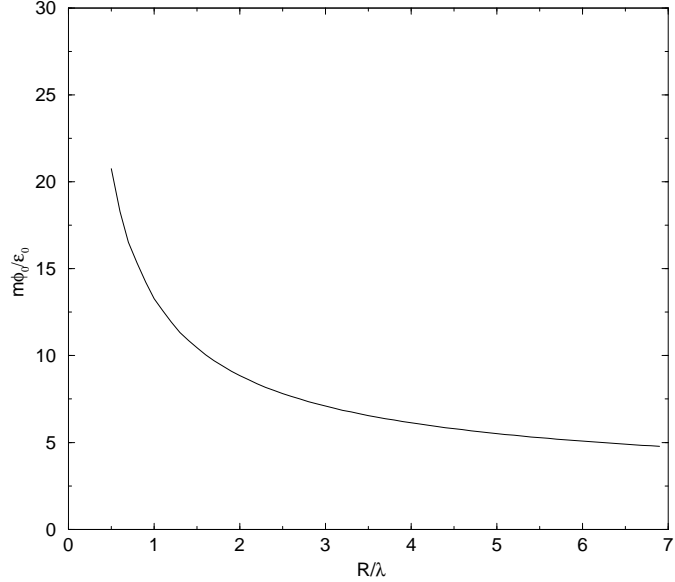}
\caption{\label{parphase}
Phase diagram for vortices-anti-vortices
induced by the magnetic dot with in-plane magnetization.
}
\end{center}
\end{figure}

\noindent The critical curve in the plane of two dimensional
ratios $\frac{m_0\Phi_0}{\epsilon_0}$ and $\frac{R}{\lambda}$ is
plotted numerically in figure  (\ref{parphase}). The creation of 
vortex-anti-vortex is energy unfavorable in the region below this 
curve and favorable above it. The 
phase diagram suggests that the smaller is the radius $R$ of the
dot the larger value $\frac{m_0\Phi_0}{\epsilon_0}$ is necessary
to create the vortex-anti-vortex pair. At large values of $R$ and
$m_0\Phi_0\geq\epsilon_0$, the vortex is separated by a large
distance from the antivortex. Therefore, their energy is
approximately equal to that of two free vortices. This positive
energy is compensated by the attraction of the vortex and
antivortex to the magnetic dot. The critical values of
$m_0\Phi_0/\epsilon_0$ seems to be numerically large even at
$R/\lambda\sim 1$. This is probably a consequence of comparably
ineffective interaction of in-plane magnetization with the vortex.

Magnetic dots with a finite thickness were considered by Milosevic
{\it et al.} \cite{peeters1,peeters2,peeters3}. No qualitative changes of
the phase diagram or magnetic fields were reported.

\subsection{Array of Magnetic Dots and  Superconducting Film  \label{dots}}

\subsubsection{Vortex Pinning by Magnetic Dots \label{dotpin}}

\begin{figure}[t]
\begin{center}
\includegraphics[width=3.in]{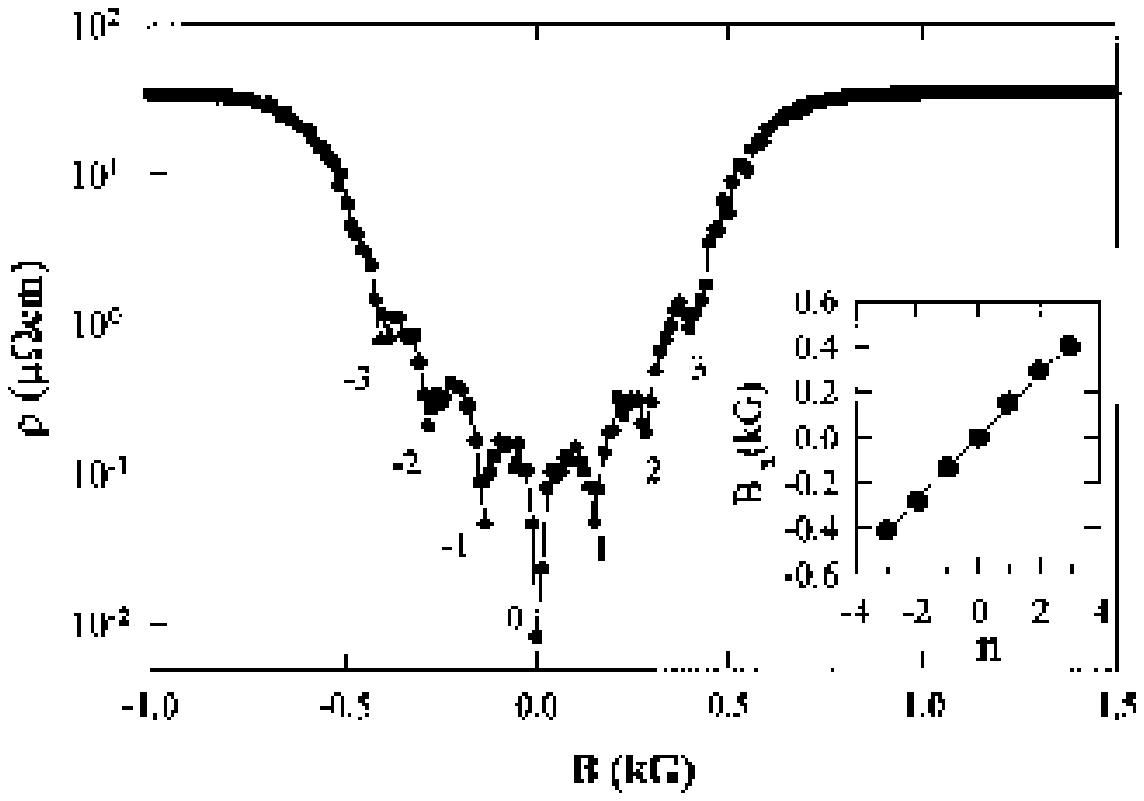}
\includegraphics[width=3.in]{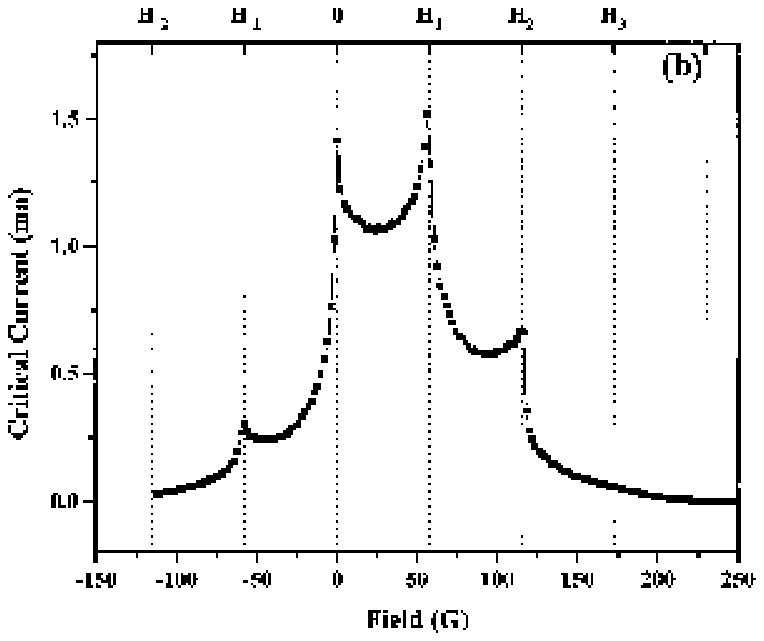}
\caption{\label{first}
Left: Field dependence of the resistivity of a Nb thin film
with a triangular array of Ni dots.
(From  Martin {\it et al.} \cite{shull}). \newline
Right: Critical current as a function of field for the
high density triangular array at T = 8.52 K  T$_c$= 8.56 K    
(From  Morgan and Ketterson \cite{ketterson}).
}
\end{center}
\end{figure}

Vortex pinning in superconductors  is of the great practical
importance. First time the artificial vortex pinning was created
by S-film thickness modulation in seventies. Martinoli {\it et
al.} \cite{martinoli} have used grooves on the film surface to pin
vortices and Hebard {\it et al.} \cite{hebard1,hebard2} have used
triangular arrays of holes. 
Magnetic structures provide additional
possibilities to pin vortices. First experiments were performed in
the Louis Neel lab in Grenoble \cite{parallel,parallel1}. These
experiments were performed with dots several microns wide with the
magnetization  parallel to the superconducting film. They observed
oscillations of the magnetization vs magnetic field. These
oscillation was attributed to a simple matching effect: pinning
becomes stronger when vortex lattice is commensurate with the
lattice of pinning centers. This can be measured in terms of
external, normal to the film magnetic field needed to generate
integer number of vortices per unit cell of the pinning array.

Flux pinning by a triangular array of submicron size dots with
typical spacing 400-600nm and diameters close to 200nm magnetized
in-plane was first reported by Martin {\it et al.} \cite{shull}.
Oscillations of the resistivity  with increasing flux were
observed with period corresponding to one flux quanta per unit
cell of magnetic dot lattice (see figure  \ref{first}Left.
This can be  explained by the matching effect. Though
matching effect is not specific to magnetic pinning arrays,
enhanced pinning with magnetic dots with magnetization
parallel to the film was observed by Martin {\it et
al.} \cite{shull}.

Dots array with out-of-plane magnetization component was first
created and studied by Morgan and Ketterson \cite{ketterson}. They
have measured critical current as a function of the external magnetic
field and found strong asymmetry of the pinning properties vs magnetic
field direction (see figure \ref{first}Right).
This experiment has given the first direct experimental evidence
that the physics of vortex pinning by magnetic dots is different
from that of common pinning centers.

Pinning properties of the magnetic dots array depends on several
factors: magnetic moment orientation, the strength of the stray
field, the ratio of the dot size and the dot lattice constant to
the effective penetration depth, array magnetization, the strength
and direction of the external field, etc. The use of magnetic imaging
technique, namely Scanning Hall Probe Microscope (SHPM) and 
Magnetic Force Microscope (MFM) has revealed  exciting pictures
of vortex ``world''. Such studies in combination with traditional
measurements gives new insight in  vortex physics. This work
was done mainly by the group at the University of 
Leuven.   Below we briefly discuss only a
few cases studied in great details by this group.

\begin{figure}
\begin{center}
\includegraphics[width=3.5in]{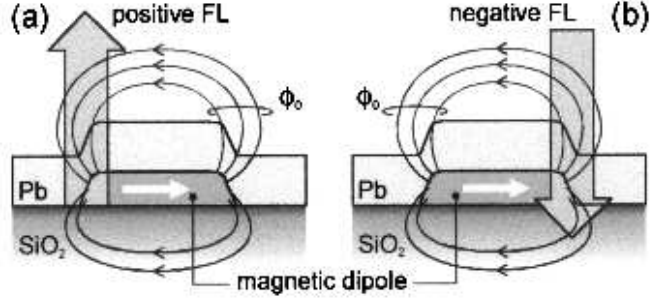}
\caption{\label{dotparal}
Schematic presentation of the polarity dependent flux
pinning, presenting the cross section of a Pb film deposited over
a magnetic dipole with in-plane magnetization: (a) A positive
FL (wide gray arrow) is attached to the dot at the pole where
a negative flux quantum is induced by the stray field (black
arrows), and (b) a negative FL is pinned at the pole where a
positive flux quantum is induced by the stray field.
(From  Van Bael {\it et al.}\cite{mosh13})
}
\end{center}
\end{figure}

{\it Dots with Parallel Magnetization}. Van Bael {\it et al.}\cite{mosh13}
studied with Scanning Hall Probe Microscope (SHPM) the
magnetization and vortex distribution in a square array (1.5$\mu$m
period) of rectangular (540nmX360nm)cobalt trilayer
Au(7.5nm)/Co(20nm)/Au(7.5nm) dots with magnetization along the
edges of the dots lattice. SHPM images have revealed magnetic
field redistribution due to superconducting transition in the
covered 50nm thin lead superconducting film. These data were
interpreted by Van Bael {\it et al.}\cite{mosh13} as formation of
vortices of opposite sign on both sides of the dot. By applying
external magnetic field Van Bael {\it et al.}\cite{mosh13} have
demonstrated the commensurate lattice of vortices residing on the
``end'' of magnetized dots. This location is in agreement with
theoretical prediction \cite{EKLP}. Remarkably, they were able to
observe ``compensation'' of the vortices created by the dots stray
field with vortices of the opposite sign due to applied normal
field (see figure  \ref{dotparal}).

\begin{figure}
\begin{center}
\includegraphics[width=3.5in]{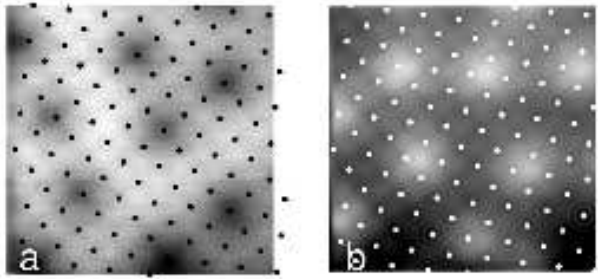}
\caption{\label{vorperp}
SHPM images of a (10.5$\mu$m)$^2$ area of the sample in
H=-1.6 Oe (left panel) and H=1.6 Oe (right panel), at 
T=6.8 K (field-cooled). The tiny black/white dots indicate the posi-
tions of the Co/Pt dots, which are all aligned in the negative sense
(m $<$ 0). The flux lines emerge as diffuse dark (H $<$ 0) or bright
(H $>$ 0) spots in the SHPM images.
(From  Van Bael {\it et al.}\cite{mosh11})
}
\end{center}
\end{figure}

{\it Dots with Normal Magnetization}. Van Bael {\it et al.}\cite{mosh11} have
elucidated with the SHPM images the nature of previously reported (see e.g.
work  \cite{ketterson})
anisotropy in the vortices pinning by the array of dots with
normal magnetization. They have used 1$\mu$m period lattice  of
square, 400nm side length and 14nm thin, Co/Pt multilayer dots
covered with 50nm thin lead film. Zero field SHPM images show the
checkerboard-like distribution of magnetic field (see Sec. 3.3.4)
The stray field from the dots were not sufficient to create
vortices. In a very weak (1.6 Oe) external field the average
distance between vortices was about 4 lattice spacings. In the
case of the field parallel to the dots magnetization vortices
reside on the dots, as the SHPM image shows (see figure \ref{vorperp}a). 
In the case of the
same field with opposite direction, the SHPM shows vortices
located at interstitial positions in the magnetic dots lattice
 (see figure  \ref{vorperp}b). It
is plausible that the pinning barriers are lower in the second
case.

Figure \ref{mhperp} shows dependance of superconduction film
magnetization versus applied magnetic field normal to the film.
Moshchalkov {\it et al.}\cite{krittok} have shown that magnetic
field dependence of film magnetization of the superconducting
film is very similar to the critical current dependence on magnetic
field. Figure   \ref{mhperp} shows strong anisotropy of the pinning
properties on the external magnetic field direction. Magnetic field 
parallel to the dots magnetic moment shows much stronger vortex
pinning than antiparallel.  
\begin{figure}
\begin{center}
\includegraphics[width=3.5in]{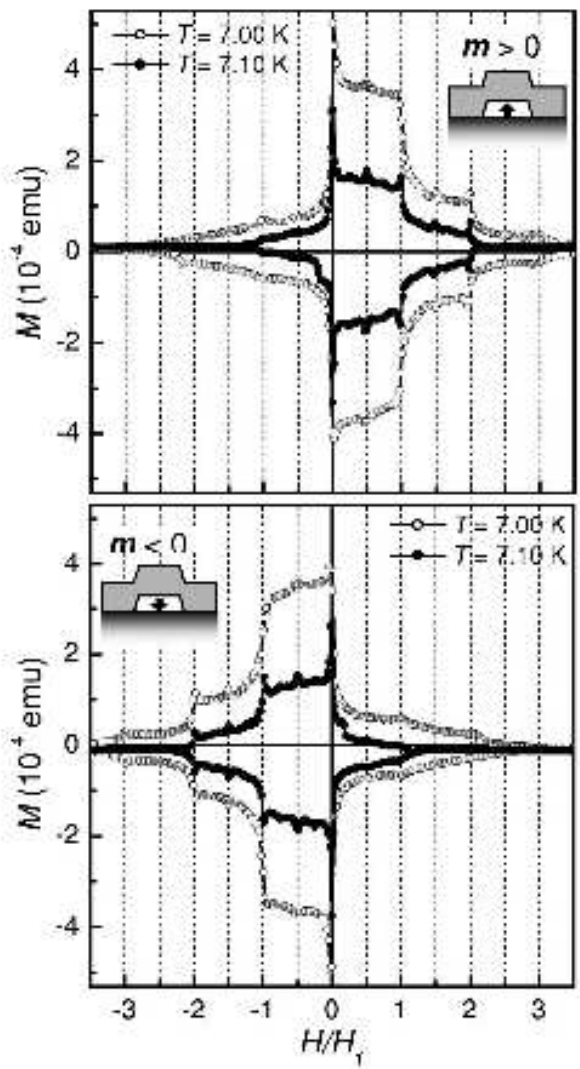}
\caption{\label{mhperp}
 M (H/H$_1$) magnetization curves at different temperatures 
near T$_c$  (7.00 K open symbols, 7.10 K filled symbols) showing
the superconducting response of the Pb layer on top of the Co/Pt dot
array with all dots aligned in a positive upper panel and negative
lower panel sense. H$_1$=20.68 Oe is the first matching field.
(From  Van Bael {\it et al.}\cite{mosh11})
}
\end{center}
\end{figure}

\subsubsection{Magnetic Field Induced Superconductivity \label{lange}}

\begin{figure}[t]
\begin{center}
\includegraphics[width=3.5in]{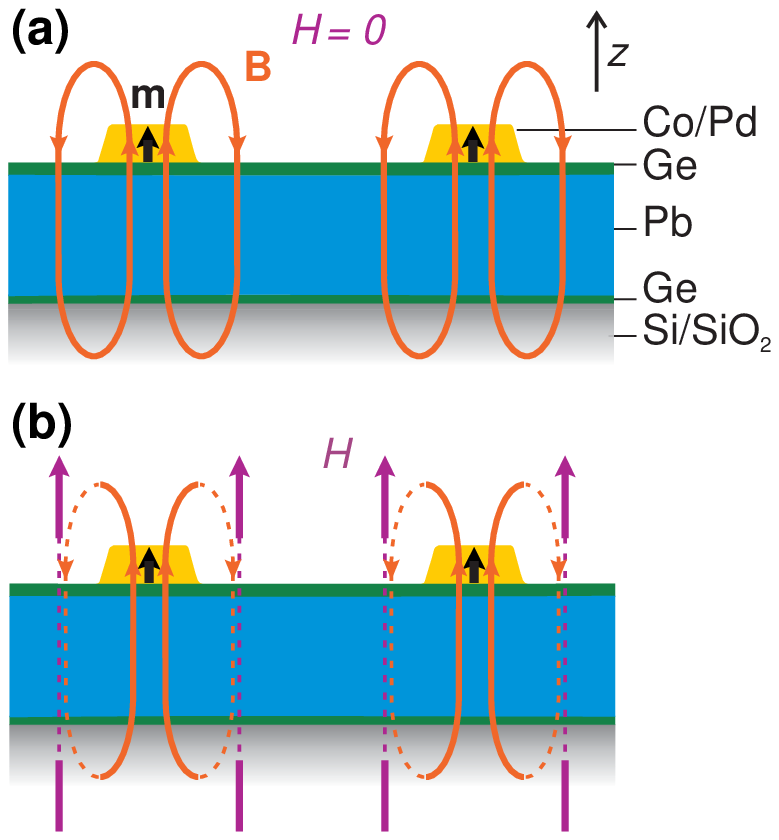}
\caption{\label{flux}
Schematical magnetic field distribution in the
the array of dots with normal to the superconducting film
magnetization. (From  Lange {\it et al.} cond-mat/0209101)
}
\end{center}
\end{figure}

Consider a regular array of magnetic dots placed upon a
superconducting film with magnetization normal to the film. For
simplicity we consider very thin magnetic dots. Namely this
situation is realized magnetic films with normal magnetization
used in experiment  \cite{mosh10}). The net flux from the magnetic
dot through any plane including the surface of the superconducting
film (see figure \ref{flux}) is exactly zero. Suppose that on the top
of the magnetic dot the z-component of the magnetic field is
positive as shown in the mentioned figure. Due to the requirement
of zero net flux the z-component of the magnetic field between the
dots must be negative. Thus, S-film occurs in a negative magnetic
field normal to the film. It can be partly or fully compensated by
an external magnetic field parallel to the dot magnetization (see
figure \ref{flux}). Such a compensation can be even more effective in
for a regular array of magnetic wires embedded in alumina template
\cite{ln,ln3,trap}. 
Lange {\it et al.}\cite{mosh10} have proposed this trick and
reached a positive shift of the S-transition temperature in an
external magnetic field, the result looking counterintuitive if
one forgets about the field generated by the dots. In this
experiment a thin superconducting film was covered with a square
array of the CoPd magnetic dots with normal to the film
magnetization. The dots had square shape with the side 0.8$\mu$m,
the thickness 22nm and the dot array period 1.5$\mu$m. The H-T
phase diagrams presented in \cite{mosh10} for zero and finite dots
magnetization demonstrate appearance of the superconductivity by
applying magnetic field parallel to the dot magnetization. At
T=7.20K the system with magnetized dots is in normal state. It
undergoes transition to the superconducting state in the field
0.6mT and back to the normal state at 3.3mT. From the data in
figure 3 in work by Lange {\it et al.}\cite{mosh10} one can conclude that
the compensating field is about 2mT.

\subsubsection{Magnetization  Controlled Superconductivity \label{dima}}

   \begin{figure}
   \begin{center}
\includegraphics[width=2.5in]{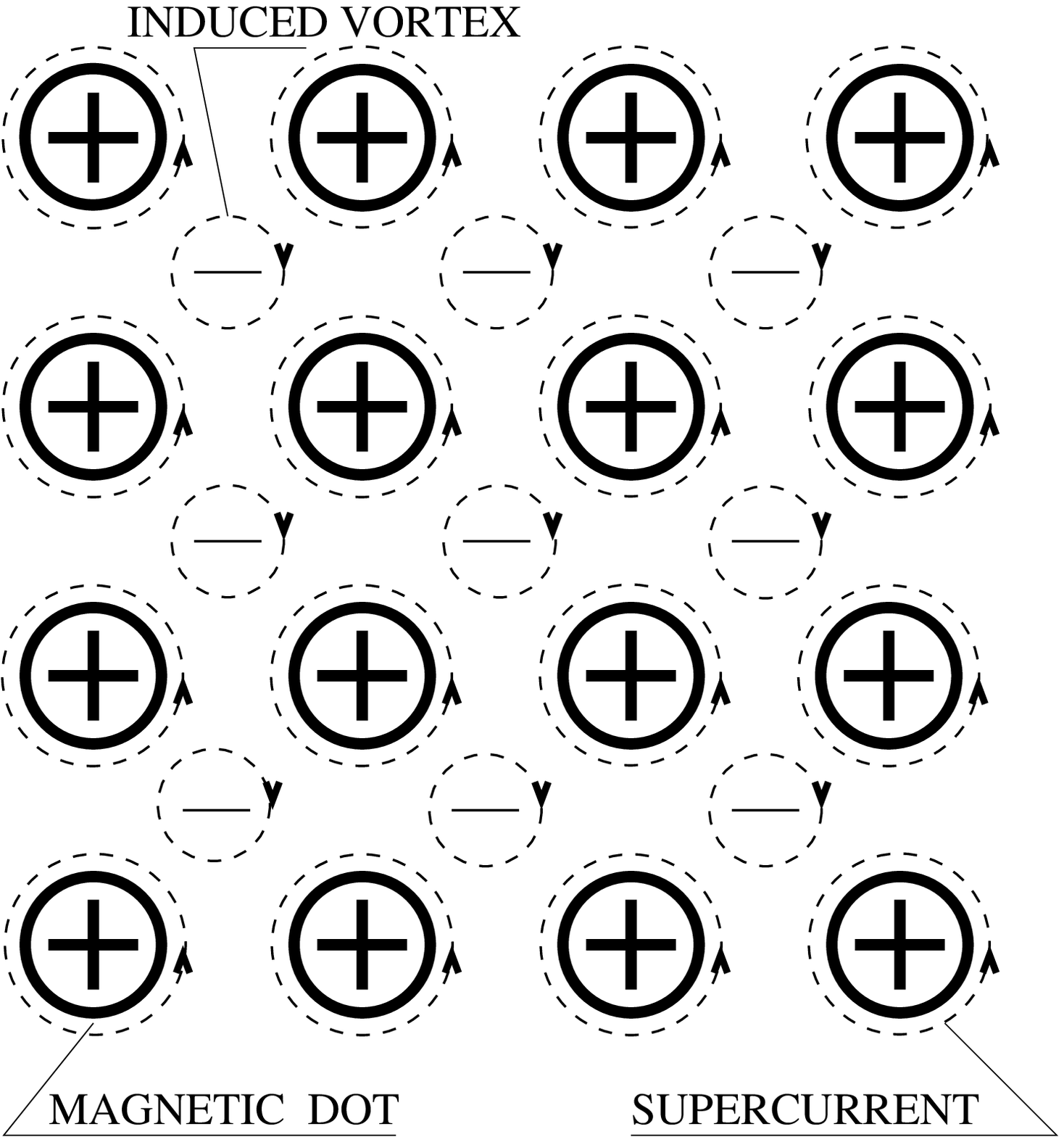}
\hskip 2.truecm
\includegraphics[width=2.5in]{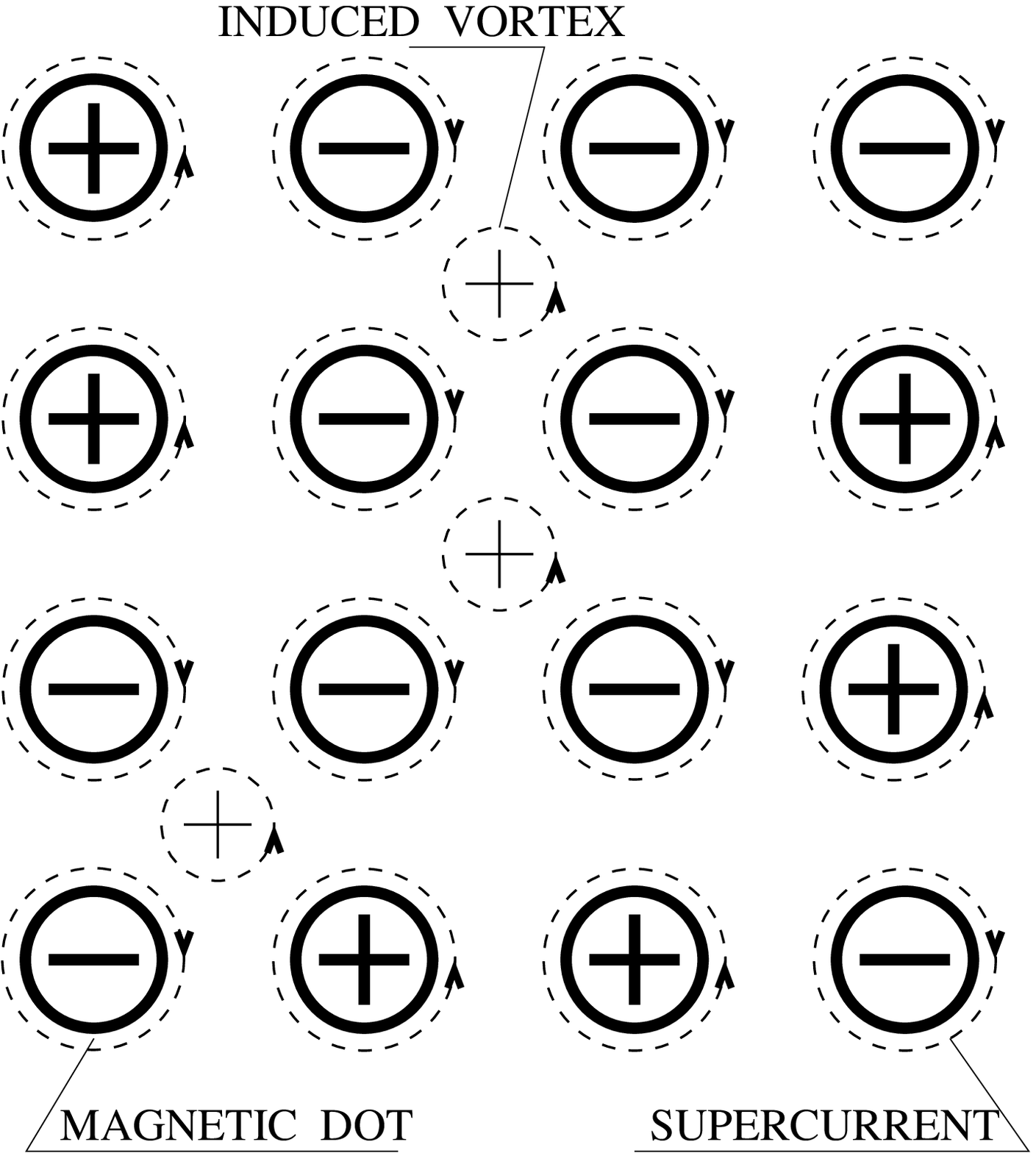}
\caption{ \label{dots0} Left: Magnetized magnetic dots array.
Vortices of different signs are shown schematically by supercurrent 
direction (dashed lines). The magnetic moment direction is indicated by 
$\pm$. Both vortices bound by dots and created spontaneously are shown.
Magnetized array of dots create regular lattice 
of vortices and antivortices and provide strong pinning.
 Right:  Demagnetized magnetic dots array results in  strongly fluctuating
random potential which creates unbound antivortices/vortices, thus
transforming superconducting film into resistive state. 
}
   \end{center}
   \end{figure}

Above (Sec. \ref{lange}) we have discussed example when application of
magnetic field can transform FSH system from normal to superconducting
state. This was due compensation of the dots stray magnetic field 
with external magnetic field.

Earlier  Lyuksyutov and Pokrovsky \cite{dotlp}
have discussed theoretically situation when demagnetized array of
magnetic dots with normal  magnetization create resistive
state in the coupled superconducting film. However, superconducting
state can be restored by magnetization of the dots array. 
This 
counter intuitive phenomena can be explained on qualitative level.
In the case when single dot creates one vortex, 
magnetized array of dots results in periodical vortex/antivortex
structure with anti-vortices localized at the centers of the unit
cells of  the square  lattice of dots as shown 
in figure \ref{dots0}Left. Such order provides strong 
pinning. 
More interesting is demagnetized state in which
the induced vortices and antivortices create a
random field for a probe vortex. If the lattice constant of the
array $a$ is less than the effective penetration depth $\lambda$,
the random fields from vortices are logarithmic. The effective
number of random logarithmic potentials acting on a probe vortex
is $N=(\lambda /a)^2$ and the effective depth of potential well
for a vortex (antivortex) is $\sqrt{N}\epsilon_v$. At proper
conditions, for example near the S-transition point, the potential
wells can be very deep enabling the spontaneous generation of the
vortex-antivortex pairs at the edges between potential valleys and
hills. The vortices and antivortices will screen these deep wells
and hills similarly to the screening in the plasma. The difference
is that, in contrast to plasma, the screening "charges" do not
exist without external potential. In such a flattened
self-consistent potential relief the vortices have percolated
infinite trajectories passing through the saddle points
\cite{rdot}. The drift motion of the delocalized vortices and
antivortices in the external field generates dissipation and
transfer the S-film into the resistive state 
(see figure \ref{dots0}Right). 
Replacing slow
varying logarithmic potential by a constant at distances less than
$\lambda$ and zero at larger distances, Feldman {\it et al.} have
found thermodynamic and transport characteristics of this system.
Below we briefly outline their main results.
   \begin{figure}[t]
   \begin{center}
\includegraphics[width=2.5in]{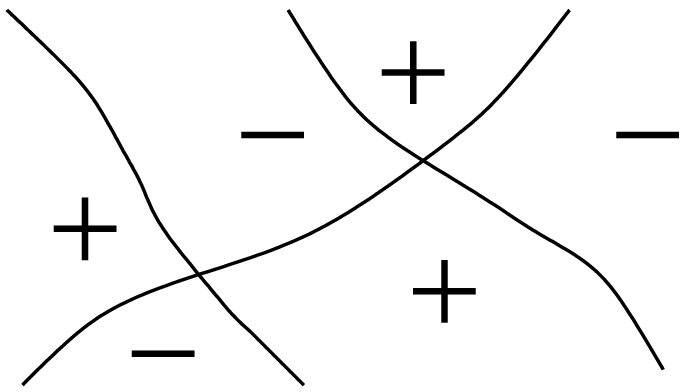}
\hskip 2.truecm
\includegraphics[width=2.5in]{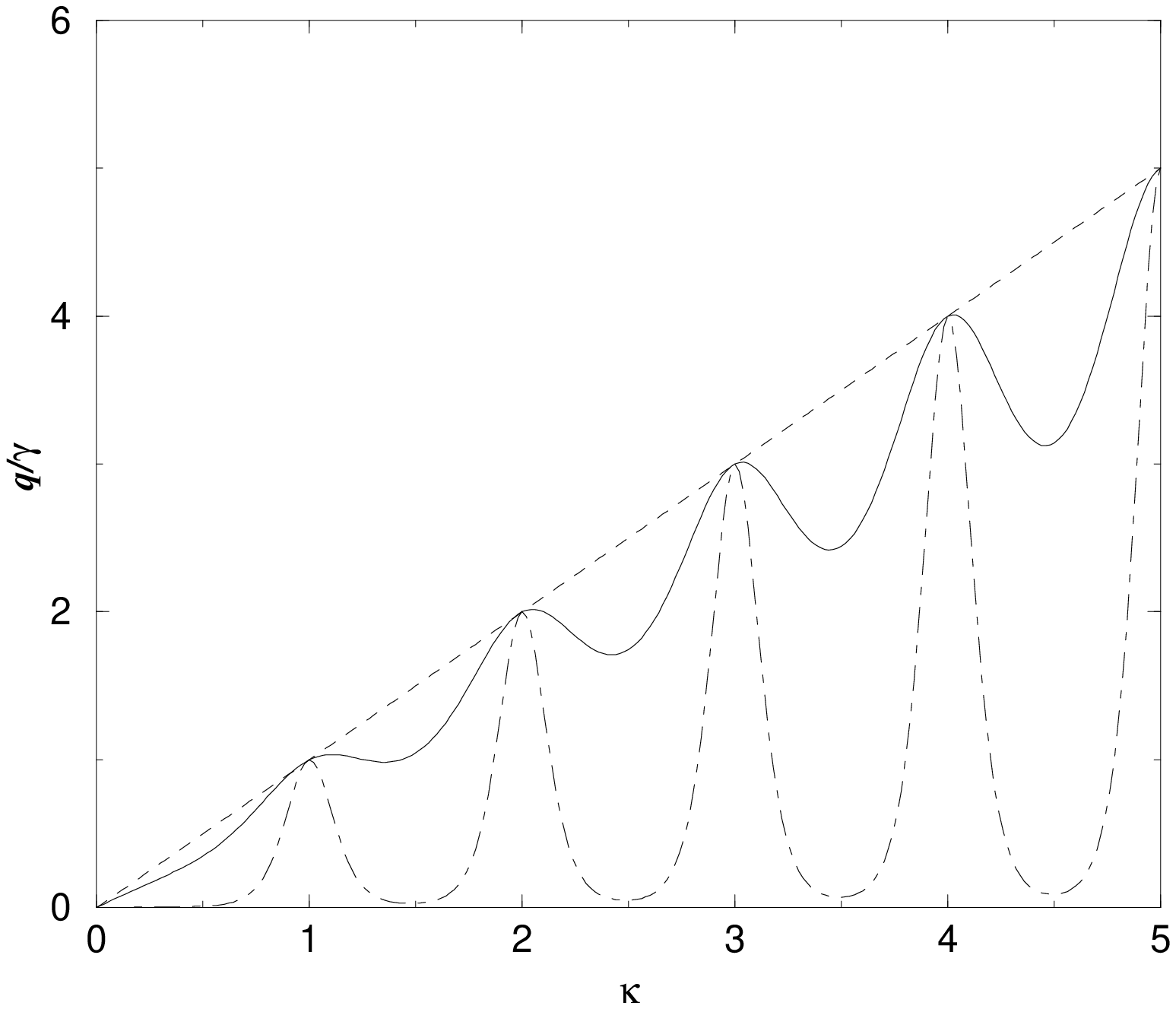}
   \end{center}
\caption{Left:The checker-board average
structure of the vortex plasma.
Right: The average number of the unbound vortices
in the cell of size $a$ via the parameter
$\kappa$ proportional to the dot magnetic moment.
Dot-dashed line corresponds to $T/\epsilon_0=0.15$,
solid line corresponds to $T/\epsilon_0=0.4$,
dashed line corresponds to $T/\epsilon_0=2$.
\label{dots2} 
}
\end{figure}
For the sake of simplicity we replace
this slow varying potential $V({\bf r})$ 
by a potential having a constant
value within the single cell:
$V_{0}=2\epsilon_0$ at the distance $r<\lambda$ 
and zero at $r>\lambda$, where
$\epsilon_0=\Phi_0^2/(16\pi^2\lambda)$, $\Phi_0$ is the magnetic flux
quantum. Considering the film as a set of
almost unbound cells of the linear size $\lambda$ 
we arrive at the following
Hamiltonian for such a cell:

\begin{equation}
\label{ham}
H=-U\sum_i\sigma_i n_i +\epsilon\sum_i n_i^2
+2\epsilon_0\sum_{i>j}n_i n_j,
\end{equation}
where $n_i$ is integer vorticity on either a dot and or a
site of the dual lattice (between the dots) which we
conventionally associate with location of unbound vortices.
$\sigma_i=\pm 1$,
where subscript $i$ relates to the dot, 
describes the random sign of the dot
magnetic moments.
$\sigma_i=0$ on the sites of dual lattice.
The first term of the Hamiltonian (\ref{ham}) 
describes the binding energy
of the vortex at the magnetic dot 
and $U\approx\epsilon_0\Phi_d/\Phi_0$,
with $\Phi_d$ being the magnetic flux through a single dot. The second
term in the Hamiltonian is the sum of single vortex energies,
$\epsilon=\epsilon_0\ln (\lambda_{\rm eff}/a)$,
where $a$ is the period of the dot array, $\xi$ is the
superconducting coherence length. The third term mimics the
intervortex interaction. Redefining the constant $\epsilon$,
one can replace the last term of equation  (\ref{ham}) by
$\epsilon_0(\sum n_i)^2$.
The sign of the vorticity on a dot follows two possible (`up'- and
`down-') orientations of its magnetization.
The vortices located between the dots ($n_i$ on the
dual lattice) are correlated on the scales of order $\lambda$
and form the above-mentioned irregular checker-board potential relief.

\begin{figure}
   \begin{center}
\includegraphics[width=2.5in]{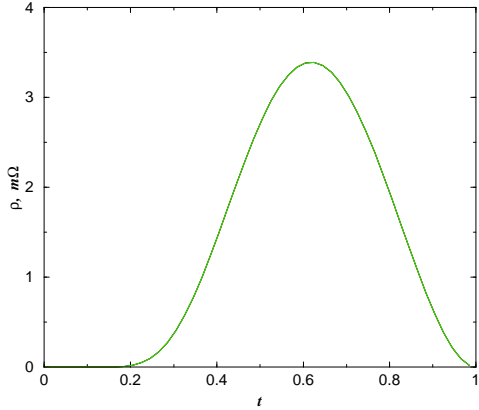}
\caption{The static resistance $\rho$ 
of the film vs dimensionless
temperature $t=T/T_c$ at
typical values of parameters.
\label{F1}
}
\end{center}
\end{figure}

To find the ground state, we consider
a cell with large number of the dots of
each sign $\sim(\lambda/a)^2\gg 1$. The energy (\ref{ham}) is minimal
when the "neutrality" condition
$Q\equiv\sum n_i=0$ is satisfied. Indeed, if $Q\neq 0$ the interaction
energy grows as $Q^2$, whereas the first term
of the Hamiltonian behaves as $|Q|$ and can not compensate the last
one unless $Q\sim 1$. The neutrality constraint means that the unbound
vortices screen almost completely the ``charge"
of those bound by dots, that is
$K\sim (N_+-N_-)\sim\sqrt{N_\pm}\sim\lambda/a$
where $K$ is the difference between the numbers of the positive
and negative dots and $N_{\pm}$ are the numbers of the positive and
negative vortices, respectively. Neglecting the total charge $|Q|$ as
compared with
$\lambda/a$, we minimize the energy (\ref{ham}) 
accounting for the neutrality constraint.
At $Q=0$ the
Hamiltonian (\ref{ham}) can be written as the sum of one-vortex
energies:
\begin{equation}
\label{singleold}
H=\sum H_i; \qquad H_i=-U\sigma_in_i + \epsilon n_i^2.
\end{equation}

The minima for any $H_i$ is achieved by choosing
$n_i=n_i^{0}$, an integer closest to the magnitude 
$\sigma_i\kappa=\sigma_i U/(2\epsilon)$. 
The global minimum consistent with the neutrality
is realized by values of $n_i$ that differ from the
local minima values $n^0_i$ not more than over $\pm 1$.
Indeed, in the configuration with $n_i=n^0_i$,
the total charge is $|\sum n_i^0|\sim\kappa|\sum\sigma_i|=\kappa K$.
Hence, if $\kappa\ll \lambda/a$, then the change of the vorticity at a
 small
part of sites by $\pm 1$ restores neutrality. To be more specific
let us consider $K>0$. Let $\bar n$ be the integer closest to $\kappa$,
and consider the case $\kappa<\bar n$. Then the minimal
energy corresponds to a configuration with the vorticity $n_i=-\bar n$
at each negative dot and with the vorticity $\bar n$ or $\bar n-1$ at
positive dots. The neutrality constraint implies that the
number of positive dots with the vorticity $\bar n-1$ is $M=K\bar
n$. In the opposite case $\kappa>\bar n$ the occupancies of all the
positive dots are $\bar n$; whereas, the occupancies of the negative
dots are either $\bar n$ or $\bar n+1$. Note that in our model
the unbound vortices are absent in the ground state
unless $\kappa$ is an integer. Indeed,
the transfer of a vortex from a dot with the occupancy $n$ to
a dual site changes the energy by $\Delta E=2\epsilon(\kappa - n+2)$.
Hence, the energy transfer is zero if and 
only if $\kappa$ is an integer,
otherwise the energy change upon 
the vortex transfer is positive. At integer
$\kappa$, the number of the unbound vortices
can vary from $0$ to $K\bar n$ without change of energy. The
ground state is degenerate at any non-integer $\kappa$ since,
while the total number of the dots with the different vorticities
are fixed, the vortex exchange between two dots with
the vorticities $n$ and $n\pm 1$ does not change the total energy.
Thus, our model predicts a step-like dependence of dot occupancies
on $\kappa$ at the zero temperature
and peaks in the concentration of unbound vortices
as shown in figure \ref{dots2}.
The data for finite temperature were  calculated
in the  Ref.\cite{rdot}
The dependencies of the unbound vortex concentration on $\kappa$
for several values of $x=\epsilon/T$ are shown in figure  \ref{dots2}.
Oscillations are well pronounced for $x\gg 1$ and are suppressed
at small $x$ (large temperatures).
At low temperatures, $x\gg 1$, the half-widths of
the peaks in the density of the unbound vortices are 
$\Delta\kappa\approx 1/x$ and the heights of peaks 
are $\approx \gamma n$, where $\gamma =K/N$.

{\it Vortex transport--} At moderate external currents
$j$ the vortex transport and dissipation are
controlled by unbound vortices.
The typical energy barrier associated with the
vortex motion is $\epsilon_0$. The unbound vortex
density is $m\sim a^{-2}\gamma\sim(a\lambda)^{-1}$ and oscillates with
$\kappa$ as it was shown above. The average distance
between the unbound vortices is $l\sim\sqrt{a\lambda}$.
The transport current exerts the Magnus (Lorentz) force
$F_M=j\Phi_0/c$ acting on a vortex. Since the
condition $T\ll \epsilon_0$ is satisfied in the vortex state everywhere
except for the regions too close
to $T_c$, the vortex motion
occurs via  thermally activated jumps with the rate:

\begin{equation}
\label{dr}
\nu=\nu_0\exp(-\epsilon_0/T)=(\mu
j\Phi_0/cl)\exp(-\epsilon_0/T),
\end{equation}
where $\mu=(\xi^2\sigma_n)/(4\pi e^2)$ is the Bardeen-Stephen vortex
mobility \cite{BS}. The induced electric field  is accordingly

\begin{equation}
\label{field}
E_c=l\dot B/c=m\Phi_0 \nu l/c,
\end{equation}
The Ohmic losses per unbound vortex are 
$W_c=jE_c\lambda a=j\Phi_0 \nu l/c$
giving rise to the dc resistivity as
\begin{equation}
\label{res}
\rho_{\rm dc}=\frac{W_c}{j^2\lambda^2}=
\frac{\mu\Phi_0^2}{c^2\lambda^2}
\exp[-\epsilon_0(T)/T]
\end{equation}
Note the non-monotonic dependence of $\rho_{\rm dc}$
on temperature $T$  figure \ref{F1}.
The density  of the unbound vortices
is the oscillating function of the flux through a dot.
The resistivity of such a system is determined 
by thermally activated jumps of vortices
through the corners of the irregular 
checkerboard formed by the positive
or negative unbound vortices
and oscillates with $\Phi_d$. These
oscillations can be observed by additional deposition (or removal)
of the magnetic material to the dots.

\subsection{Ferromagnet - Superconductor Bilayer  \label{fsb}}

\subsubsection{Topological Instability in the FSB \label{order}}

Let us consider a F/S bilayer with both layers infinite and
homogeneous. An infinite magnetic film with ideal parallel
surfaces and homogeneous magnetization generates no magnetic field
outside. Indeed, it can be considered as a magnetic capacitor, the
magnetic analog of an electric capacitor, and therefore its
magnetic field confined inside. Thus, there is no direct
interaction between the homogeneously magnetized F-layer and a
homogeneous S-layer in the absence of currents in it. However,
Lyuksyutov and Pokrovsky argued \cite{vmlp} that such a system is
unstable with respect to spontaneous formation of vortices in the
S-layer. Below we reproduce these arguments.

Assume the magnetic anisotropy to be sufficiently strong to keep
magnetization perpendicular to the film (in the $z$-direction). As
we have demonstrated above, the homogeneous F-film creates no
magnetic field outside itself. However, if a Pearl vortex somehow 
appears in the superconducting film, it generates magnetic field
interacting with the magnetization ${\bf m}$ per unit area of the
F-film. At a proper circulation direction in the vortex and the
rigid magnetization ${\bf m}$ this field decreases the total
energy over the amount $m\int B_z({\bf r})d^2x=m\Phi$, where
$\Phi$ is the total flux. We remind that each Pearl vortex carries
the flux equal to the famous flux quantum $\Phi_0=\pi\hbar c/e$.
The energy necessary to create the Pearl vortex in the isolated
S-film is $\epsilon_{v}^{(0)} =\epsilon_0 \ln (\lambda/\xi)$
\cite{pearl}, where $\epsilon_0 = \Phi_0^2/16\pi^2 \lambda$,
$\lambda = \lambda_{L}^{2}/d$ is the effective penetration
depth\cite{abrikosov}, $\lambda_{L}$ is the London penetration
depth, and  $\xi$ is the coherence length. Thus, the total energy
of a single vortex in the FSB is:
\begin{equation}
\epsilon_v = \epsilon_{v}^{(0)} - m \Phi_0, \label{ven}
\end{equation}
\noindent and the FSB becomes unstable with respect to spontaneous
formation vortices as soon as $\epsilon_v$ turns negative.  Note
that close enough to the S-transition temperature $T_s$, 
$\epsilon_v$ is definitely negative since the S-electron density 
$n_s$ and, therefore, $\epsilon_{v}^{(0)}$ is zero at $T_s$. If 
$m$ is so small that $\epsilon_v>0$ at $T=0$, the instability 
exists in a temperature interval $T_v<T<T_s$, where $T_v$ is defined 
by equation $\epsilon_v(T_v)=0$. Otherwise instability persists 
till $T=0$.

\begin{figure}[t]
\begin{center}
\includegraphics[width=3.5in]{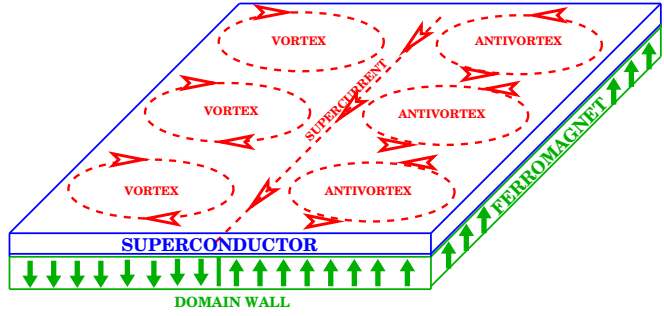}
\caption{\label{dwvor} Magnetic domain wall and coupled arrays of
superconducting vortices with opposite vorticity. Arrows show the
direction of the supercurrent. }
\end{center}
\end{figure}

A newly appearing vortex phase cannot consist of the vortices of
one sign. More accurate statement is that any finite, independent
on the size of the film $L_f$ density of vortices is energetically 
unfavorable in the thermodynamic limit $L_f\rightarrow\infty$.
Indeed, any system with the non-zero average vortex density $n_v$
generates a constant magnetic field $B_z = n_v \Phi_0$ along the
$z$ direction. The energy of this field for a large but finite
film of the linear size $L_f$ grows as $L_f^3$ exceeding the gain
in energy due to creation of vortices proportional to $L_f^2$ in
thermodynamic limit. Thus, paradoxically the vortices appear, but
can not proliferate to a finite density. This is a manifestation
of the long-range character of magnetic forces. The way from this
controversy is similar to that in ferromagnet: the film should
split in domains with alternating magnetization and vortex
circulation directions. Note that these are combined topological 
defects: vortices in the S-layer and domain walls in the F-layer.
They attract each other. The vortex density is higher near the
domain walls. The described texture represents a new class of 
topological defects which does not appear in isolated S and F
layers. We show below that if the domain linear size $L$ is much 
greater than the effective penetration length $\lambda$, the most
favorable arrangement is the stripe domain structure (see
(figure \ref{dwvor})). The quantitative theory of this structure was
given by Erdin {\it et al.} \cite{ELPV}.

The total energy of the bilayer can be represented by a sum: 
\begin{equation}
U\,=\,U_{sv}+U_{vv}+U_{vm}+U_{mm}+U_{dw} \label{energy}
\end{equation}
where $U_{sv}$ is the sum of energies of single vortices; $U_{vv}$ 
is the vortex-vortex interaction energy; $U_{vm}$ is the energy of
interaction between the vortices and magnetic field generated by
domain walls; $U_{mm}$ is the self-interaction energy of magnetic
layer; $U_{dw}$ is the linear tension energy of domain walls. We
assume the 2d periodic domain structure consisting of two 
equivalent sublattices. The magnetization $m_z({\bf r})$ and 
density of vortices $n({\bf r})$ alternate when passing from one 
sublattice to another. Magnetization is supposed to have a
constant absolute value: $m_z({\bf r})=ms({\bf r})$, where $s({\bf
r})$ is the periodic step function equal to +1 at one sublattice
and -1 at the other one. We consider a dilute vortex system in 
which the vortex spacing is much larger than $\lambda$. Then the 
single-vortex energy is:
\begin{equation}
U_{sv}\,=\,\epsilon_v\int n({\bf r})s({\bf r})d^2x;\,\,\,\,
\epsilon_v = \epsilon_v^{(0)} - m \Phi_0 \label{single}
\end{equation}
\noindent The vortex-vortex interaction energy is:
\begin{equation}
U_{vv}\,=\,\frac{1}{2}\int n({\bf r})V({{\bf r-r}^{\prime}})n({\bf
r}^{\prime})d^2xd^2x^{\prime}, \label{vv}
\end{equation}
where $V({\bf r - r}^{\prime})$ is the pair interaction energy
between vortices located at points ${\bf r}$ and ${\bf
r}^{\prime}$.  Its asymptotics at large distances $\mid{\bf r -
r}^{\prime}\mid\gg \lambda$ is $V({\bf r -
r}^{\prime})=\Phi_0^2/(4\pi^2\mid{\bf r - r}^{\prime}\mid)$
\cite{degennes}. This long-range interaction is induced by
magnetic field generated by the Pearl vortices and their slowly 
decaying currents\footnote{From this long-range interaction of the 
Pearl vortices it is ready to derive that the energy of a system 
of vortices with the same circulation, located with the permanent 
density $n_v$ on a film having the lateral size $L$, is 
proportional to $n_v^2L^3$}. The energy of vortex interaction with 
the magnetic field generated by the magnetic film looks as follows
\cite{EKLP}:
\begin{equation}
U_{vm}\,=\,-\frac{\Phi_0}{8 \pi^2 \lambda}\int \nabla \varphi
({\bf r} - {\bf r^\prime}) n({\bf r}^\prime)\cdot {\bf
a}^{(m)}({\bf r})d^2xd^2x^{\prime} \label{vm}
\end{equation}
Here $\varphi ({\bf r} - {\bf
r^\prime})=\arctan\frac{y-y^{\prime}}{x-x^{\prime}}$ is a phase
shift created at a point ${\bf r}$ by a vortex centered at a point
${\bf r}^{\prime}$ and ${\bf a}^{(m)}({\bf r})$ is the value of
the vector-potential induced by the F-film upon the S-film. This 
part of energy similarly to what we did for one vortex can be
reduced to the renormalization of the single vortex energy with
the final result already shown in equation (\ref{single}). The
magnetic self-interaction reads:
\begin{equation}
U_{mm}\,=\,-\frac{m}{2}\int B_z^{(m)}({\bf r})s({\bf r}) d^2x
\label{mm}
\end{equation}
Finally, the domain walls linear energy is
$U_{dw}=\epsilon_{dw}L_{dw}$ where $\epsilon_{dw}$ is the linear
tension of the domain wall and $L_{dw}$ is the total length of the
domain walls.

Erdin {\it et al.}\cite{ELPV} have compared energies of stripe,
square and triangular domain wall lattices, and found that stripe 
structure has the lowest energy. Details of calculation can be
found in  \cite{ELPV} (see correction in  \cite{PW}.
The equilibrium domain width and the equilibrium energy for the
stripe structure are:
\begin{equation}
L_{s}=\frac{\lambda }{4}\exp \left( \frac{\varepsilon_{dw}}
{4\tilde{m}^{2}}-C+1\right) \label{L}
\end{equation}

\begin{equation}
U_{s}=-\frac{16\widetilde{m}^{2}} {\lambda }\exp \left(
-\frac{\varepsilon _{dw}}{4\widetilde{m}^{2}} +C-1\right)\label{U}
\end{equation}
where $\tilde{m}=m-\epsilon_v^0/\Phi_0$ and C=0.57721 is the Euler
constant.The vortex density for the stripe domain case is:
\begin{equation}
n(x)\,=\, -\frac{4 \pi \tilde{\epsilon}_v}{\Phi_0^2 L_s}
\frac{1}{\sin({\pi x}/{L_s})} \label{dens-x}
\end{equation}
Note a strong singularity of the vortex density near the domain
walls. Our approximation is invalid at distances of the order of
$\lambda$, and the singularities must be smeared out in the band
of the width $\lambda$ around the domain wall.

The domains become infinitely wide at $T=T_s$ and at $T=T_v$. If
$\epsilon_{dw}\leq 4\tilde{m}^2$, the continuous approximation
becomes invalid (see sec. 3.2.3) and instead a discrete lattice of
vortices must be considered.
It is possible that the long nucleation time can interfere with
the observation of described textures. We expect, however that the
vortices that appear first will reduce the barriers for domain
walls and, subsequently, expedite domain nucleation.\newline

Despite of theoretical simplicity the ideal bilayer is not easy to
realize experimentally. The most popular material with the
perpendicular to film magnetization is a multilayerer made from Co
and Pt ultrathin films (see Sec.\ref{expfsb}). This material has very large
coercive field and rather chaotic morphology. Therefore, the
domain walls in such a multilayer are chaotic and almost unmovable
at low temperatures (see Sec.\ref{expfsb}). We hope,
however, that these experimental difficulties will be overcome and
spontaneous vortex structures will be discovered before long.

\subsubsection{Superconducting transition temperature of the FSB}

The superconducting phase transition in  
ferromagnet-superconductor bilayer was studied by
Pokrovsky and Wei \cite{PW}.
They have  demonstrated that in the FSB the transition
proceeds discontinuously as a result of competition 
between the stripe domain structure in
a FM layer at suppressed superconductivity and the 
combined vortex-domain structure in the FSB. Spontaneous
vortex-domain structures in the FSB tend to increase the
transition temperature, whereas the effect of the FM 
self-interaction decreases it. The final shift of 
transition temperature T$_c$ depends on several 
parameters characterizing the
SC and FM films and varies typically between    -0.03T$_c$ and
0.03T$_c$ .

As it was discussed earlier, the homogeneous state of the FSB with
the magnetization perpendicular to the layer is unstable with
respect to formation of a stripe domain structure, in which both,
the direction of the magnetization in the FM film and the
circulation of the vortices in the SC film alternate together.
The energy of the stripe structure per unit area $U$ and the stripe
equilibrium width $L_{s}$ is given in
equations  \ref{U}, \ref{L}. To find the transition
temperature, we combine the energy given by equation  \ref{U} with the
Ginzburg-Landau free energy. The total free energy per unit area
reads:
\begin{equation}
F =U+F_{GL}
=\frac{-16{\tilde{m}}^{2}}{\lambda _{e}}
\exp (\frac{-\epsilon _{dw}}{4{\tilde{m}}^{2}}+C-1)
+n_{s}d_{s}[\alpha (T-T_{c})+\frac{\beta }{2}n_{s}]\,.
\label{U-tot}
\end{equation}
Here $\alpha $ and $\beta $ are the Ginzburg-Landau parameters. We
omit the gradient term in the Ginzburg-Landau equation since the
gradients of the phase are included in the energy (\ref{U}),
whereas the gradients of the superconducting electrons density can
be neglected everywhere beyond the vortex cores.
Minimizing the total free energy
Pokrovsky and Wei \cite{PW} have found that near $T_c$ the FSB free
energy can be represented as
\begin{equation}
F_s=-\frac{\alpha^{2}(T-T_r)^{2}}{2\beta}d_s
\label{free-sup}
\end{equation}
where $T_r$ is given by the equation:
\begin{equation}
T_{r}=T_{c} + \frac{64\pi m^{2}e^{2}}{\alpha m_{s}c^{2}}%
\exp (\frac{-\epsilon _{dw}}{4m^{2}}+C-1)\,
\label{shift0}
\end{equation}

The SC phase is stable if its free energy equation \ref{free-sup} 
is less than the
free energy of a single FM film with the stripe domain 
structure, which has the following form \cite{pw20,pw21}: 
\begin{equation}
F_{m}=-\frac{4m^{2}}{L_f}
\label{free-mag}
\end{equation}
where $L_f$ is the stripe width of the single FM film. Near the
SC transition point the temperature dependence of the 
variation of this magnetic energy is negligible. Hence, when T
increases, the SC film transforms into a normal state at some
temperature T$^{*}_c$ below T$_r$. This is the 
first-order phase transition. At transition 
point both energies $F_s$ and  $F_m$ 
are equal to each other.
The shift of the transition temperature is determined by
a following equation:

\begin{equation}
\Delta T_{c}\equiv T_{c}^{*}-T_{c}
=\frac{64\pi m^{2}e^{2}}{\alpha m_{s}c^{2}}
\exp (\frac{-\epsilon _{dw}}{4m^{2}}+C-1)\,
-\sqrt{\frac{8\beta m^2 }{\alpha^2d_sL_f}}
\label{shift}
\end{equation}

Two terms in equation \ref{shift}  play opposite roles. The first one is due
to the appearance of spontaneous vortices which lowers the
free energy of the system and tends to increase the transition
temperature. The second term is the contribution of the
purely magnetic energy, which tends to decrease 
the transition temperature. The values of 
parameters entering equation \ref{shift} 
can be estimated as follows. The dimensionless 
Ginzburg-Landau parameter is $\alpha= 7.04T_c /\epsilon_F$, 
where $\epsilon_F$ is the Fermi energy. 
A typical value of  $\alpha$ is about 10$^{-3}$ 
for low-temperature superconductors. The second Ginzburg-Landau
parameter is   $\beta = \alpha T_c/n_e$, 
where $n_e$ is the electron density.
For estimates Pokrovsky and Wei \cite{PW}  
take $T_c \sim 3$K, $n_e \sim 10^{23}$ cm$^3$. 
The magnetization per unit area $m$ is the product 
of the magnetization per unit volume $M$ 
and the thickness of the FM film $d_m$.
For typical values of $M\sim 10^2$ Oe and $d_m\sim 10$nm. 
$m\sim 10^{-4}$ Gs/cm$^2$. In an ultrathin 
magnetic film the observed values of $L_f$ vary 
in the range 1 to 100$\mu$m \cite{pw22,pw23}.
 If $L_f\sim 1\mu$m, $d_s= d_m = 10$nm, 
and $\exp(-\tilde\epsilon_{dw}/4m^2 + C - 1)\approx 10^{-3}$, 
 $\Delta T_c /T_c \sim -0.03$. 
For $L_f=100\mu$m,  $d_s = 50$nm, 
and and $\exp(-\tilde\epsilon_{dw}/4m^2 + C - 1)\approx 10^{-2}$, 
 $\Delta T_c /T_c \sim 0.02$.

\subsubsection{Transport properties of the FSB}

The spontaneous domain structure violates initial rotational
symmetry of the FSB. Therefore, it makes transport properties of
the FSB anisotropic. Kayali and Pokrovsky \cite{KP} have
calculated the periodic pinning force in the stripe vortex
structure resulting from a highly inhomogeneous distribution of
the vortices and antivortices in the FSB.
The transport properties of the FSB are associated with the
driving force acting on the vortex lattice from an external
electric current. In the FSB the pinning force is due to the
interaction of the domain walls with the vortices and antivortices
and the vortex-vortex interaction $U_{vv}$ Periodic pinning forces
in the direction parallel to the stripes do not appear in
continuously distributed vortices. In the work
\cite{KP} the discreteness effects were incorporated.
Therefore,
one need to modify the theory \cite{ELPV} to incorporate the
discreteness effects.

\begin{figure}
\begin{center}
\includegraphics[width=3.5in]{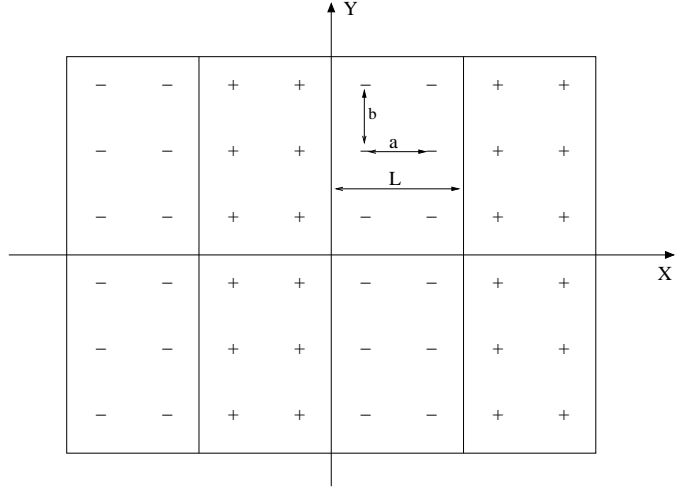}
\caption{\label{figure2} Schematic vortex distribution in the
FSB.
   The sign $\pm$ refers to the vorticity of the trapped flux.
}
\end{center}
\end{figure}

Kayali and  Pokrovsky \cite{KP}
have showed that, in the absence of a driving force, the vortices
and antivortices lines themselves up in straight chains and that 
the force between two chains of vortices falls off exponentially
as a function of the distance separating the chains. They also
argued that pinning force in the direction parallel to the domains 
drops faster in the vicinity of the superconducting transition
temperature $T_s$ and vortex disappearance temperature
$T_v$.

 In the presence of a permanent current there are three kinds of 
forces acting onto a vortex. They are i) The Magnus force
proportional to the vector product of the current density and the
velocity of the vortex; ii) The viscous force directed oppositely
to the vortex velocity; iii) Periodic pinning force acting on a
vortex from other vortices and domain walls. The pinning force
have perpendicular and parallel to domain walls components. In the
continuous limit the parallel component obviously vanishes. It
means that it is exponentially small if the distances between
vortices are much less than the domain wall width. The sum of all
three forces must be zero. This equation determines the dynamics
of the vortices. It was solved under a simplifying assumptions that 
vortices inside one domain move with the same velocity. The
critical current have been calculated for for parallel and
perpendicular orientation. Theory predicts a strong anisotropy of
the critical current. The ratio of the parallel to perpendicular
critical current is expected to be in the range $10^2 \div 10^4$
close to the superconducting transition temperature $T_s$ and to
the vortex disappearance temperature $T_v$. The anisotropy
decreases rapidly when the temperature goes from the ends of this
interval reaching its minimum somewhere inside it. The anisotropy
is associated with the fact that the motion of vortices is very
different in this two cases. At perpendicular to the domains 
direction of the permanent current all the vortices are involved 
by the friction force into a drift in the direction of the
current, whereas the Magnus force induces the motion of vortices 
(antivortices) in neighboring domains in opposite directions, both 
perpendicular to the current. The motion of all vortices
perpendicular to the domains captures domain walls, which also 
move in the same direction. This is a Goldstone mode, no 
perpendicular pinning force appears in this case. The periodic
pinning in the parallel direction and together with it the
perpendicular critical current is exponentially small. In the case
of parallel current the viscous force involves all vortices into
the parallel motion along the domain walls and in alternating
motion perpendicularly to them. The domain walls remain unmoving
and provide very strong periodic pinning force in the
perpendicular direction. This anisotropic transport behavior could
serve as a diagnostic tool to discover spontaneous topological
structures in magnetic-superconducting systems.

\begin{figure}
\begin{center}
\includegraphics[width=3.5in]{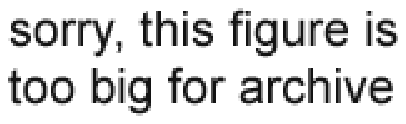}
%{lange2.eps}
\caption{\label{buble}
Magnetic properties of the Co/Pt multilayer: 
(a)~Hysteresis loop measured by
magneto-optical Kerr effect with $H$ perpendicular 
to the sample surface. MFM images
(5~$\times$~5~$\mu\mathrm{m}^{2}$) show 
that the domain structure of the sample
consists of band domains after out-of-plane 
demagnetization (b), bubble domains in the
$s=0.3$ (c) and $s=0.93$ (d) states.
 (From  Lange {\it et al.} cond-mat/0310132).
}
\end{center}
\end{figure}

\subsubsection{Experimental studies of the FSB \label{expfsb}}

In the preceding theoretical discussion
we assumed that the magnetic film changes
its magnetization direction in a weak external field and
achieves the equilibrium state.
All experimental works have been done with the
Co/Pt, Co/Pd  multilayers, which have large coercive field and are
virtually "frozen" at the experiment temperature.  Lange {\it et
al.} \cite{fsbexp,mosh16,mosh17} have studied phase diagram and
pinning properties of such magnetically "frozen" FSB. In these
works the average magnetization is characterized by the parameter
$s$, the fraction of spins directed up. Magnetic domains in Co/Pd(Pt)
multilayers look like meandering irregular bands at $s=0.5$ (zero
magnetization) (see figure  \ref{buble}b) and as 
"bubble" domains  (see figure  \ref{buble}d) with typical size
0.25$\mu$m-0.35$\mu$m near fully magnetized states ($s=0$ or
$s=1$). 
The stray field from domains is maximal at $s=0.5$ and
decreases the superconducting transition temperature  $T_{c}$ of the Pb
film by 0.2K (see figure  \ref{phdiag}). 
The effective penetration
depth is about 0.76$\mu$m at 6.9K. 

\begin{figure}
\begin{center}
\includegraphics[width=3.5in]{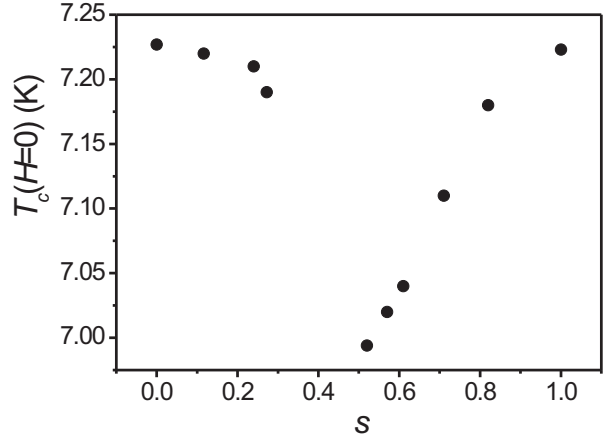}
\caption{\label{phdiag}
Dependence of the critical temperature 
at zero field $T_{c}($H=0$)$ on the
parameter $s$. The minimum value of 
$T_c$ is observed for $s=0.5$.
(From  Lange {\it et al.} cond-mat/0310132)
}
\end{center}
\end{figure}

Close to $s=0$ or $s=1$ Lange {\it et al.} \cite{fsbexp,mosh16,mosh17} 
have observed behavior in the applied magnetic field which 
is similar to the array of magnetic
dots with normal to the film magnetization (see Sec. \ref{dotpin}). 
They have found asymmetry
in the applied magnetic field for $T_c(H)$ dependence and for pinning
properties.
The bubble domains 
have a perpendicular magnetic moment. If the thickness and 
magnetization is sufficient, they can pin vortices which appear in 
the applied external magnetic field. In this respect they are 
similar to randomly distributed dots with normal magnetization. 
Thus, in the range of filling factor $s\approx 0\, \textrm{or} 1$ 
the critical current must be large enough. Contrary to this 
situation at $s\approx 0.5$ the randomly bent band domains destroy 
a possible order of the vortex lattice and provide percolation 
"routes" for the vortex motion. Thus the pinning is weaker and
corresponds either to smaller critical current or to a resistive 
state. This qualitative difference between magnetized and
demagnetized state has been observed in the experiments by Lange
et al(\cite{fsbexp,mosh16,mosh17}). The above qualitative picture
of vortex pinning is close to that developed by Lyuksyutov and
Pokrovsky \cite{dotlp} and by Feldman et al \cite{rdot} for the
transport properties of the regular array of magnetic dots with
the random normal magnetization (see Sec.\ref{dima}). In this model the 
demagnetized state of the dot array is associated with the vortex creep 
through the percolating network. The strongly magnetized state, on 
the contrary, provides more regular vortex structure and enhances 
pinning. 
 
\subsubsection{Thick  Films}

\begin{figure}
\begin{center}
\includegraphics[width=3.5in]{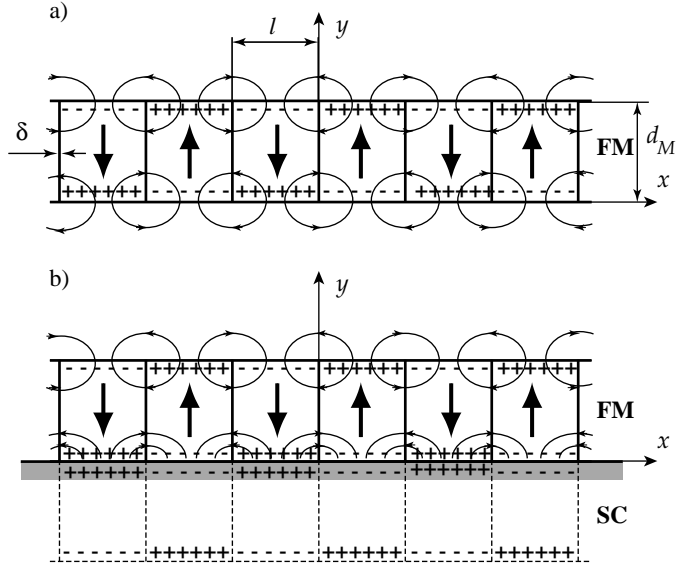}
\caption{\label{sonin}Magnetic charges (+ and -) and magnetic 
flux (thin lines with arrows)    in a
ferromagnetic film (FM) without (a) and with (b) 
a superconducting substrate (SC).
The magnetization vectors in domains are shown by thick arrows.
(From Sonin cond-mat/0102102)
}
\end{center}
\end{figure}

Above we have discussed the case when both magnetic and superconducting
films are thin, namely, $d_s \ll \lambda_L$ and $d_m \ll L_f$. 
In this subsection we briefly discuss, following works
by Sonin \cite{sonin1},  situation when 
both films are thick   $d_s \gg \lambda_L$ and $d_m \gg L_f$.
Below we neglect the domain wall width.
Consider first the ferromagnetic film without superconductor.
This problem has been solved exactly by Sonin \cite{sonin4}.    
Figure \ref{sonin}a shows schematically magnetic field 
distribution around thick ferromagnetic film.
The problem can be solved by calculating field from ``magnetic
charges'' on the magnetic film surface \cite{sonin1}.

The magnetic field, without a superconducting substrate, 
at the ferromagnetic film 
boundary $y=0^+$ is given by \cite{sonin1}:
\begin{equation}
H_{x}(x)=-4M \ln\left|\tan{\pi x \over 2L_f}\right|~.
 \label{x0} 
\end{equation}
\begin{equation}
H_{y}(x)=\mp 2\pi M \mbox{sign} \left(\tan{\pi x \over 2L_f}\right)
\mbox{at}~  ~y \rightarrow \pm 0~.
 \label{y0} 
\end{equation}
The  field pattern is periodic with 
the period $2L_f$ along the axis $x$. 
The magnetic charge on the film boundary 
 $y=0$ is 
\begin{equation}
\rho_{M}=- M  \delta(y)\mbox{sign}\left(\tan{\pi x \over 2L_f}\right)~.
\label{ch} 
\end{equation}

Sonin has argued \cite{sonin1} that in the case of bulk superconductor
and with additional requirement $\lambda_L / L_f \to 0 $,
the magnetic flux from the magnetic film is practically 
expelled from superconductor and problem can be solved 
by using images of magnetic charges on the magnetic film
surface as shown in figure  \ref{sonin}b. 
Sonin has calculated energy change due to presence of the superconducting
substrate and concluded that 
the substrate increases the total magnetic energy by 1.5 times.
The energy of the domain walls per unit length along the axis $x$ is inversely
proportional to  domain width $L_{fs}$  and the energy of 
the stray fields is proportional to  $L_{fs}$. The  domain width $L_{fs}$
is determined by minimization of the total energy per unit length. 
The growth of the magnetic energy  decreases 
the domain width  $L_{fs}$ by $\sqrt{1.5}$ times. 
Relative correction
to the energy for finite  $\lambda_L/L_{fs}$  
is of the order of  $\lambda_L/L_{fs}$ \cite{sonin1}.

\section{Proximity Effects in Layered Ferromagnet - Superconductor
Systems}

\subsection{Oscillations of the order parameter \label{oscillations}}

All oscillatory phenomena
theoretically predicted and partly observed in the S/F layered
systems are based on the Larkin-Ovchinnikov-Fulde-Ferrel (LOFF)
effect first proposed for homogeneous systems with coexisting
superconductivity and ferromagnetism \cite{LO, FF}. They predicted 
that the energy favorable superconducting order parameter in the
presence of exchange field should oscillate in space. The physical
picture of this oscillation is as follows. In a singlet Cooper
pair the electron with the spin projection parallel to the
exchange field acquires the energy $-h$, whereas the electron with
the antiparallel spin acquires the energy $+h$. Their Fermi
momenta therefore split onto the value $q=2h/v_F$. The Cooper pair 
acquires such a momentum and therefore its wave function is 
modulated. The direction of this modulation vector in the bulk 
superconductor is arbitrary, but in the S/F bilayer the
preferential direction of the modulation is determined by the
normal to the interface ($z$-axis). There exist two kinds of 
Cooper pairs differing with the direction of the momentum of the
electron whose spin is parallel to the exchange field. The
interference of the wave functions for these two kinds of pairs
leads to the standing wave:
\begin{equation}\label{LOFF}
    F(z)=F_0\cos qz
\end{equation}
A modification of this consideration for the case when the Cooper
pair penetrates to a ferromagnet from a superconductor was
proposed by Demler {\it et al.}\cite{demler}. They argued that the
energy of the singlet pair is bigger than the energy of 2
electrons in the bulk ferromagnet by the value $2h$ (the
difference of exchange energy between spin up and spin down
electrons). It can be compensated if the electrons slightly change
their momentum so that the pair will acquire the same total
momentum $q=2h/v_F$. The value $l_m=v_F/h$ called the magnetic 
length is a natural length scale for the LOFF oscillations in a 
clean ferromagnet. Anyway, equation (\ref{LOFF}) shows that the 
sign of the order parameter changes in the ferromagnet. This 
oscillation leads to a series of interesting phenomena that will
be listed here and considered in some details in next subsections.
\begin{enumerate}
    \item Periodic transitions from 0- to $\pi$ phase in the S/F/S
    Josephson  junction when varying thickness $d_f$ of the ferromagnetic
    layer and temperature $T$.
    \item Oscillations of the critical current vs. $d_f$ and $T$.
    \item Oscillations of the critical temperature vs. thickness
    of magnetic layer.
\end{enumerate}
The penetration of the magnetized electrons into superconductors
strongly suppresses the superconductivity. This obvious effect is
accompanied with the appearance of magnetization in the
superconductor. 
It penetrates on the depth of the
coherence length and is directed opposite to magnetization of the
F-layer. Another important effect which does not have oscillatory
character and will be considered later is the preferential
antiparallel orientation of the two F-layers in the S/F/S
trilayer.

 The described simple physical picture can be also treated in terms 
of the Andreev reflection at the boundaries\cite{andreev}, long
known to form the in-gap bound states \cite{degennes},
\cite{blonder}. Due to the exchange field the phases of Andreev
reflection in the S/F/S junction are different than in junctions
S/I/S or S/N/S (with non-magnetic normal metal N). Indeed, let
consider a point P inside the F-layer at a distance $z$ from one
of the interfaces \cite{FTG}. The pair of electrons emitted from
this point at the angle $\pm\theta, \pm(\pi-\theta)$ to the
$z$-axis will be reflected as a hole along the same lines and
returns to the same point (figure \ref{reflections}).
\begin{figure}[t]
\begin{center}
\includegraphics[width=3.5in]{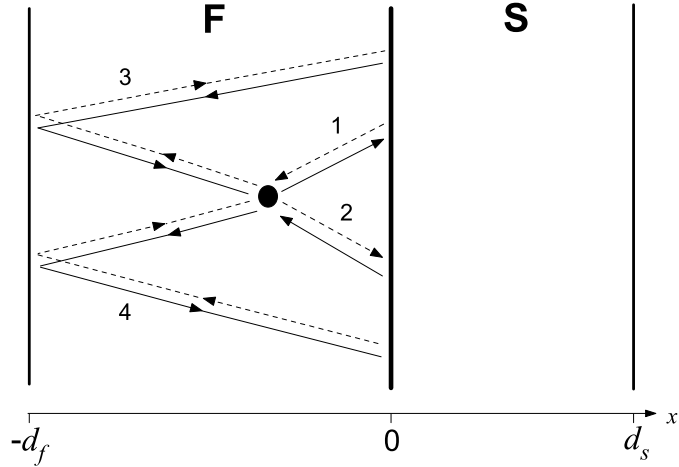}
\caption{\label{reflections}
Four types of trajectories contributing 
(in the sense of Feynman's path integral) 
to the anomalous wave function of correlated 
quasiparticles in the ferromagnetic region. The
solid lines correspond to electrons, the dashed 
lines --- to holes; the arrows indicate the direction of
the velocity. (From Fominov {\it et al.} cond-mat/0202280)
}
\end{center}
\end{figure}
The interference of the Feynman amplitudes for these 4
trajectories creates an oscillating wave function of the Cooper
pair. The main contribution to the total wave function arises from
a small vicinity of $\theta=0$. Taking only this direction, we
find for the phases: $S_1=-S_2=-qz; S_3=-S_4=-q(2d_f-z)$. Summing
up all Feynman's amplitudes $e^{iS_k}; k=1...4$, we find the
spatial dependence of the order parameter:
\begin{equation}\label{andreev}
    F\propto \cos qd_f \cos q(d_f-z)
\end{equation}
At the interface $F\propto (\cos qd_f)^2$. It oscillates as a 
function of magnetic layer thickness with the period $\Delta 
d_f=\pi/q=2\pi v_F/h$ and decays due to the interference of
trajectories with different $\theta$. \newline
   In a real experimental setup the LOFF oscillations are strongly
suppressed by the elastic impurity scattering. The trajectories
are diffusive random paths and simple geometrical picture is not
more valid. However, as long as the exchange field $h$ exceeds or 
is of the same order of magnitude as the scattering rate in the
ferromagnet $1/\tau_f$, the oscillations do not disappear 
completely. Unfortunately, the experiments with strong magnets
possessing large exchange fields are not reliable since the period
of oscillations goes to the atomic scale. Two layers with
different thickness when they are so thin can have different
structural and electronic properties. In this situation it is very
difficult to ascribe unambiguously the oscillations of properties
to quantum interference.

\subsection{Non-monotonic behavior of the transition
temperature.\label{Tc}}

This effect was first predicted by Radovic {\it et al.}
\cite{radovic1}. Its reason is the LOFF oscillations described in 
subsection \ref{oscillations}. If the transparency of
the S/F interface is low, one can expect that the order parameter 
in the superconductor is not strongly influenced by the
ferromagnet. On the other hand, the condensate wave function at
the interface in the F-layer $F\propto (\cos 2d_f/\lambda_m)^2$
becomes zero at $d_f=\pi\lambda_m(n+1/2)/2$ ($n$ is an integer).
At this values of thickness the discontinuity of the order
parameter at the boundary and together with it the current of
Cooper pairs into the ferromagnet has a maximum. Therefore one can
expect that the transition temperature is minimal \cite{proshin}.
Experimental attempts to observe this effect were made many times
on the S/F multilayers Nb/Gd \cite{Jiang}, Nb/Fe \cite{Muehge},
V/V-Fe \cite{Aarts}, V/Fe \cite{Lazar}. More references and
details about these experiments and their theoretical description
can be found in the cited reviews \cite{izyumov,garifullin-review}.
Unfortunately, in these experiments the magnetic component was a
strong ferromagnet and, therefore, they faced all the difficulties
mentioned in subsection \ref{order}: the F-layer must be too thin
and its variation produce uncontrollable changes in the sample,
the influence of the growth defects is too strong. Besides, in
multilayers the reason of the non-monotonous dependence of $T_c$
on $d_f$ may be the $0-\pi$ transition. Therefore, the reliable
experiment should be performed with a bilayer possessing a
sufficiently thick F-layer. Such experiments were performed
recently \cite{rusanov, ryazanovT}. The idea was to use a weak
ferromagnet (the dilute ferromagnetic alloy Cu-Ni) with rather
small exchange field $h$ to increase the magnetic length
$l_m=\sqrt{D_f/h}$. They performed the
experiments with S/F bilayers to be sure that the non-monotonic 
behavior is not originated from the $0-\pi$-transition. In these 
experiments the transparency of the interface was not too small or
too large, the exchange field was of the same order as the 
temperature and the thickness of the F-layer was of the same order 
of magnitude as magnetic length. Therefore, for the quantitative
description of the experiment theory should not be restricted by
limiting cases only. Such a theory was developed by Fominov {\it
et al.}\cite{FTG}. In the pioneering work by Radovic {\it et al.}
\cite{radovic1} the exchange field was assumed to be very
strong.\newline

As always when it goes about critical temperature, the energy gap
and anomalous Green function $F$ are infinitely small. Therefore
one needs to solve linearized equations of superconductivity. The
approach by Fominov {\it et al.} is based on solution of the
linearized Usadel equation and is valid in the diffusion limit
$\tau_s T_c\ll 1; \tau_f T_c\ll 1; \tau_f h\ll 1$. Namely this
situation was realized in the cited experiments \cite{rusanov,
ryazanovT}. The work by Fominov {\it et al.} \cite{FTG} covers 
numerous works by their predecessors \cite{tagirov1, tagirov2,
BuzdinT, demler, proshin} clarifying and improving their methods.
Therefore in the presentation of this subsection we follow
presumably the cited work \cite{FTG} and briefly describe specific
results of other works. \newline

The starting point is the linearized Usadel equations for singlet
pairing for anomalous Green functions $F_s$ in the superconductor
and $F_f$ in the ferromagnet:
\begin{equation}\label{Fs}
    D_s\frac{\partial^2F_s}{\partial z^2}-|\omega_n|F_s+\Delta=0;
0<z<d_s.
\end{equation}
\begin{equation}\label{Ff}
    D_f\frac{\partial^2F_f}{\partial
z^2}-(|\omega_n|+ih\textrm{sgn}\omega_n)F_f=0; -d_f<z<0.
\end{equation}
Thus, we accept a simplified model in which $\Delta=0$ in the
ferromagnetic layer and $h=0$ in the superconducting one. The
geometry is schematically shown in figure  \ref{FS-geom}.

\begin{figure}[t]
\begin{center}
\includegraphics[width=3.5in]{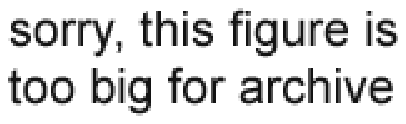}
%{FSgeom.eps}
\caption{\label{FS-geom} FS bilayer. The F and S layers occupy 
the regions $-d_f<z<0$ and $0<z<d_s$, respectively.
}
\end{center}
\end{figure}

Equations
(\ref{Ff},\ref{Fs}) must be complemented with the self-consistency
equation:
\begin{equation}\label{self}
    \Delta({\textbf{r}})\ln\frac{T_{cs}}{T}=\pi T\sum_n
\left(\frac{\Delta({\textbf{r}})}{|\omega_n|} -
F_s(\omega_n,{\textbf{r}})\right),
\end{equation}
where $T_{cs}$ is the bulk SC transition temperature, and with
linearized boundary conditions at the interface:
\begin{eqnarray}
  \sigma_s\frac{dF_s}{dz} &=& \sigma_f\frac{dF_f}{dz} \label{curr-cont}\\
  {\cal A}\sigma_f\frac{dF_f}{dz} &=&
G_b\left(F_s(0)-F_f(0)\right),\label{jump}
\end{eqnarray}
where $\sigma_{s,f}$ is the conductivity of the superconducting
(ferromagnetic) layer in the normal state; $G_b$ is the
conductance of the interface and $\cal{A}$ is the area of the
interface. We assume that the normal derivative of the anomalous
Green function is equal to zero at the interface with the vacuum:
\begin{equation}\label{vacuum}
    \frac{dF_f}{dz}|_{z=-d_f}=\frac{dF_s}{dz}|_{z=d_s}=0
\end{equation}
The condition of solvability of linear equations
(\ref{Fs},\ref{Ff},\ref{self}) with the boundary conditions
(\ref{curr-cont},\ref{jump},\ref{vacuum}) determines the value of
transition temperature $T_c$ for the F/S bilayer. \newline

The solution $F_f(\omega_n,z)$ in the F-layer satisfying the
boundary condition (\ref{vacuum}) reads:
\begin{equation}\label{sol-f}
    F_f(\omega_n,z)=C(\omega_n)\cosh [k_{fn}(z+d_f)];\,\,\,
k_{fn}=\sqrt{\frac{|\omega_n|+ih\textrm{sgn}\omega_n}{D_f}},
\end{equation}
where $C(\omega_n)$ is the integration constant to be determined 
from the matching condition at the F/S interface $z=0$. From the
two boundary conditions at $z=0$ (\ref{curr-cont},\ref{jump}) it
is possible to eliminate $F_f$ and $dF_f/dz$ and reduce the
problem to finding the function $F_s$ from equation (\ref{Fs}) and 
the effective boundary condition at $z=0$:
\begin{equation}\label{eff-bc}
    \xi_s\frac{dF_s}{dz}=\frac{\gamma}{\gamma_b+B_f(\omega_n)}F_s;
\end{equation}
where $\xi_s=\sqrt{D_s/(2\pi T_{cs}}; \gamma=\sigma_f/\sigma_s;
\gamma_b=({\cal A}\sigma_f)/(\xi_sG_b)$ and $B_f(\omega_n)=\left(
k_{fn}\xi_s\tanh(k_{fn }d_f)\right)^{-1}$. Since $k_{fn}$ is
complex the parameter $B_f(\omega_n)$ and consequently the
function $F_s$ is complex. The coefficients of the Usadel equation
(\ref{Fs}) are real. Therefore, it is possible to solve it for the
real part of the function $F_s$ traditionally denoted as
$F_s^{+}(\omega_n,z)\equiv\frac{1}{2}
\left(F_s(\omega_n,z)+F_s(-\omega_n,z)\right)$. The boundary
condition for this function reads:
\begin{equation}\label{bc+}
    \xi_s\frac{dF_s^+}{dz}=W(\omega_n)F_s^+|_{z=0};\,\,\,
W(\omega_n)=\frac{A_{sn}(\gamma_b+\Re
B_f)+\gamma}{A_{sn}|\gamma_b+B_f|^2+\gamma (\gamma_b+\Re B_f)},
\end{equation}
where $A_{sn}=k_{sn}d_s\tanh(k_{sn}d_s)$ and
$k_{sn}=\sqrt{\frac{D_s}{|\omega_n|}}$. To derive this boundary
condition we accept the function $\Delta(z)$ to be real (it will
be justified later). Then the imaginary part of the anomalous
Green function $F_s^-(\omega_n,z)$ obeys the homogeneous linear
differential equation;
$$ \frac{d^2F_s^-}{dz^2}=k_{sn}^2F_s^-$$
and the boundary condition $\frac{dF_s^-}{dz}=0$ at $z=d_s$. Its
solution is $F_s^-(\omega_n,z)=E(\omega_n)\cosh [k_{sn}(z-d_s)]$.
At the interface $z=0$ its derivative $\frac{dF_s^-}{dz}|_{z=0}$
is equal to $-k_{sn}\tanh(k_{sn}d_s)F_s^-(\omega_n,z=0)$.
Eliminating $F_s^-$ and its derivative from real and imaginary
parts of the boundary condition (\ref{eff-bc}), we arrive at the
boundary condition (\ref{bc+}). Note that only $F_s^+$
participates in the self-consistence equation (\ref{self}). This
fact serves as justification of our assumption on reality of the
order parameter $\Delta(z)$.\newline
  Simple analytic solutions of the problem are available for
different limiting cases. Though these cases are unrealistic at
the current state of experimental art, they help to understand the
properties of the solutions and how do they change when parameters
vary. Let us consider the case of very thin S-layer $d_s\ll
\xi_s$. In this case the order parameter $\Delta$ is almost a
constant. The solution of equation (\ref{Fs}) in such a situation
is $F_s^+(\omega_n,z)=-\frac{\Delta}{|\omega_n|}+\alpha_n\cosh{
k_{sn}(z-d_s)}$, where $\alpha_n$ is an integration constant. From
the boundary condition (\ref{bc+}) we find:
\begin{equation}\label{alpha}
    \alpha_n=-\frac{2\Delta
W(\omega_n)}{|\omega_n|\left(A_{sn}+W(\omega_n)\right)},
\end{equation}
where the coefficients $A_{sn}$ are the same as in equation
(\ref{bc+}). We assume that $k_{sn}d_s\ll 1$. Then $A_{sn}\approx
k_{sn}^2\xi_sd_s=\frac{d_s}{\xi_s}(n+1/2)$. The function
$F_s^+(\omega_n,z)$ almost does not depend on $z$. The
self-consistence equation reads:
\begin{equation}\label{Tcthin}
    \ln\frac{T_{cs}}{T_c}=2\sum_{n\geq 0}\frac{W(\omega_n)}
{(n+\frac{1}{2})\left(\frac{d_s}{\xi_s}(n+\frac{1}{2})+W(\omega_n)\right)}
\end{equation}

\begin{figure}[t]
\begin{center}
\includegraphics[width=3.5in]{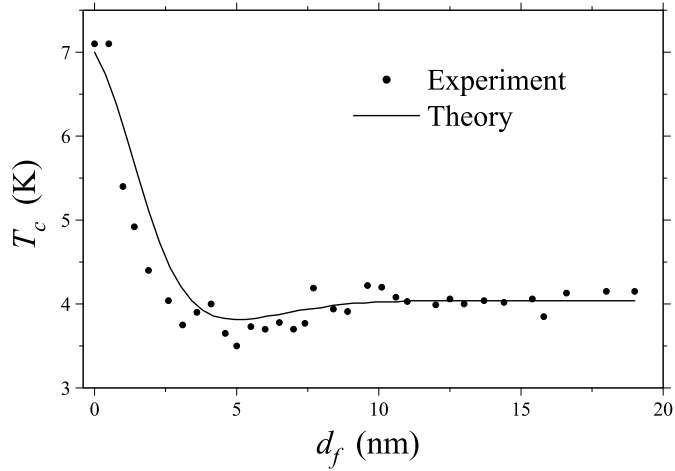}
\caption{\label{fominov-ryaz}Theoretical fit to the experimental 
data. (From Fominov {\it et al.} cond-mat/0202280).
% of Ref.~\cite{ryazanov}.
%In the experiment, Nb was the superconductor 
%(with $d_s=11$\,nm, $T_{cs}=7$\,K) and Cu$_{0.43}$Ni$_{0.57}$
%was the weak ferromagnet. From our fit we estimate 
%$E_\mathrm{ex}\approx 130$\,K and $\gamma_b\approx 0.3$.
}
\end{center}
\end{figure}

The summation can be performed explicitly in terms of
digamma-functions (${\mathsf F}$):
\begin{equation}\label{Tcthindigamma}
    \ln\frac{T_{cs}}{T_c}=\frac{\gamma\xi_s}{2(\gamma_b+\Re B_f)d_s}
\Re\left\{\left(1-\frac{i(\gamma_b+\Re B_f)}
{\Im B_f}\right)\left[
%\digamma
{\mathsf F}
\left(\frac{1}{2}
+\frac{\gamma\xi_s}{d_s}\left(1-\frac{i\Im B_f}
{\gamma_b+\Re B_f}\right)\right)
%-\digamma
-{\mathsf F}
(\frac{1}{2})\right]\right\}.
\end{equation}
Possible oscillations are associated with the coefficients $B_f$.
If the magnetic length $\lambda_m=\sqrt{D_f/h}$ is much less than
$d_f$, then $B_f\approx \sqrt{\frac{h}{4\pi
T_{cs}}}\exp{(-\frac{2id_f}{\lambda_m}-\frac{i\pi}{4})}$. In the
opposite limiting case $\lambda_m\gg d_f$ there are no
oscillations of the transition temperature. Note that $\ln
(T_{cs}/T_c)$ can be rather large $\sim \xi_s/d_s$, i.e the
transition temperature in the F/S bilayer with very thin S-layer
can be exponentially suppressed. This tendency is reduced if the
resistance of the interface is large ($\gamma_b\gg 1$).\newline
  In a more realistic situation considered in the work \cite{FTG}
neither of parameters $d_s/\xi_s, d_f/\lambda_m, \gamma, \gamma_b$
is very small or very large and an exact method of solution should
be elaborated. The authors propose to separate explicitly the
oscillating part of the functions $F_s^+(\omega_n,z)$ and
$\Delta(z)$ and the reminders:
\begin{equation}\label{separationF}
    F_s^+(\omega_n,z)=f_n\frac{\cos q(z-d_s)}{\cos qd_s}+
\sum_{m=1}^{\infty}f_{nm}\frac{\cosh q_m(z-d_s)}{\cosh q_md_s}
\end{equation}
\begin{equation}\label{separationD}
    \Delta(z)=\delta \frac{\cos q(z-d_s)}{\cos qd_s}+
\sum_{m=1}^{\infty}\delta_m\frac{\cosh q_m(z-d_s)}{\cosh q_md_s}
\end{equation}
where the wave-vectors $q$ and $q_m$ as well as the coefficients
of the expansion must be found from the boundary conditions and
self-consistence equation. Equation (\ref{bc+}) results in
relations between coefficients of the expansion:
\begin{equation}\label{f-delta}
    f_n=\frac{\delta}{|\omega_n|+D_sq^2};\,\,\,\,f_{nm}=\frac{\delta_m}{|\omega_n|-D_sq_m^2}.
\end{equation}
Substituting the values of coefficients $f_n, f_{nm}$ from
equation (\ref{f-delta}) to the boundary condition (\ref{bc+}), we
find an infinite system of homogeneous linear equations for
coefficients $\delta$ and $\delta_m$:
\begin{equation}\label{bc-delta}
    \delta\frac{q\tan qd_s-W(\omega_n)}{|\omega_n|+D_sq^2}+
\sum_{m=1}^{\infty}\delta_m\frac{q_m\tanh
q_md_s-W(\omega_n)}{|\omega_n|-D_sq_m^2}
\end{equation}
Equating the determinant of this system $\cal D$ to zero, we find
a relation between $q$ and $q_m$. It is worthwhile to mention a
popular approximation adopted by several theorists \cite{tagirov1,
demler, proshin} the so-called single-mode approximation. In our
terms it means that all coefficients $\delta_m, m=1,2...$ are zero
and only the coefficient $\delta$ survives. The system
(\ref{bc-delta}) implies that it is  only possible when the
coefficients $W(\omega_n)$ do not depend on their argument
$\omega_n$. It happens indeed in the limit $d_s/\xi_s\ll 1$ and
$h\gg T$. For a more realistic regime the equation ${\cal D}=0$
must be solved numerically together with the self-consistence
condition, which turns into a system of equations:
\begin{eqnarray}
  \ln\frac{T_{cs}}{T_c} &=& 
%\digamma 
{\mathsf F}
\left(\frac{1}{2}+\frac{D_sq^2}{T_c}\right)
%-\digamma 
-{\mathsf F}
(\frac{1}{2})\nonumber \\
  \ln\frac{T_{cs}}{T_c} &=& 
%\digamma 
{\mathsf F}
\left(\frac{1}{2}-\frac{D_sq_m^2}{T_c}\right)
%-\digamma
-{\mathsf F}
(\frac{1}{2})\label{self-system}
\end{eqnarray}
These systems were truncated and solved with all data extracted
from the experimental setup used by Ryazanov {\it et al.}
\cite{ryazanovT}. The only 2 fitting parameters were $h=130K$ and
$\gamma_b=0.3$.

Figure  \ref{fominov-ryaz} demonstrates rather good
agreement between theory and experiment. Various types of the
curves $T_c(d_f)$ are shown in figure  \ref{Tc-curves}. Note that the
minimum on these curves eventually turns into a plateau at
$T_c=0$, the reentrant phase transition into the superconducting
state. Some of the curves have a well-pronounced discontinuity,
which can be treated as the first order phase transition. The
possibility of the first order transition to superconducting state
in the F/S bilayer was first indicated by Radovic {\it et al.}
\cite{radovic1}.

\begin{figure}[t]
\begin{center}
\includegraphics[width=2.5in]{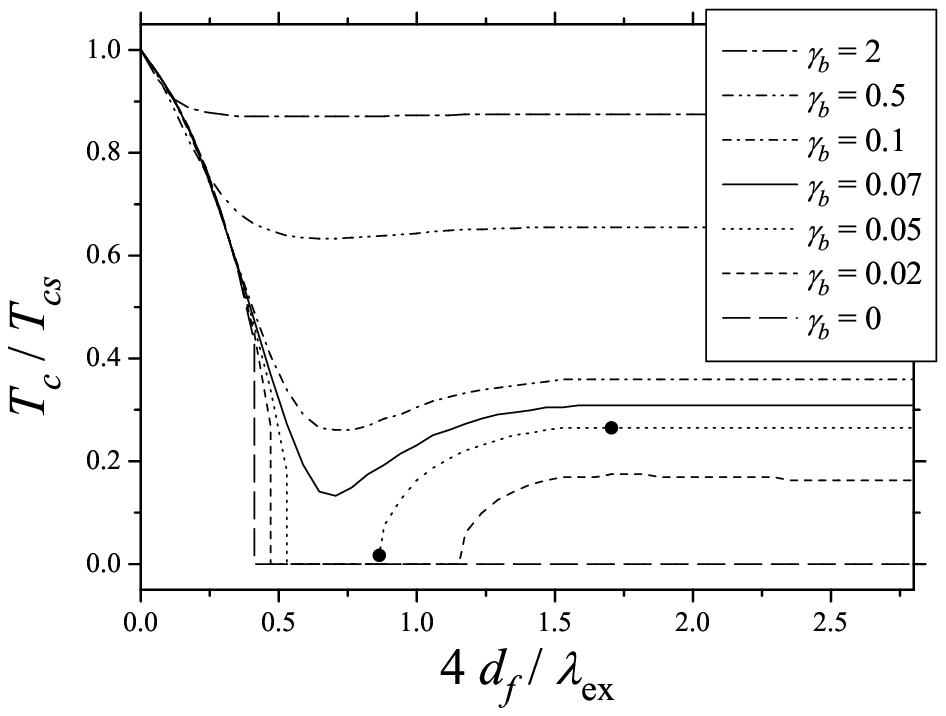}
\includegraphics[width=2.5in]{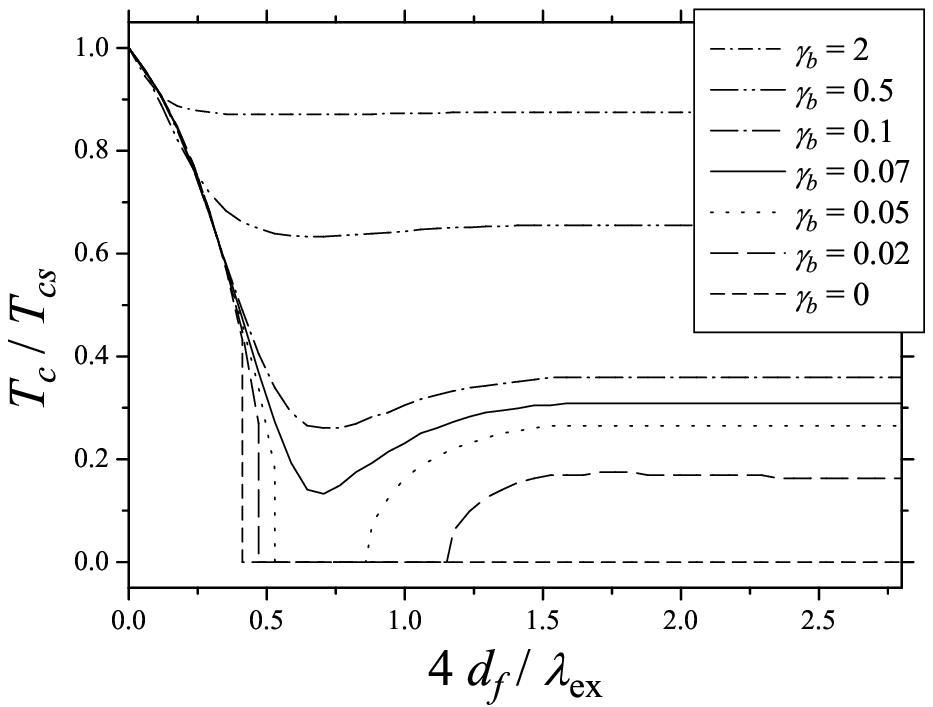}
\caption{\label{Tc-curves}Theoretical fit to the experimental 
data. (From Fominov {\it et al.} cond-mat/0202280).
}
\end{center}
\end{figure}

\subsection{Josephson effect in S/F/S junctions}
As we already mentioned the exchange field produces oscillations
of the order parameter inside the F-layer. This effect in turn can
change the sign of the Josephson current in the S/F/S junction
compared to the standard S/I/S or S/N/S junctions. As a result the
relative phase of the S-layers in the ground state is equal to
$\pi$ (the so-called $\pi$-junction). In the closed
superconducting loop with such a junction spontaneous magnetic
flux and spontaneous current appear in the ground state. These
phenomena were first predicted by Bulaevsky {\it et al.}
\cite{BulKuz} for $\pi$-junction independently on the way of its
realization. Buzdin {\it et al.} \cite{BulBuz} have first argued
that such a situation can be realized in the S/F/S junction at a
proper choice of its length. Ryazanov {\it et al.}
\cite{ryazanov01-1, ryazanov01-2} have realized such a situation
employing the weak ferromagnet $\textrm{Cu}_x\textrm{Ni}_{1-x}$ as
a ferromagnetic layer. A similar approach was used by Kontos {\it
et al.} \cite{kontos}, who used a diluted alloy PdNi. The
weakness of exchange field allowed them to drive the oscillations
and in particular the 0-$\pi$ transition by the temperature at a
fixed magnetic field. The success of this experiment have
generated an extended literature. The theoretical and experimental
study of this and related phenomena still are active. In what
follows we present a brief description of relevant theoretical
ideas and the experiments.
\subsubsection{Simplified approach and experiment}
Here we present a simplified picture of the S/F/S junction based
on the following assumptions:\newline i)The transparency of the
S/F interfaces is small. Therefore the anomalous Green function in
the F-layer is small and it is possible to use the linearized
Usadel equation.\newline ii) The energy gap $\Delta$ inside each 
of the S- layers is constant and equal to $\Delta_0 e^{\mp
i\varphi/2}$ (the sign $-$ relates to the left S-layer, $+$ to the
right one).\newline iii) $\Delta =0$ in the F-layer and $h=0$ in
the S-layers.\newline
 The geometry of the system is shown in figure  (\ref{FS-geom}). From the
second assumption it follows that the anomalous Green function $F$
is also constant within each of S-layers:
$F=\frac{\Delta}{\sqrt{|\omega_n|^2+\Delta_0^2}}$. The linearized
Usadel equation in the F-layer (\ref{Ff}) has a following general
solution:
\begin{equation}\label{general}
    F(\omega_n,z)=\alpha_n e^{k_{fn}z} + \beta_n e^{-k_{fn}z},
\end{equation}
where\begin{equation}\label{kfn}
    k_{fn}=\sqrt{\frac{|\omega_n|+ih\textrm{sgn}\omega_n}{D_f}}
\end{equation}
(compare equation (\ref{sol-f})). The boundary condition at the
two interfaces follows from the second boundary condition of the
previous section (\ref{jump}) in which $F_f$ is neglected:
\begin{equation}\label{boundary}
    \xi_f\frac{dF_f}{dz}=\mp\gamma_bF_s
\end{equation}
The coefficients $\alpha_n$ and $\beta_n$ are completely
determined by the boundary conditions (\ref{boundary}):
\begin{eqnarray}
  \alpha_n &=& Q_n \frac{\cos(\frac{\varphi-ik_{fn}d_f}{2})}{sinh(k_{fn}d_f)} \\
  \beta_n &=& Q_n \frac{\cos(\frac{\varphi+ik_{fn}d_f}{2})}{sinh(k_{fn}d_f)}
\end{eqnarray}
where
$Q_n=\frac{\Delta_0}{\gamma_b\xi_fk_{fn}\sqrt{|\omega_n|^2+\Delta_0^2}}$.
Equation (\ref{current}) for the electric current must be slightly
modified to incorporate the exchange field $h$:
\begin{equation}\label{current-h}
    {\bf j}=ie\pi
TN(0)D\sum_{n}(\tilde{F}\hat{\partial}F-F\hat{\partial}\tilde{F}),
\end{equation}
where $\tilde{F}(\omega_n,z)=F^*(-\omega_n,z)$. Note that at this
transformation the wave vectors $k_{fn}$ remain invariant. After
substitution of the solution (\ref{general}) we find that
$j=j_c\sin\varphi$ with the following expression for the critical
current \cite{BuzdinT, schoen, ryazanov01-1}:
\begin{equation}\label{critcurrent}
    j_c=\frac{4\pi T\Delta_0^2}{eR_N\Gamma_b}\Re\left[\sum_{\omega_n>0}\left((\omega_n^2+\Delta_0^2)
    k_{fn}d_fsinh(k_{fn}d_f)\right)^{-1}\right],
\end{equation}
where $R_N$ is the normal resistance of the ferromagnetic layer
and $\Gamma_b$ is the dimensionless parameter characterizing the
ratio of the interface resistance to that of the F-layer.
Kupriyanov and Lukichev \cite{KuLu} have found the relationship
between $\Gamma_b$ and the the barrier transmission coefficient
$D_b(\theta)$ ($\theta$ is the angle between the electron velocity
and the normal to the interface):
\begin{equation}\label{Gamma}
    \Gamma_b=\frac{2l_f}{3d_f}\langle\frac{\cos\theta
    D_b(\theta)}{1-D_b(\theta)}\rangle.
\end{equation}
As we explained earlier, the oscillations appear since $k_{fn}$ are 
complex values. If $h\gg 2\pi T$ $\omega_n$, then 
$k_{fn}\approx (1+i) \sqrt{\frac{h}{2D_f}}$ and oscillations are
driven only by the thickness. It was very important to use a weak
ferromagnet with exchange field $h$ comparable to $\pi T$. Then
the temperature also drives the oscillations. In the Cu-Ni alloys
used in the experiment \cite{ryazanov01-1} the Curie point $T_m$ 
was between 20 and 50K. Nevertheless, the ratio $h/\pi T$ was in 
the range of 10 even for the lowest $T_m$. In this situation
$k_{fn}$ does not depend on $n$ for the large number of terms in
the sum (\ref{critcurrent}). This is the reason why the sum in
total is periodic function of $d_f$ with the period
$\lambda_m=\pi\sqrt{2D_f/h}$. The dependence on temperature is
generally weak. However, if the thickness is close to the value at
which $j_c$ turns into zero at $T=0$, the variation of temperature
can change the sign of $j_c$.

\begin{figure}
\begin{center}
\includegraphics[width=3.5in]{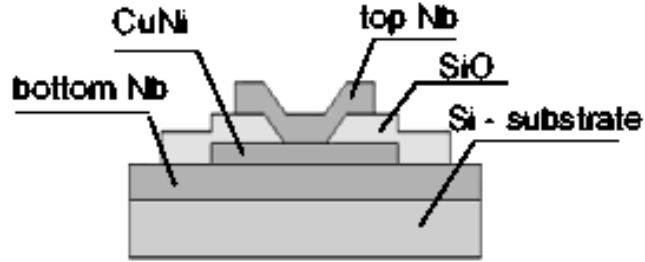}
\includegraphics[width=3.5in]{sorry.eps}
%{ryazanov0b.eps}
\caption{\label{ryazanov}
(Upper) Schematic cross-section of
the sample. 
(Lower) Left : critical current Ic as function of
temperature for Cu$_{0.48}$Ni$_{0.52}$ junctions with differ-
ent F-layer thicknesses between 23 nm and 27 nm as
indicated. Right : model calculations of the tem-
perature dependence of the critical current in an
SFS junction.
(From Ryazanov {\it et al.} cond-mat/0008364)
}
\end{center}
\end{figure}

In figure  (\ref{ryazanov}b)
theoretical curves $j_c(T)$ from cited work \cite{FTG} are
compared with the experimental data by Ryazanov {\it et al.}
\cite{ryazanov01-1, ryazanovusp}. The curves are plots of the
modulus of $j_c$ vs $T$. Therefore, the change of sign of $j_c$ is
seen as a cusp on such a curve. At temperature of the cusp the
transition from 0- to $\pi$-state of the junction proceeds. The
change of sign is clearly seen on the curve corresponding to
$d_f=27nm$. The experimental S/F/S junction is schematically shown
in figure  (\ref{ryazanov}a). The details of the experiment are
described in original paper \cite{ryazanov01-1} and in reviews
\cite{ryazanovusp, izyumov}. Not less impressive agreement between
theory and experiment is reached by Kontos {\it et al.}
\cite{kontos} (theory was given by T. Kontos) (see figure \ref{kontos}).

\begin{figure}
\begin{center}
\includegraphics[width=3.5in]{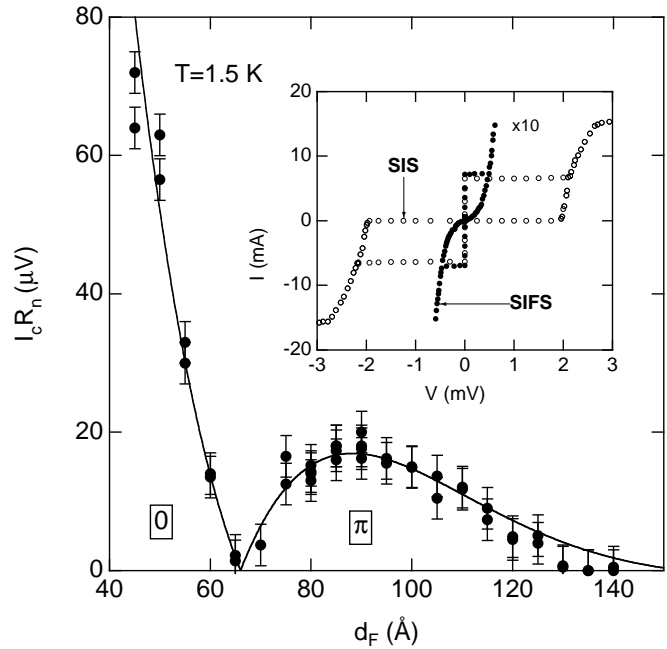}
\caption{\label{kontos}Josephson coupling as a function 
of thickness of the PdNi layer (full circles). 
The critical current cancels
out at $d_{F} \simeq$ 65 {\AA} indicating the transition from "$0$" to
 "$\pi$"-coupling. The full line is the best fit
obtained from the theory  as described in the text. Insert shows 
typical I-V characteristics of two junctions
with (full circles), and without (empty circles) PdNi layer.
(From Kontos {\it et al.} cond-mat/0201104).
}
\end{center}
\end{figure}

Very good agreement with the same experiment was
reached in a recent theoretical work by Buzdin and Baladie
\cite{BuzBal} who solved the Eilenberger equation.\newline
 Zyuzin {\it et al.}  \cite{zyuzin} have found that in a dirty 
sample the amplitude of the Josephson current $j_c$ is a random 
value with an indefinite sign. They estimated the average square 
fluctuations of this amplitude for the interval of the F-layer
thickness $\xi_s<d_f<\sqrt{D/T}$ as:
\begin{equation}\label{fluctuation} 
    \langle j_c^2\rangle={\cal A}\xi_s^2\left(\frac{g}{8\pi 
    N(0)D_f}\right)^4\left(\frac{D_f}{2\pi^2Td_f^2}\right)^2 
\end{equation} 
where ${\cal A}$ is the area of interface and $g$ is its 
conductance per unit area. The fluctuations are significant when 
$d_f$ becomes smaller than the diffusive thermal length 
$\sqrt{D/T}$.

\subsubsection{Josephson effect in a clean system}
In a recent work by Radovic {\it et al} \cite{radovic2} considered
the same effect in a clean S/F/S trilayer. A similar, but somewhat
different in details approach was developed by Halterman and
Olives \cite{halterman}. The motivation for this consideration is
the simplicity of the model and very clear representation of the
solution. Though in the existing experimental systems the
oscillations are not disguised by impurity scattering, it is
useful to have an idea what maximal effect could be reached and
what role plays the finite transparency of the interface. The
authors employed the simplest version of theory,
Bogolyubov-DeGennes equations:
\begin{equation}\label{BDG}
    \hat{H}\left(%
\begin{array}{c}
  u_{\sigma} \\
  v_{\bar{\sigma}} \\
\end{array}%
\right)=E\left(%
\begin{array}{c}
  u_{\sigma} \\
  v_{\bar{\sigma}} \\
\end{array}\right),
\end{equation}
where $\bar{\sigma}$ means $-\sigma$ and the effective Hamiltonian
reads:
\begin{equation}\label{eff}
    \hat{H}=\left(%
\begin{array}{cc}
  H_0({\bf r})-\sigma h({\bf r}) & \Delta ({\bf r}) \\
  \Delta^* ({\bf r}) & -H_0+\bar{\sigma h}({\bf r}) \\
\end{array}%
\right),
\end{equation}
\begin{equation}\label{H0}
    H_0({\bf r})=-\frac{\hbar^2}{2m}\nabla^2-\mu +W({\bf r})
\end{equation}
In the last equation $\mu$ is the chemical potential and $W({\bf
r})$ is the barrier potential:
\begin{equation}\label{barrier}
    W({\bf r})=W[\delta (z+d/2)+\delta (z-d/2)].
\end{equation}
The assumption about exchange field $h({\bf r})$ and the order
parameter $\Delta({\bf r})$ are the same as in the previous
subsubsection. We additionally assume that the left and right
S-layers are identical and semi-infinite extending from
$z=-\infty$ to $z=-d/2$ and from $z=d/2$ to $z=\infty$. Due to
translational invariance in the ($x,y$)-plane the dependence of
the solution on the lateral coordinates is a plane wave:
\begin{equation}\label{plane}
    \left(%
\begin{array}{c}
  u_{\sigma} \\
  v_{\sigma} \\
\end{array}%
\right)=e^{i{\bf k}_{\|}{\bf r}}\Psi(z)
\end{equation}
There are 8 fundamental solutions of these equations corresponding
to the injection of the quasiparticle or quasihole from the left
or from the right with spin up or down. We will write explicitly
one of them $\Psi_1(z)$, corresponding to the injection of the
quasiparticle from the right. In the superconducting area $z<-d/2$ 
we will see the incident quasiparticle wave with the coefficient 1
and the normal wave vector $k^+$, the reflected quasiparticle with
the reflection coefficient $b^+$ and the normal wave vector
$-k^+$; the reflected quasihole (Andreev reflection) with the
reflection coefficient $a_1$ and the wave vector $k_-$, where
$(k^{\pm})^2=\frac{2m}{\hbar^2}(E_F\pm \xi)^2-{\bf k}_{\|}^2$,
$E_F$ is the Fermi energy and $\xi =\sqrt{E^2-|\Delta|^2}$. Thus 
the solution $\Psi_1(z)$ at $z<-d/2$ reads:
\begin{equation}\label{S1}
    \Psi_1(z)=(e^{ik^+z}+b_1e^{-ik^+z})\left(
\begin{array}{c}
  ue^{-i\varphi/2} \\
  ve^{i\varphi/2} \\
\end{array}%
\right) +a_1e^{ik^-z})\left(%
\begin{array}{c}
  v e^{-i\varphi/2}\\
  u e^{i\varphi/2}\\
\end{array}%
\right)
\end{equation}
where $u$ and $v$ are the bulk Bogolyubov-Valatin coefficients:
$u=\sqrt{(1+\xi/E)/2}; v=\sqrt{(1-\xi/E)/2}$. In the F-layer
$-d/2<z<d/2$ there appear transmitted and reflected electron and
transmitted and reflected hole. Since according to our assumption
$\Delta =0$ in the F-layer, there is no mixing of the electron and 
hole. With this explanation we can write directly the solution
$\Psi_1(z)$ in the F-layer:
\begin{equation}\label{F}
    \Psi_1(z)=(C_1e^{iq^+z}+C_2e^{iq^+z})\left(%
\begin{array}{c}
  1 \\
  0 \\
\end{array}%
\right) + (C_3e^{iq_\sigma^-z}+C_4e^{iq_\sigma^-z})\left(%
\begin{array}{c}
  0 \\
  1 \\
\end{array}%
\right),
\end{equation}
where $q_\sigma^{\pm}=\sqrt{\frac{2m}{\hbar^2}(E_F^f+\sigma h\pm
E)-{\bf k}_{\|}^2}$. Finally in the right S-layer $z>d/2$ only the
transmitted quasiparticle and quasihole propagate:
\begin{equation}\label{S2}
    \Psi_1(z)=c_1e^{ik^+z}\left(%
\begin{array}{c}
  ue^{i\varphi/2} \\
  ve^{-i\varphi/2} \\
\end{array}%
\right)+d_1e^{-ik^-z}\left(%
\begin{array}{c}
  ve^{i\varphi/2} \\
  ue^{-i\varphi/2} \\
\end{array}%
\right)
\end{equation}
The value of all coefficients can be established by matching of
solutions at the interfaces:
\begin{equation}\label{psibc}
    \Psi (\pm \frac{d}{2}-0)=\Psi (\pm
    \frac{d}{2}+0);\,\,\,\frac{d\Psi}{dz}|_{\pm
    \frac{d}{2}+0}-\frac{d\Psi}{dz}|_{\pm
    \frac{d}{2}-0}=\frac{2mW}{\hbar^1}\Psi
\end{equation}
Other fundamental solutions can be found by symmetry relations:
\begin{equation}\label{symmetry}
    a_2(\varphi)=a_1(-\varphi); a_3=a_2; a_4=a_1; b_3=b_1;
    b_4=b_2,
\end{equation}
where index 2 relates to the hole incident from the left, indices
3,4 relate to the electron and hole incident from the right. Each
mode generates the current independently on others. The critical
current reads:
\begin{equation}\label{current-clean}
    j_c=i\frac{e\Delta T}{\hbar}\sum_{\sigma,\omega_n,{\bf k}_{\|}}
    \frac{k_n^{+}+k_n^{-}}{2\xi_n}\left(\frac{a_{1n}}{k_n^{+}}-\frac{a_{2n}}{k_n^{-}}\right).
\end{equation}
Here all the values with the index $n$ mean functions of energy
$E$ denoted by the same symbols in which $E$ is substituted by
$i\omega_n$, for example $\xi_n=i\sqrt{\omega_n^2+\Delta^2}$.
 We will not demonstrate here straightforward,
but somewhat cumbersome calculations and transit to conclusions.
The critical current displays oscillations originated from two
different types of the bound states. One of them appears if the
barrier transmission coefficient is small. This is the geometrical
resonance. The superconductivity is irrelevant for it. Another one
appears even in the case of ideal transmission: this is the
resonance due to the Andreev reflection. When the transmission
coefficient is not small and not close to 1, it is not easy to
separate these two type of resonances and the oscillations picture
becomes rather chaotic. The LOFF oscillations are better seen when
transmission coefficient is close to 1 since geometrical
resonances do not interfere. Varying the thickness, one observes
periodic transitions from 0 to $\pi$-state with the period equal
to $\lambda_f/2=2\pi v_F/h$. The lowest value of $d$ at which
$0-\pi$-transition takes place is approximately $\lambda_f/4$. The
temperature changes this picture only slightly, but near the
thickness corresponding the $0-\pi$-transition the non-monotonic
behavior of $j_c$ vs. temperature including temperature driven
$0-\pi$-transition can be found.\newline
  An intermediate case between the diffusion and clean limits was
considered by Bergeret {\it et al.} \cite{BVEJos}. They assumed
that the F-layer is so clean that $h\tau_f\gg 1$, whereas
$T_c\tau_s\ll 1$. Therefore Usadel equation is not valid for the
F-layer and they solved the Eilenberger equation. They have found
that the superconducting condensate oscillates as function of the
thickness with period $\lambda_f$ and penetrates into the F-layer
over the depth equal to the electron mean free path $l_f$. The
period of oscillations of the critical current is $\lambda_f/2$.
No qualitative differences with considered cases appear unless the
magnetization is inhomogeneous. Even very small inhomogeneity can
completely suppress the $0-\pi$-transitions. This is a consequence
of the generation of the triplet pairing, which will be considered
later.

\subsubsection{Half-integer Shapiro steps at the $0-\pi$ transition}
Recently Sellier {\it et al.} \cite{Sellier} have reported the
observation of the Shapiro steps at the voltage equal to
half-integer of the standard values $V_n=n\hbar\omega/2e$, where
$\omega$ is the frequency of the applied ac current. Let us remind
that the standard (integer) Shapiro steps appear as a consequence
of the resonance between the external ac field and the
time-dependent Josephson energy $E_J=-\frac{\hbar
j_c}{ed_f}cos\varphi(t)$ where the phase is proportional to time
due to external permanent voltage through the contact:
$\varphi(t)=2eVt/\hbar$. Just in the $0-\pi$ transition point
$j_c$ turns into zero. Then the next term in the Fourier-expansion
of the Josephson energy proportional to $\cos(2\varphi)$
dominates. That means that the Josephson current is proportional
to $\sin(2\varphi)$. Such a term leads to the Shapiro steps not
only at integer, but also at half-integer values since the
resonance now happens at $(4eV/\hbar)=\omega$. Normally the term
with $\sin(2\varphi)$ is so small that it was always assumed to
vanish completely. The resonance hf method used by the authors had
sufficient sensitivity to discover this term.\newline
 The authors prepared the
 $\textrm{Nb}/\textrm{Cu}_{52}\textrm{Ni}_{48}/\textrm{Nb}$ junction by the
photolitography method. Curie temperature of the F-layer is 20K.
The 2 samples they used had the thicknesses 17 and 19 nm. The
$0-\pi$ transition was driven by temperature. The transition
temperature in the first and second sample were 1.12 and 5.36K,
respectively. The external ac current had the frequency
$\omega=800$ kHz and amplitude about 18 $\mu$A. The voltage
current curves for $d_f=17nm$ and temperatures close to 1.12 K are
shown in figure  (\ref{shapiro}). The fact that the half-integer
steps disappear at very small deviation from the transition
temperature proves convincingly that it is associated with the
$0-\pi$ transition.

\begin{figure}[t]
\begin{center}
\includegraphics[width=2.5in]{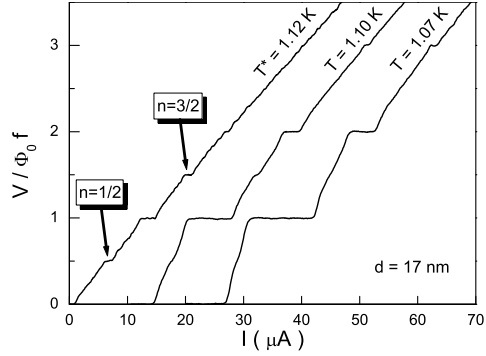}
\caption{\label{shapiro}
Shapiro steps in the voltage-current curve of a 17 nm
thick junction with an excitation at 800 kHz (amplitude about 18
$\mu$A). Half-integer steps (n=1/2 and n=3/2) appear at the
0\,--\,$\pi$ crossover temperature $T^*$. Curves at 1.10 and 1.07
K are shifted by 10 and 20 $\mu$A for clarity.
(From  Sellier {\it et al.} cond-mat/0406236).
}
\end{center}
\end{figure}

\subsubsection{Spontaneous current and flux in a closed loop}
Bulaevsky {\it et al.} \cite{BulKuz} argued that a closed loop
containing the $\pi$-junction may carry a spontaneous current and
flux in the ground state. Below we reproduce their arguments. The
energy of the closed superconducting loop depends on the total
flux $\Phi$ through the loop:
\begin{equation}\label{loop-en}
    E(\Phi)=-\frac{\hbar}{2e}J_c\cos\varphi
+\frac{\Phi_0^2\varphi^2}{8\pi^2Lc^2},
\end{equation}

where $\varphi=\frac{2\pi\Phi}{\Phi_0}$, $J_c$ is the critical
current and $L$ is the inductance of the loop. The first term in
equation (\ref{loop-en}) is the Josephson energy, the second is
the energy of magnetic field. The location of the energy minimum 
depends on the the parameter $k=\frac{\Phi_0}{4\pi LJ_cc}$. If $k$ 
is positive, there is only one minimum at $\varphi=0$. If $k<-1$, 
the only minimum is located again at $\varphi=0$. If $-1<k<0$, the 
minimum is located at the nonzero root of equation
$\sin\varphi/\varphi=|k|$; the value $\varphi=0$ corresponds to a
maximum of energy. Thus, the spontaneous flux appears at
sufficiently large inductance of the loop. It is possible to avoid
this limitation measuring the dependence of the current inside the
loop on the external flux through it \cite{ryazanov01-2}.

\begin{figure}
\begin{center}
\includegraphics[width=3.5in]{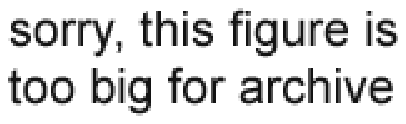}
%{ryazanov1.eps}
\caption{\label{triangular}Real (upper) 
and schematic (low) picture 
of the network of five SFS junctions 
$Nb-Cu_{0.46}Ni_{0.54}-Nb$ ($d_{F}=19$~nm), 
which was used in
the phase-sensitive experiment. 
(From From Ryazanov {\it et al.} cond-mat/0103240).
}
\end{center}
\end{figure}

They 
used triangular bridge array with $\pi$-junctions in each shoulder 
(see Figs. \ref{triangular}).
Due to the 
central $\pi$-junction the phases of the current in two sub-loops 
of the bridge differ by $\pi$. Therefore the critical current 
between the two contacts of the bridge is equal to zero in the 
absence of magnetic field. If the flux inside the loop reaches 
half of flux quantum, it compensates the indicated phase 
difference and the currents from both sub-loops are in the same 
phase. Thus, the shift of the current maximum from $\Phi=0$ to 
$\Phi=\Phi_0/2$ is the direct evidence of the $0-\pi$ transition.
Such experimental evidence was first obtained in the same work
\cite{ryazanov01-1}.

%\newpage
\begin{figure}
\begin{center}
\includegraphics[width=3.5in]{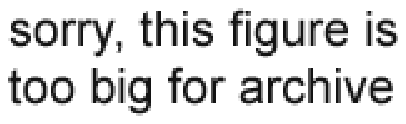}
%{ryazanov2.eps}
\caption{\label{I(H)}Magnetic field dependences of 
the critical transport current for the structure
depicted in figure  \ref{triangular} at temperature above (a) 
and below (b) $T_{cr}$.
(From From Ryazanov {\it et al.} cond-mat/0103240).
}
\end{center}
\end{figure}

The graphs of the current vs magnetic field for
two different temperatures (figure  \ref{I(H)}) clearly demonstrates
the shift of the current maximum from zero to non-zero magnetic
field. The next graph figure  \ref{Phi(T)} shows the shift of the
flux through the loop 0 to 1/2 of the flux quantum at the 
temperature driven $0-\pi$ transition.

\begin{figure}
\begin{center}
\includegraphics[width=3.5in]{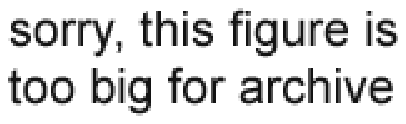}
%{ryazanov3.eps}
\caption{\label{Phi(T)}
(a) Temperature dependence of the critical transport current
for the structure depicted in figure  \ref{triangular}
in the absence of magnetic field; 
(b) temperature dependence (jump) of the position of the maximal 
peak on the curves $I_{m}(H)$, corresponding to the two limiting 
temperatures depicted in figure  \ref{I(H)}.
(From From Ryazanov {\it et al.} cond-mat/0103240).
}
\end{center}
\end{figure}

\subsection{F/S/F junctions}

The trilayers F/N/F (N is normal non-magnetic metal) have
attracted much attention starting from the discovery by Gr\"unberg
\cite{grunberg} of the Giant Magnetoresistance (GMR). The
direction of magnetization of ferromagnetic layers in these
systems may be either parallel or antiparallel in the ground state
oscillating with the thickness of the normal layer on the scale of
few nanometers. The mutual orientation can be changed from
antiparallel to parallel by a rather weak magnetic field.
Simultaneously the resistance changes by the relative value
reaching 50\%. This phenomenon has already obtained a
technological application in the magnetic transistors and valves
used in computers \cite{parkin}. A natural question is what
happens if the central layer is superconducting: will it produce
the spin-valve effect (a preferential mutual orientation of
F-layers magnetization) and how does it depend on thicknesses of S
and F-layers? This question was considered theoretically by
several authors \cite{deMelo, BuzVed, tagirov1, BalBuzVed,
BalBuzFSF}. Recently the spin-valve effect was experimentally
observed by Tagirov {\it et al.} \cite{tagirov-exp}.\newline
 Even without calculations it is clear that, independently on the
thicknesses of S and F layers, the antiparallel orientation of
magnetizations in F-layers has always lower energy than the
parallel one. It happens because the exchange field always
suppresses superconductivity. When the fields from different
layers are parallel, they enhance this effect and increase the
energy, and vice versa. The effect strongly depends on the
interfaces transparency. If it is very small, the effect is weak.
In the case of almost ideally transparent interfaces the majority
electrons with the preferential spin orientation can not penetrate
from the F-layer to the S-layer deeper than to the coherence
length $\xi_s$. Therefore, it is reasonable to work with the
S-layer whose thickness does not exceed $\xi_s$. The choice of the
material and thickness of F-layers is dictated by the requirement
that they could be reoriented by sufficiently weak magnetic field.
Thus, the coercive force must be small enough. We refer the reader
to the original works for quantitative details.\newline
 An alternative approach is to study the thermodynamics of the
F/S/F trilayer at a fixed mutual orientation of magnetic moments.
Such a study was performed by Baladie and Buzdin \cite{BalBuzFSF}
for the case of very thin superconducting layer $d_s\ll \xi_s$.
They considered $F_s$ almost as a constant, but incorporated small
linear and quadratic deviations and solved the linearized Usadel
equation as it was shown in subsection \ref{Tc} to find the
critical temperature vs. thickness of the ferromagnetic layers.
They have found that at large $\gamma_b$ (low interface
transparency) the transition temperature monotonically decreases
with $d_f$ increasing from its value in the absence of the
F-layers to some saturation value and there is no substantial
difference between parallel and antiparallel orientations. At
smaller values of $\gamma_b$ the suppression of $T_c$ increases
and at parallel orientation the reentrant transition occurs at
$d_f\sim\xi_f$, but still the transition temperature saturates at
large $d_f$. At $\gamma_b$ smaller than a critical value the
transition temperature becomes zero at a finite thickness $d_f$
for both parallel and antiparallel orientation. The authors also
have found some evidences that at low $\gamma_b$ the SC transition
becomes discontinuous for the parallel orientation. This
conclusion was confirmed by a recent theoretical study by Tollis
\cite{tollis}, who has proved that the SC transition for the
antiparallel orientation is always of the second order, whereas
for the parallel orientation it becomes of the first order for
small \cite{tollis}. Baladie and Buzdin \cite{BalBuzFSF} have
considered also the energy gap at low temperature. For the case of
thick ferromagnetic layers $d_f\gg\xi_f$ they have found that the
energy gap is the monotonically decreasing function of the
dimensionless collision frequency $(\tau_f\Delta_0)^{-1}$, where
$\Delta_0$ is the value of the energy gap in the absence of the
ferromagnetic layers. It turns into zero at
$(\tau_f\Delta_0)^{-1}=0.25$ for the parallel and 0.175 for the
antiparallel orientation.

\subsection{Triplet pairing}

If the direction of the magnetization in F-layer is inhomogeneous
due to a domain wall or artificially, the singlet Cooper pairs
penetrating into the F- from S-layer will be partly transformed
into the triplet pairs. This effect was first predicted by
Kadigrobov {\it et al.} \cite{KSJ} and by Bergeret {\it et al.}
\cite{BEV1}. The triplet pairs cannot penetrate to the
superconductors over the length larger than magnetic length
$l_m=\sqrt{D_f/h}$ (or $v_F/h$ for the clean ferromagnet), but in 
the ferromagnet they are neither exchange interaction nor the 
elastic scattering suppresses them. Therefore, they can penetrate 
over much longer distance $\xi_T=\sqrt{D_f/T}$. Even if the
triplet pairing is weak, it provides the long-range coupling
between two superconducting layers in a S/F/S junction. Moreover,
if the thickness $d_f$ exceeds $l_m$ significantly, only triplet
pairs survive at distances much larger than $l_m$ completely
changing the symmetry properties of the superconducting
condensate.\newline
 The exchange field rotating in the $y-z$-plane is 
naturally described by the operator in the spin space
$\hat{h}=h(\hat{\sigma}_3\cos\alpha+\hat{\sigma}_2\sin\alpha)$, 
where $h$ is a scalar function of coordinates, $\hat{\sigma}_2$ 
and $\hat{\sigma}_3$ are the Pauli matrices and the angle $\alpha$
is a function of coordinates. It is clear that the non-diagonal
part of $h$ flips one of spins of the pair transforming the
singlet into the triplet. It does not appear if the magnetization
is collinear ($\alpha=0$). To make things more explicit, let
consider the Usadel equation in the F-layer, i.e. equation
(\ref{usadel-exch}) of the Section (\ref{basic}). First we
simplify them by linearization, which is valid if either the
transparency of the interface barrier is small \cite{BEV2003}.
Then the condensate Green tensor $\check{f}$ in F-layer is small.
The linearized Usadel equation reads:
\begin{equation}\label{magn-linear}
    \frac{D_f}{2}\frac{\partial^2\check{f}}{\partial
    z^2}-|\omega|\check{f}+ih\left[\hat{\tau}_0\{\hat{\sigma}_3,\check{f}\cos\alpha
    +\hat{\tau}_3[\hat{\sigma}_2,\check{f}]\sin\alpha\}\right]=0,
\end{equation}
where $\{A,B\}$ means the anticommutator of operators $A$ and $B$.
If $\alpha =const$, equations (\ref{magn-linear}) have an 
exponential solution $\check{f}=e^{kz}\check{f}_0$. The secular
equation for $k$ is:
\begin{equation}\label{secular}
    (k^2-k_{\omega}^2)^2\left[(k^2-k_{\omega}^2)^2+\frac{2h}{D_f}\right]=0,
\end{equation}
where $k_{\omega}^2=2|\omega|/D_f$. Note that the secular equation 
does not depend on $\alpha$. It is a consequence of rotational
invariance of the exchange interaction. At $\alpha=0$ the two-fold
eigenvalue $k^2=k_{\omega}^2$ corresponds to $f_{1,2}$ (triplet 
pairing with projection $\pm 1$ onto the magnetic field). Since
$\omega_n$ is proportional to $T$, these modes are long-range. Two
other modes have wave vectors $k=k_h$ and $k=k_h^*$, where
$k_h^2=2(|\omega|+ih\textrm{sign}\omega)/D_f$. They penetrate not
deeper than on the magnetic length. These short-range modes are
linear combinations of the singlet and triplet with spin
projection zero, i.e. orthogonal to the magnetic field.\newline
  Bergeret {\it et al.} considered two different geometries. In
the first one \cite{BEV1} they considered S/F bilayer. The angle
$\alpha$ was a linear function of coordinate starting from 0 at
the S/F interface, reaching a value $\alpha_w$ at the distance $w$
from the interface and remaining constant at larger distances.
They have solved the linearized Usadel equation 
(\ref{magn-linear}) with the boundary condition
$\xi_f\frac{dF_f}{dz}=\gamma_bF_s$ proper at small transparency of
the interface by a clever unitary transformation
$\check{f}\rightarrow \hat{U}(z)\check{f}[\hat{U}(z)]^{-1}$ with
$\hat{U}(z)=\exp(iQ\hat{\sigma}_1z/2)$ and
$Q=\frac{d\alpha}{dz}=\frac{\alpha_w}{w}$. This transformation
turns the rotating magnetic field into the constant one, directed
along $z$-axis, but differential term generates perturbations
proportional to $Q$ and $Q^2$. By this trick the initial equations
with the coordinate dependent $\check{h}(z)$ is transformed into
an ordinary differential equation with constant (operator)
coefficients. The generation of the triplet component is weak if
$\gamma_b$ is large and it acquires an additional small factor if
the ratio $\xi_f/w$ is small ($w$ mimics the domain wall width),
but, as we have demonstrated, this component has a large
penetration depth. Experimentally it could produce a strong
enhancement of the F-layer conductivity. Such an enhancement was
observed in the experiment by Petrashov {\it et al.}\cite{petrashov}
in 1999, two years afore the theoretical works. They studied an
F/S bilayer made from 40nm thick Ni and 55nm thick Al films. The
interface was about 100x100nm$^2$. The samples were prepared by
$e$-beam lithography. They measured the resistivity and the
barrier resistance directly. They have found also the diffusion
coefficients $D_s=100cm^2/s$ and $D_f=10cm^2/s$, which we cite
here to give an idea about the order of magnitudes. They have
found a large drop of the resistance of the sample, which could
not be explained by the existing singlet pairing mechanism. We are
not aware about the detailed comparison of the theory \cite{BEV1}
and the experiment \cite{petrashov}. One more evidence of long
range penetration of the superconducting order parameter through
the ferromagnet was reported in \cite{giroud}. The authors
measured the resistance of 0.5$\mu m$ Ni loop connected with
superconducting Al wire. They extracted the decay length for
proximity effect in ferromagnet from differential resistance and
concluded that it is much larger than it could be expected for
singlet pairing.
\newline
\begin{figure}[t]
\begin{center}
\includegraphics[width=3.5in]{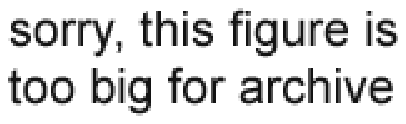}
%{6-layer.eps}
\caption{\label{6-layer} 6-layer structure.
}
\end{center}
\end{figure}

 In their second work on the triplet pairing \cite{BEV2003} the
authors have proposed an interesting 6-layer structure presented
in figure  \ref{6-layer}. The assume that the magnetization in each
layer is constant, but its direction is different in different
layers. It is supposed to lay in the $y-z$-plane and thus it can
characterized by one angle. Let this angle is $-\alpha$ in the
layer $F_1$, 0 in the layer $F_2$ and $\pm\alpha$ in the layer
$F_3$. They speak about the positive chirality if the sign is $+$
and negative chirality if the sign is $-$. They prove that, if the 
thickness of F-layers is larger than $l_m$, the superconducting
layers $S_A$ and $S_B$ are connected by 0-junction if the
chirality is positive and by $\pi$-junction if the chirality is
negative. This phenomenon is completely due to the triplet pairing
since it dominates on this distance.
 Kulic and Kulic \cite{kulic2} considered two bulk magnetic
superconductors with rotating magnetization separated by an
insulating layer. They also have found that the sign of the
Josephson current can be negative depending on the relative
chirality. In this system singlet and triplet pairs coexist in the
bulk, whereas in the system proposed by Bergeret {\it et al.} the
triplet dominates. We will give a brief description how did they
derive their results. They solved the Usadel equation in each
layer separately (it can be done without linearization, since the
coefficients of the differential equations are constant) and match
these solutions using the Kupriyanov-Lukichev boundary conditions.
The current density in the $F_2$ layer can be calculated using the
modified Eilenberger-Usadel expression:
\begin{equation}\label{current-mod}
    j=\sigma_f\textrm{Tr}
\left(\hat{\tau_3}\hat{\sigma_0}\pi
T\sum_{\omega_n}\check{f}\frac{d\check{f}}{dz}\right)
\end{equation}
The maximal effect is reached when magnetic moment of the central
layer is perpendicular to two others.

\section{Conclusions}
This short review shows that though the studies 
of Ferromagnet-Superconductor Hybrids are coming of age,
we are at the beginning of interesting voyage into this
emerging field. The most active development undoubtedly take
place in the field of proximity based phenomena in layered
ferromagnet-superconductor systems. The strong point of this
thrust is fruitful collaboration between experiment and 
theory. 
This progress was achieved due to a new idea due to Ryazanov and coworkers to 
use the weak ferromagnets in th experiment. This idea allowed to increase the 
thickness of ferromagnetic layers to a macrosopic scale and simultaneously 
allowing to drive the non-monotonous behavior of the Josephson current by 
temperature. On this way experimenters have reliably found several interesting 
phenomena predicted many years ago, as $0-\pi$-transition and oscillations of 
critical temperature vs. the thickness of the F-layer and also some new 
phenomena as the valve effect in F/S/F junction and the Shapiro steps at half-
integer frequencies.

The experimental studies of ordering/transport in FSH have greatly
benefited with introduction of imaging technique (SHPM,MFM) in the 
field. We expect that several experimental groups will get access 
to this technique in the near future which will result in more exciting
experiments. The theoretical and experimental studies of ordering/transport
in FSH have surprisingly little overlap, especially in comparison
with studies of proximity based phenomena. 
The materials used in the experiment are far from being regular, whereas the 
theorist so far preferred simple problems with regular, homogeneous or 
periodical systems. Even the simplest idea about topological instability in 
the S/F-bilayer was not checked experimentally. It would be very instructive 
to find experimentally the phase diagram of a single magnetic dot using the 
SQUID magnetometer or the MFM. Finally the transport properties of the S/F-
bilayer and the S-films supplied with regular or randomly magnetized arrays of 
F-dots should be measured.
On the other hand the experiment dictates new problems for theory: a 
description of random set of strongly pinned domain walls, their magnetic 
field and its effect on the S-films.
We think that both 
experimental and theoretical communities can find systems 
of common interests. Another  possibility for interesting 
development in the FSH field we expect with introduction of new
types of FSH, e.g. arrays of magnetic nanowires in alumina templates,
covered with superconducting film.
Such arrays provides alternative to magnetic dots source of 
alternating magnetic field of high strength and short scale 
variation.

\section{Acknowledgements}
The authors acknowledge the support by NSF under the grants
DMR-0103455 and DMR-0321572, by DOE under the grant
DE-FG03-96ER45598,  by Telecommunications and Informatics Task
Force at Texas A\&M University 
and by Deutsche Forschungsgemeinschaft. 
V.P. acknowledges the support from
the Humboldt Foundation, Germany. He is indebted to the University
of Cologne and to Prof. T. Nattermann for the hospitality during
his stay in Cologne, where a part of this work was performed. 
I.L. is grateful to Prof. H. Pfn\"ur, for kind hospitality
during stay at Hannover University, where part of the work
has been done.

{}

\end{document}